
%

\newcount\driver \driver=1          

\newcount\mgnf\newcount\tipi\newcount\tipoformule
\mgnf=1          
\tipi=2          
\tipoformule=1   
\ifnum\mgnf=0
   \magnification=\magstep0\hoffset=-0.5cm
   \voffset=-0.5truecm\hsize=15truecm\vsize=24.truecm
   \parindent=12.pt\fi
\ifnum\mgnf=1
   \magnification=\magstep1\hoffset=0.truecm
   \voffset=-0.5truecm\hsize=16.5truecm\vsize=24.truecm
   \baselineskip=14pt plus0.1pt minus0.1pt \parindent=12pt
   \lineskip=4pt\lineskiplimit=0.1pt      \parskip=0.1pt plus1pt\fi
%
%
\let\a=\alpha       \let\d=\delta \let\e=\varepsilon
\let\z=\zeta  \let\h=\eta   \let\th=\vartheta \let\l=\lambda
\let\m=\mu    \let\n=\nu    \let\x=\xi           \let\r=\rho
 \let\t=\tau    \let\ch=\chi
\let\ps=\psi  \let\o=\omega 
 \let\D=\Delta    
            
 \let\c=\chi

{\count255=\time\divide\count255 by 60 \xdef\oramin{\number\count255}
        \multiply\count255 by-60\advance\count255 by\time
   \xdef\oramin{\oramin:\ifnum\count255<10 0\fi\the\count255}}

\def\ora{\oramin }

\def\data{\number\day/\ifcase\month\or gennaio \or febbraio \or marzo \or
aprile \or maggio \or giugno \or luglio \or agosto \or settembre
\or ottobre \or novembre \or dicembre \fi/\number\year;\ \ora}

\setbox200\hbox{$\scriptscriptstyle \data $}

\newcount\pgn \pgn=1
\def\foglio{\number\numsec:\number\pgn
\global\advance\pgn by 1}
\def\foglioa{A\number\numsec:\number\pgn
\global\advance\pgn by 1}

%


\global\newcount\numsec\global\newcount\numfor
\global\newcount\numfig
\gdef\profonditastruttura{\dp\strutbox}
\def\senondefinito#1{\expandafter\ifx\csname#1\endcsname\relax}
\def\SIA #1,#2,#3 {\senondefinito{#1#2}
\expandafter\xdef\csname #1#2\endcsname{#3} \else
\write16{???? ma #1,#2 e' gia' stato definito !!!!} \fi}
\def\etichetta(#1){(\veroparagrafo.\veraformula)
\SIA e,#1,(\veroparagrafo.\veraformula)
 \global\advance\numfor by 1
 \write16{ EQ \equ(#1) == #1  }}
\def \FU(#1)#2{\SIA fu,#1,#2 }
\def\etichettaa(#1){(A\veroparagrafo.\veraformula)
 \SIA e,#1,(A\veroparagrafo.\veraformula)
 \global\advance\numfor by 1
 \write16{ EQ \equ(#1) == #1  }}
\def\getichetta(#1){Fig. \verafigura
 \SIA e,#1,{\verafigura}
 \global\advance\numfig by 1
 \write16{ Fig. \equ(#1) ha simbolo  #1  }}
\newdimen\gwidth
\def\BOZZA{
\def\alato(##1){
 {\vtop to \profonditastruttura{\baselineskip
 \profonditastruttura\vss
 \rlap{\kern-\hsize\kern-1.2truecm{$\scriptstyle##1$}}}}}
\def\galato(##1){ \gwidth=\hsize \divide\gwidth by 2
 {\vtop to \profonditastruttura{\baselineskip
 \profonditastruttura\vss
 \rlap{\kern-\gwidth\kern-1.2truecm{$\scriptstyle##1$}}}}}
\footline={\rlap{\hbox{\copy200}\ $\st[\number\pageno]$}\hss\tenrm
\foglio\hss}
}
\def\alato(#1){}
\def\galato(#1){}
\def\veroparagrafo{\number\numsec}\def\veraformula{\number\numfor}
\def\verafigura{\number\numfig}
\def\geq(#1){\getichetta(#1)\galato(#1)}
\def\Eq(#1){\eqno{\etichetta(#1)\alato(#1)}}
\def\eq(#1){\etichetta(#1)\alato(#1)}
\def\Eqa(#1){\eqno{\etichettaa(#1)\alato(#1)}}
\def\eqa(#1){\etichettaa(#1)\alato(#1)}
\def\eqv(#1){\senondefinito{fu#1}$\clubsuit$#1\write16{No translation for #1}%
\else\csname fu#1\endcsname\fi}
\def\equ(#1){\senondefinito{e#1}\eqv(#1)\else\csname e#1\endcsname\fi}

\ifnum\tipoformule=1\let\Eq=\eqno\def\eq{}\let\Eqa=\eqno\def\eqa{}
\def\equ{{}}\fi

\def\include#1{
\openin13=#1.aux \ifeof13 \relax \else
\input #1.aux \closein13 \fi}
\openin14=\jobname.aux \ifeof14 \relax \else
\input \jobname.aux \closein14 \fi
%
%
%
\newdimen\xshift \newdimen\xwidth
%
%
\def\ins#1#2#3{\vbox to0pt{\kern-#2 \hbox{\kern#1 #3}\vss}\nointerlineskip}
%
%
\def\insertplot#1#2#3#4{
    \par \xwidth=#1 \xshift=\hsize \advance\xshift
     by-\xwidth \divide\xshift by 2 \vbox{
  \line{} \hbox{ \hskip\xshift  \vbox to #2{\vfil
 \ifnum\driver=0 #3  
                 \special{ps: plotfile #4.ps} 
 \fi
 \ifnum\driver=1  #3    \includegraphics{#4.ps}       \fi
 \ifnum\driver=2  #3   \ifnum\mgnf=0
                       \special{#4.ps 1. 1. scale}\fi
                       \ifnum\mgnf=1
                       \special{#4.ps 1.2 1.2 scale}\fi\fi
 \ifnum\driver=3 \ifnum\mgnf=0
                 \psfig{figure=#4.ps,height=#2,width=#1,scale=1.}
                 \kern-\baselineskip #3\fi
                 \ifnum\mgnf=1
                 \psfig{figure=#4.ps,height=#2,width=#1,scale=1.2}
                 \kern-\baselineskip #3\fi
 \ifnum\driver=5  #3  \fi
\fi}
\hfil}}}

\newskip\ttglue
\def\TIPI{
\font\ottorm=cmr8   \font\ottoi=cmmi8
\font\ottosy=cmsy8  \font\ottobf=cmbx8
\font\ottott=cmtt8  
\font\ottoit=cmti8
\def \ottopunti{\def\rm{\fam0\ottorm}
\textfont0=\ottorm  \textfont1=\ottoi
\textfont2=\ottosy  \textfont3=\ottoit
\textfont4=\ottott
\textfont\itfam=\ottoit  \def\it{\fam\itfam\ottoit}%
\textfont\ttfam=\ottott  \def\tt{\fam\ttfam\ottott}%
\textfont\bffam=\ottobf
\normalbaselineskip=9pt\normalbaselines\rm}
\let\nota=\ottopunti}
\def\TIPIO{
\font\setterm=amr7 
\font\settesy=amsy7 \font\settebf=ambx7 
\def \settepunti{\def\rm{\fam0\setterm}
\textfont0=\setterm   
\textfont2=\settesy   
\textfont\bffam=\settebf  \def\bf{\fam\bffam\settebf}
\normalbaselineskip=9pt\normalbaselines\rm
}\let\nota=\settepunti}

\def\TIPITOT{
\font\twelverm=cmr12
\font\twelvei=cmmi12
\font\twelvesy=cmsy10 scaled\magstep1
\font\twelveex=cmex10 scaled\magstep1
\font\twelveit=cmti12
\font\twelvett=cmtt12
\font\twelvebf=cmbx12
\font\twelvesl=cmsl12
\font\ninerm=cmr9
\font\ninesy=cmsy9
\font\eightrm=cmr8
\font\eighti=cmmi8
\font\eightsy=cmsy8
\font\eightbf=cmbx8
\font\eighttt=cmtt8
\font\eightsl=cmsl8
\font\eightit=cmti8
\font\sixrm=cmr6
\font\sixbf=cmbx6
\font\sixi=cmmi6
\font\sixsy=cmsy6
\font\twelvetruecmr=cmr10 scaled\magstep1
\font\twelvetruecmsy=cmsy10 scaled\magstep1
\font\tentruecmr=cmr10
\font\tentruecmsy=cmsy10
\font\eighttruecmr=cmr8
\font\eighttruecmsy=cmsy8
\font\seventruecmr=cmr7
\font\seventruecmsy=cmsy7
\font\sixtruecmr=cmr6
\font\sixtruecmsy=cmsy6
\font\fivetruecmr=cmr5
\font\fivetruecmsy=cmsy5
\textfont\truecmr=\tentruecmr
\scriptfont\truecmr=\seventruecmr
\scriptscriptfont\truecmr=\fivetruecmr
\textfont\truecmsy=\tentruecmsy
\scriptfont\truecmsy=\seventruecmsy
\scriptscriptfont\truecmr=\fivetruecmr
\scriptscriptfont\truecmsy=\fivetruecmsy
\def \eightpoint{\def\rm{\fam0\eightrm}
\textfont0=\eightrm \scriptfont0=\sixrm \scriptscriptfont0=\fiverm
\textfont1=\eighti \scriptfont1=\sixi   \scriptscriptfont1=\fivei
\textfont2=\eightsy \scriptfont2=\sixsy   \scriptscriptfont2=\fivesy
\textfont3=\tenex \scriptfont3=\tenex   \scriptscriptfont3=\tenex
\textfont\itfam=\eightit  \def\it{\fam\itfam\eightit}%
\textfont\slfam=\eightsl  \def\sl{\fam\slfam\eightsl}%
\textfont\ttfam=\eighttt  \def\tt{\fam\ttfam\eighttt}%
\textfont\bffam=\eightbf  \scriptfont\bffam=\sixbf
\scriptscriptfont\bffam=\fivebf  \def\bf{\fam\bffam\eightbf}%
\tt \ttglue=.5em plus.25em minus.15em
\setbox\strutbox=\hbox{\vrule height7pt depth2pt width0pt}%
\normalbaselineskip=9pt
\let\sc=\sixrm  \let\big=\eightbig  \normalbaselines\rm
\textfont\truecmr=\eighttruecmr
\scriptfont\truecmr=\sixtruecmr
\scriptscriptfont\truecmr=\fivetruecmr
\textfont\truecmsy=\eighttruecmsy
\scriptfont\truecmsy=\sixtruecmsy
}\let\nota=\eightpoint}

\newfam\msbfam   
\newfam\truecmr  
\newfam\truecmsy 
\newskip\ttglue
\ifnum\tipi=0\TIPIO \else\ifnum\tipi=1 \TIPI\else \TIPITOT\fi\fi

\def\didascalia#1{\vbox{\nota\0#1\hfill}\vskip0.3truecm}

%
\def\V#1{\vec#1}
\def\T#1{#1\kern-4pt\lower9pt\hbox{$\widetilde{}$}\kern4pt{}}

\let\dpr=\partial\let\io=\infty\let\ig=\int

\let\0=\noindent

\def\guida{\leaders\hbox to 1em{\hss.\hss}\hfill}
\def\tende#1{\vtop{\ialign{##\crcr\rightarrowfill\crcr
              \noalign{\kern-1pt\nointerlineskip}
              \hglue3.pt${\scriptstyle #1}$\hglue3.pt\crcr}}}
\def\otto{{\kern-1.truept\leftarrow\kern-5.truept\to\kern-1.truept}}

\def\pagina{\vfill\eject}\def\acapo{\hfill\break}
\def\qed{\raise1pt\hbox{\vrule height5pt width5pt depth0pt}}

\let\ciao=\bye

\def\etc{\hbox{\sl etc}}\def\eg{\hbox{\sl e.g.\ }}

\def\ie{\hbox{\sl i.e.\ }}


\def\AA{{\V A}}\def\aa{{\V\a}}\def\nn{{\V\n}}\def\oo{{\V\o}}

\def\nn{{\V\n}}
\def\NN{{\cal N}}\def\FF{{\cal F}}\def\VV{{\cal V}}
\def\CC{{\cal C}}\def\RR{{\cal R}}\def\LL{{\cal L}}\def\BB{{\cal B}}
\def\TT{{\cal T}}\def\MM{{\cal M}}\def\WW{{\cal W}}\def\GG{{\cal G}}
\def\DD{{\cal D}}
\def\={{ \; \equiv \; }}
\def\HH{{\cal H}}\def\PP{{\cal P}}
\let\ot=\leftarrow

\def\pps{{\V\ps{\,}}}


\font\cs=cmcsc10
\font\ss=cmss10
\font\sss=cmss8
\font\crs=cmbx8

\ifnum\mgnf=0
\def\openone{\leavevmode\hbox{\ninerm 1\kern-3.3pt\tenrm1}}%
\fi
\ifnum\mgnf=1
\def\openone{\leavevmode\hbox{\ninerm 1\kern-3.6pt\tenrm1}}%
\fi

\ifnum\mgnf=0
\def\ZZZ{\hbox{{\ss Z}\kern-3.3pt{\ss Z}}}
\def\zzz{\hbox{{\sss Z}\kern-3.3pt{\sss Z}}}
\fi
\ifnum\mgnf=1
\def\ZZZ{\hbox{{\ss Z}\kern-3.6pt{\ss Z}}}
\def\zzz{\hbox{{\sss Z}\kern-3.6pt{\sss Z}}}
\fi

\ifnum\mgnf=0
\def\RRR{\hbox{I\kern-1.4pt{\ss R}}}
\def\rrr{\hbox{{\crs I}\kern-1.3pt{\sss R}}}
\fi
\ifnum\mgnf=1
\def\RRR{\hbox{I\kern-1.7pt{\ss R}}}
\def\rrr{\hbox{{\crs I}\kern-1.6pt{\sss R}}}
\fi

\ifnum\mgnf=0
\def\TTT{\hbox{{\ss T}\kern-5.0pt{\ss T}}}
\def\ttt{\hbox{{\sss T}\kern-4.2pt{\sss T}}}
\fi
\ifnum\mgnf=1
\def\TTT{\hbox{{\ss T}\kern-5.3pt{\ss T}}}
\def\ttt{\hbox{{\sss T}\kern-3.9pt{\sss T}}}
\fi

\ifnum\mgnf=0
\def\NNN{\hbox{I\kern-1.4pt{\ss N}}}
\def\nnn{\hbox{{\crs I}\kern-1.3pt{\sss N}}}
\fi
\ifnum\mgnf=1
\def\NNN{\hbox{I\kern-1.7pt{\ss N}}}
\def\nnn{\hbox{{\crs I}\kern-1.6pt{\sss N}}}
\fi


\def\2{{1\over2}}
\def\OO{{\cal O}}

\def\st{\scriptscriptstyle}
\let\\=\noindent

\def\*{\vskip0.3truecm}


\catcode`\%=12\catcode`\}=12\catcode`\{=12
              \catcode`\<=1\catcode`\>=2

\openout13=fig1.ps
\write13<gsave>
\write13</punto { gsave 2 0 360 newpath arc fill stroke grestore} def>
\write13<0 100 punto     >
\write13<70 100 punto    >
\write13<120 70 punto   >
\write13<160 140 punto  >
\write13<200 120 punto  >
\write13<240 180 punto  >
\write13<240 140 punto  >
\write13<240 100 punto   >
\write13<240 10 punto    >
\write13<240 40 punto   >
\write13<210 80 punto   >
\write13<240 80 punto   >
\write13<240 60 punto   >
\write13<0 100 moveto 70 100 lineto>
\write13<70 100 moveto 120 70 lineto>
\write13<70 100 moveto 160 140 lineto>
\write13<160 140 moveto 200 120 lineto>
\write13<160 140 moveto 240 180 lineto>
\write13<200 120 moveto 240 140 lineto>
\write13<200 120 moveto 240 100 lineto>
\write13<120 70 moveto 240 10 lineto>
\write13<120 70 moveto 240 40 lineto>
\write13<120 70 moveto 210 80 lineto>
\write13<210 80 moveto 240 80 lineto>
\write13<210 80 moveto 240 60 lineto>
\write13<stroke>
\write13<grestore>
\closeout13
\catcode`\%=14\catcode`\{=1
\catcode`\}=2\catcode`\<=12\catcode`\>=12



\catcode`\%=12\catcode`\}=12\catcode`\{=12
              \catcode`\<=1\catcode`\>=2

\openout13=fig2.ps
\write13<gsave>
\write13</punto { gsave 2 0 360 newpath arc fill stroke grestore} def>
\write13<0 40 punto     >
\write13<40 40 punto    >
\write13<60 55 punto   >
\write13<80 10 punto  >
\write13<80 40 punto  >
\write13<80 70 punto  >
\write13<140 40 punto     >
\write13<180 40 punto    >
\write13<200 25 punto   >
\write13<220 10 punto  >
\write13<220 40 punto  >
\write13<220 70 punto  >
\write13<0 40 moveto 40 40 lineto>
\write13<40 40 moveto 60 55 lineto>
\write13<40 40 moveto 80 10 lineto>
\write13<60 55 moveto 80 40 lineto>
\write13<60 55 moveto 80 70 lineto>
\write13<140 40 moveto 180 40 lineto>
\write13<180 40 moveto 220 70 lineto>
\write13<180 40 moveto 200 25 lineto>
\write13<200 25 moveto 220 10 lineto>
\write13<200 25 moveto 220 40 lineto>
\write13<stroke>
\write13<grestore>
\closeout13
\catcode`\%=14\catcode`\{=1
\catcode`\}=2\catcode`\<=12\catcode`\>=12


%
\catcode`\%=12\catcode`\}=12\catcode`\{=12
              \catcode`\<=1\catcode`\>=2

\openout13=fig3.ps
\write13<
\write13<
\write13</punto { gsave 2 0 360 newpath arc fill stroke grestore} def>
\write13</cerchio { gsave 0 360 newpath arc stroke grestore} def>
\write13<240 90 80 cerchio >
\write13<270 60 30 cerchio >
\write13<240 130 35 cerchio >
\write13<245 135 25 cerchio >
\write13<120 150 40 cerchio >
\write13<165 90 165 cerchio >
\write13<100 65 42 cerchio >
\write13<10 90 punto     >
\write13<50 90 punto    >
\write13<75 90 punto   >
\write13<130 180 punto  >
\write13<90 135 punto  >
\write13<105 135 punto  >
\write13<140 150 punto  >
\write13<140 135 punto   >
\write13<140 120 punto    >
\write13<125 40 punto   >
\write13<255 146 punto   >
\write13<235 137 punto   >
\write13<175 90 punto   >
\write13<215 118 punto   >
\write13<285 40 punto   >
\write13<252 45 punto   >
\write13<280 60 punto   >
\write13<275 80 punto   >
\write13<225 106 punto   >
\write13<255 128 punto   >
\write13<10 90 moveto 50 90 lineto>
\write13<50 90 moveto 75 90 lineto>
\write13<50 90 moveto 90 135 lineto>
\write13<90 135 moveto 130 180 lineto>
\write13<90 135 moveto 105 135 lineto>
\write13<105 135 moveto 140 150 lineto>
\write13<105 135 moveto 140 135 lineto>
\write13<105 135 moveto 140 120 lineto>
\write13<75 90 moveto 125 40 lineto>
\write13<75 90 moveto 175 90 lineto>
\write13<175 90 moveto 215 118 lineto>
\write13<175 90 moveto 252 45 lineto>
\write13<215 118 moveto 235 137 lineto>
\write13<235 137 moveto 255 146 lineto>
\write13<215 118 moveto 225 106 lineto>
\write13<252 45 moveto 285 40 lineto>
\write13<252 45 moveto 280 60 lineto>
\write13<252 45 moveto 275 80 lineto>
\write13<235 137 moveto 255 128 lineto>
\write13<stroke>
\closeout13
\catcode`\%=14\catcode`\{=1
\catcode`\}=2\catcode`\<=12\catcode`\>=12
%


\catcode`\%=12\catcode`\}=12\catcode`\{=12
              \catcode`\<=1\catcode`\>=2

\openout13=fig4.ps
\write13<gsave>
\write13</punto { gsave 2 0 360 newpath arc fill stroke grestore} def>
\write13</cerchio { gsave 0 360 newpath arc stroke grestore} def>
\write13<40 30 30 cerchio >
\write13<140 30 30 cerchio >
\write13<240 30 30 cerchio >
\write13<340 30 30 cerchio >
\write13<0 50 punto     >
\write13<40 50 punto    >
\write13<40 10 punto   >
\write13<80 10 punto  >
\write13<100 50 punto     >
\write13<140 50 punto    >
\write13<140 10 punto   >
\write13<180 50 punto  >
\write13<200 10 punto     >
\write13<240 50 punto    >
\write13<240 10 punto   >
\write13<280 50 punto  >
\write13<300 10 punto     >
\write13<340 50 punto    >
\write13<340 10 punto   >
\write13<380 10 punto  >
\write13<0 50 moveto 40 50 lineto>
\write13<40 50 moveto 40 10 lineto>
\write13<40 10 moveto 80 10 lineto>
\write13<100 50 moveto 140 50 lineto>
\write13<140 50 moveto 140 10 lineto>
\write13<140 50 moveto 180 50 lineto>
\write13<200 10 moveto 240 10 lineto>
\write13<240 50 moveto 240 10 lineto>
\write13<240 50 moveto 280 50 lineto>
\write13<300 10 moveto 340 10 lineto>
\write13<340 50 moveto 340 10 lineto>
\write13<340 10 moveto 380 10 lineto>
\write13<stroke>
\write13<grestore>
\closeout13
\catcode`\%=14\catcode`\{=1
\catcode`\}=2\catcode`\<=12\catcode`\>=12





\vglue0.truecm


\font\titolo=cmbx12
\font\titolone=cmbx12 scaled\magstep 2
 3


\centerline{\titolone Methods for the analysis}
\centerline{\titolone of the Lindstedt series for KAM tori}
\vskip.2truecm
\centerline{\titolone and renormalizability in classical mechanics}
\vskip.1truecm
\centerline{\titolo A review with some applications}
%
\vskip1.truecm
\centerline{{\titolo G. Gentile}$^{\dagger}$,
{\titolo V. Mastropietro}$^{\dagger\dagger}$}

\vskip.2truecm

\0$^{\dagger}$ IHES, 35 Route de Chartres, 91440 Bures sur Yvette, France.

\0$^{\dagger\dagger}$
Dipartimento di Matematica, Universit\`a di Roma II
``Tor Vergata", 00133 Roma, Italia.
\*
\0{\cs Abstract:} {\it
This paper consists in a unified exposition of
methods and techniques of the renormalization group approach
to quantum field theory applied to classical mechanics,
and in a review of results:
(1) a proof of the KAM theorem,
by studing the perturbative expansion (Lindstedt series) for the
formal solution of the equations of motion;
(2) a proof of a conjecture by Gallavotti about the
renormalizability of isochronous hamiltonians,
\ie the possibility to add a term depending only on the actions
in a hamiltonian function not verifying the anisochrony condition
so that the resulting hamiltonian is integrable.
Such results were obtained first by Eliasson;
however the difficulties arising in the study of the perturbative
series are very similar to the problems which one has to deal with
in quantum field theory, so that the use the methods which have been
envisaged and developed in the last twenty years exactly in order to
solve them allows us to obtain unified proofs, both conceptually
and technically.
In the final part of the review, the original work of Eliasson
is analyzed and exposed in detail; its connection with other proofs
of the KAM theorem based on his method is elucidated.}
\*
\0{\sl Keywords}: {\it Classical mechanics, KAM theorem,
quantum field theory, renormalization group, multiscale analysis,
tree expansion, counterterms, cancellations}
\vskip1.truecm

\line{\vtop{
\line{\hskip7.truecm\vbox{\advance \hsize by -7.1 truecm
\it Solomon saith: {\rm There is no new thing upon the earth.}
So that as Plato had an imagination, {\rm that all knowledge was
but remembrance}; so Solomon giveth his sentence, {\rm that all novelty
is but oblivion.}}  \hfill}
\vskip.1truecm
\line{\hskip7.truecm \hfil\hfil
Francis Bacon, [Ba]\hfil} }}

\vskip1.truecm


\centerline{\titolo 1. Introduction}
\*\numsec=1\numfor=1\pgn=1

\\Quasiperiodic solutions of the equations of motion obtained
under small perturbations of integrable hamiltonian systems can be given
formal perturbative expansions, which are known as
{\sl Lindstedt series}, [L]. The existence problem
for such series was studied by Poincar\'e, [P], Vol. II, Ch. XIII, who
proved that, {\it in general}, the series diverge
({\sl Poincar\'e's triviality theorem}); he was not able to
exclude that under suitable hypotheses on the unperturbed
solution the convergence would occur, although
he found such a possibility very unlikely and
advanced the conjecture that the series could never converge,
(see in particular Vol. II, Ch. XIII, \S 149).
Doubts on Poincar\'e's conjecture were raised from Weierstrass, [W].

On the contrary the KAM theorem, stating the conservation
of quasiperiodic motions for a large class
of hamiltonian systems and initial data, proves the
convergence of the Lindstedt series.

The original proof is due to Kolmogorov, [K],
and to Arnol$'$d, [A1], [A2], who extended the theorem to cover cases
relevant for the three (or $N$) body problem.
It is based on a convergent iterative technique,
which proves the existence of a canonical transformation
conjugating, analytically in the perturbative parameter,
the motion to a simple rotation on the torus: the tecnique of proof
easily implies the convergence of the Lindstedt series.

Moser proved the corresponding theorem in the case of non analytic
(small enough) perturbation: an extension requiring deep new ideas
and tools (like Nash's implicit function theorem, [N]).
In [M1] the theorem was proven for $C^p$ area-preserving mappings
in the plane, with $p \ge 333$, and improvements were provided
later, [R2], [M5]. The extension of the theorem for
hamiltonian systems to the differentiable case ($p > 2 + 2(\t+1)
> 2(\ell+1)$, where $\t>\ell-1$ is the
diophantine constant in (1.3) below and $\ell$ is
the number of degrees of freedom) was discussed in [M2], [M4].

Only in recent times, a new proof of the convergence was provided by
Eliasson, [E1], [E3], by studying directly the Lindstedt series
and proving the existence of cancellations to all perturbative orders.
However Eliasson's work has not enjoyed a wide diffusion and
understanding as it would have deserved.
This led to a sequence of works presenting
technically and conceptually simplified proofs.
In [G7], [GM1] and [CF1] simplified models are studied,
following the attempt by Thirring, [T], Vol. 1,
Ch. 3, \S 3.6, to find a model which allows us to make
easier to explain the KAM theorem.

Our starting point is the analogy of the Lindstedt series
with the perturbative series in constructive quantum field
theory, pointed out in [FT] and [G9]. It was carried out to a
deeper extent in [GGM], where a quantum field model
is explicitly exhibited whose Feynman's graphs
are exactly the same diagrams which correspond to a
natural graphical representation of the Lindstedt series.
It is then natural to prove
the convergence of the Lindstedt series by
exploiting techniques usual in the renormalization
group approach to the quantum field theory,
like the multiscale decomposition of the propagators, the
tree expansion and the introduction of counter\-terms whose
value has to be uniquely fixed in order to make the problem
soluble, (see [G5] for a review).
A similar analysis is introduced also in [G8], [Ge1] and [Ge2]
in order to study the persistence of the stable and unstable manifolds
({\sl whiskers}) of low dimensional tori in a class
of almost integrable systems; in particular, in [Ge2], the introduction
of suitable counterterms in the hamiltonian is found to be a useful
device in order to simplify in a relevant way the proof of existence
of low dimensional tori.

In this work we extend the ideas and results
contained in [G7], [GG], [GM1] and [GM2], so recovering
completely the (analytic) KAM theorem. This means that the only
assumptions on the hamiltonian function will be: (1)
the anisochrony condition, and (2) the diophantine
property.\footnote{${}^1$}{\nota
In [CF2] the general case of the KAM theorem is discussed,
by describing explicitly the cancellations,
sketching the strategy one has to follow to complete
the proof (with the hint that it can be adapted from [CF1]).}

There are interesting technical advantages in using the methods
of quantum field theory, with respect
to [E1], [E2], [E3], [CF1] and [CF2]. In fact some
problems like the {\sl overlapping divergences} and the approximate
cancellations of the resonances are automatically bypassed,
so that there is no need to distinguish
between contributions which really require a bound
improvement and contributions which can only
apparently raise problems, (as it has been done first in [E1],
where one is led to define
several kinds of resonances, such that only
the ``critical'' ones are ``dangerous''). This is a standard
feature of the techniques used in
quantum field theory, and shows that such an approach is
``very natural'' in order to attack the problem. These and
other technical improvements are discussed further and
with more details in the next sections.

Besides the mathematical interest and motivation,
the ultimate hope is that the newly introduced methods
will allow us to solve open problems of the ``KAM theory".
In [GM2] for instance we perform a resummation
in the Lindstedt series, which is used in [GGM]
in the heuristic analysis of the universality of the breakdown
phenomenon for KAM invariant tori via the study of the
singularities of the Lindstedt series.

A further application is provided in the present paper:
if the free hamiltonian is isochronous (\ie it describes
a system of harmonic oscillators) and its frequencies
satisfy the diophantine
property, then the
Lindstedt series is not convergent. In [G2] Gallavotti
advanced the conjecture that in such a case,
if a suitable ``counterterm", analytic in the
perturbative parameter and depending only on the
action variables,
is added to the hamiltonian, then the modified hamiltonian
is integrable. Here we show that the validity of such a conjecture
is a byproduct of our method by proving that
the Lindstedt series for the modified model is convergent.

In [R1], R\"ussmann proved that, if the counterterms make
the equations of motion {\it formally} soluble, then
there there exists an analytic solution, (\ie if the
series can be formally defined, then it converges), a result
reproduced by Gallavotti, in [G2], with a different
formalism; however the problem of the formal solubility
was left unsolved.
A partial result about the conjecture
is implied by the papers of
Dinaburg and Sinai, [DS], and R\"ussmann, [R3], who proved the
conjecture to hold for special
interactions of the form $\e\AA\cdot\V f(\aa)$
(in the notations of this paper, see \S 1.1),
by a method of variation of constants, inherited from [M3]:
in fact they studied
the one-dimensional Schr\"odinger equation with a quasi-periodic
potential, but the problem can be shown to be equivalent
to a hamiltonian problem in classical mechanics of oscillators
interacting via a potential linear in the action variables,
(see also [G6]).

Gallavotti's conjecture has been first proven
in the general case also by Eliasson in
a work, [E2], which apparently had the same reception problems of the
quoted ones.
Another proof with different tools is presented in [EV],
where the notions of {\sl mould} and {\sl arborification},
introduced by Ecalle, are used in order to prove the
analyticity of the ``correction'' of any resonant
local analytic field, (the hamiltonian case being included),
under the so called {\sl Bryuno's diophantine
condition}, which is weaker than the property \equ(1.3).

In the remaining part of the introduction,
we give a more formal statement of the results which will be proven
in next sections, and introduce the basic notations.

\*

\\{\bf 1.1.} The hamiltonian function is
$$\HH(\aa,\AA) = \HH_0(\AA) +f(\aa,\AA;\e) \; , \Eq(1.1)$$
where (1) $\HH_0(\AA)$ and $f(\aa,\AA;\e)$ are analytic functions
in $\AA$, in a domain $D\subset {\RRR}^{\ell}$,
and in $\aa$, for $\hbox{Re}\,\aa\in{\TTT}^\ell$ and
$|\hbox{Im}\,\aa|<\x$ for some positive constant $\x$,
and (3) $f(\aa,\AA;\e)$ is analytical in $\e$, for $|\e|\le \e_1$,
and divisible by $\e$, (see (A2.1) below).

The {\sl free hamiltonian} $\HH_0(\AA)$ satisfies the {\sl anisochrony
condition} (or {\sl twist condition})
$$ \det \left( \dpr_{A_i}\dpr_{A_j}\HH_0(\AA_0)
\right) \neq 0 \; , \Eq(1.2) $$
where $\AA_0\in D$,
and the vector $\oo_0\=\dpr_\AA \HH_0(\AA_0)$ verifies the
{\sl diophantine property} with diophantine constants $C_0,\t>0$;
this means that
$$ C_0|\oo_0\cdot\nn|\ge |\nn|^{-\t} \; ,
\kern2.5cm\V0\ne\nn\in {\ZZZ}^\ell \; , \Eq(1.3) $$
where $\oo_0\cdot\nn=\sum_{j=1}^\ell\oo_{0j}\nu_j$ is the
scalar product in ${\RRR}^\ell$.
It is easy to see that the {\sl diophantine vectors} have
full measure in ${\RRR}^\ell$ if $\t$ is fixed
$\t>\ell-1$.

If $J_j^{-1}$, $j=1,\ldots,\ell$, are
the eigenvalues of the matrix $T(\AA_0)\=\dpr_\AA\dpr_\AA\HH_0(\AA_0)$,
such that $0<J_j^{-1}<\io$, $\forall$ $j=1,\ldots,\ell$,
(because of the twist condition), we define
$$ J_m=\min_{j=1,\ldots,\ell} J_j \; , \qquad
J_M=\max_{j=1,\ldots,\ell} J_j \; , \Eq(1.4) $$
(note that, if $\HH_0(\AA)$ is quadratic in the actions, then
the $J_j$'s are the principal momenta of inertia),
and, because of the analyticity assumption on the interaction
potential, we can write
$$ f(\aa,\AA;\e) = \sum_{\nn\in{\zzz}^\ell} e^{i\nn\cdot\aa}
\,f_\nn(\AA;\e)
\, , \kern2.5cm |f_\nn(\AA;\e)| \le F\,e^{-\x|\nn|} \; , \Eq(1.5) $$
with $\x$ defined after \equ(1.1) and
$$ F= \sup_{|\e|\le \e_1}\sup_{\AA\in W(\AA_0,\r_0)}
\sup_{\aa\in{\ttt}^{\ell}} \{ f(\aa,\AA;\e) \} \; . \Eq(1.6) $$
where $W(\AA_0,\r_0)=\{ \AA \in {\RRR}^{\ell} \; : \;
|\AA-\AA_0| \le \r_0\} \subset D$.
Finally we define
$$ E_0=\sup_{\AA\in W(\AA_0,\r_0)} \{ \HH_0(\AA) \} \, ,
\qquad E=\max\{ F\e_1^{-1} \, , \, E_0 \} \; , \Eq(1.7) $$
and introduce the dimensionless frequency $\oo=C_0\oo_0$,
and the dimensionless analyticity radius $\r=(C_0/J_m)\r_0$,
and set $r=\min\{1,\r^2\}$.

\*

In this paper we give a proof of the following result,
by verifying the convergence of the Lindstedt
series.

\*

\\{\bf 1.2.} {\cs Theorem (Kolmogorov-Arnol$'$d-Moser's theorem).}
{\it The hamiltonian model \equ(1.1), with
$\HH_0(\AA)$ verifying the anisochrony
condition \equ(1.2) and $\oo_0=\dpr_\AA \HH_0(\AA_0)$
satisfying the diophantine property \equ(1.3), admits
an $\e$-analytic family of motions starting at
$\aa(0)=\V0$ and having the form
$$ \AA(t)=\AA_0+\V H(\AA_0,\oo_0t;\e) \; , \qquad\qquad
\aa(t)=\oo_0t+\V h(\AA_0,\oo_0 t;\e) \; , \Eq(1.8) $$
where:
\acapo
(1) $\V H(\AA,\pps;\e)$, $\V h(\AA,\pps;\e)$ are analytic
in $\pps$ with} $\hbox{Re}\pps\in {\TTT}^\ell$, {\it and}
$|\hbox{Im}\pps|<\x$, {\it and in $\AA\in W(\AA_0,\r_0)$;
\acapo
(2) $\V h(\AA,\pps;\e)$ has vanishing average in ${\TTT}^\ell$;
\acapo
(3) $\V H(\AA_0,\pps;\e)$ and $\V h(\AA_0,\pps;\e)$ are
analytic for $|\e|<\e_0$
with a suitable $\e_0$ close to $0$:
$$ \e_0 = \h_0 \, [J_M\,J_m^{-2}\,C_0^2\,E\,
r^{-1} ]^{-2} \; , \Eq(1.9) $$
with $\h_0$ being a dimensionless quantity depending
only on $\ell$, $\x$ and $\t$.
This means that the set $(\AA(t),\aa(t))$
described by replacing $\oo_0 t$ in (1.8) with $\pps\in {\TTT}^\ell$
is, for $\e$ small enough,
an analytic invariant torus for \equ(1.1), which is run
quasi periodically with angular velocity vector $\oo_0$, and
coincides, for $\e=0$, with the torus
$\AA(t)=\AA_0$, $\aa(t)=\oo_0t\in {\TTT}^\ell$.}

\*

The condition in item (2) is imposed
in order to simplify the analysis of the
equations of motion, and of the recursive relations defining them.

The dependence on $\ell$ in \equ(1.9) is a factorial to some
negative power, (see comments between (5.8) and (5.9)
in \S 5), similar to the estimates derived
in the usual KAM proofs, which is of the form $(\ell !)^{-a}$,
for some constant $a>0$, (see, \eg, [G3]).

\*

\\{\bf 1.3.} In a paper by Gallavotti, [G2], the question was raised if,
given a perturbation of a free {\sl isochronous}
hamiltonian not satisfying the anisochrony condition,
it is possible to ``renormalize'' it by adding to it a function
analytic in the perturbative parameter,
in such a way that the
resulting hamiltonian turns out to be integrable.

More formally let us consider
a hamiltonian of the form \equ(1.1) with $\HH_0'\neq0$,
(if we denote by $\HH_0'$ the derivative of $\HH_0$ with
respect to its argument) and $\oo_0$ being a diophantine vector,
(see \equ(1.3)). All the functions are supposed to be analytic in
their arguments in suitable domains, as in \S 1.1.

We know, from trivial examples in [G1], [G4], (see (1.13) below),
or from the {\sl genericity theorem}, [S2], [MS],
that in such a case the hamiltonian $\HH$ is not
integrable in general. Nevertheless we can ask ourselves if it
is possible to add to $\HH$ a function $N_{f}$ depending
analytically on $\AA$ and $\e$ (in a suitable domain to
be found), $\pps$-independent and such that
the hamiltonian $\HH-N_{f}$ is integrable.

If it is so, by analogy with the renomalization problem
in quantum field theory and following [G2], we can define
``renormalized hamiltonian'' the
function $\HH-N_{f}$, and we can set
$$ :f: \; = f - N_{f} \; , $$
by calling the operator $:\,:$ so introduced the ``Wick ordering
with respect to $\HH_0$''.

Let us take the hamiltonian function as in \equ(1.1),
and choose the ``free hamiltonian'' of the form
$\HH_0(\AA)\=\oo_0\cdot\AA$, and $\oo_0$ satisfying the
diophantine property.

Then we consider the ``renormalized'' hamiltonian
$$\HH(\aa,\AA) = \oo_0\cdot\AA + f(\aa,\AA;\e) -
N_f(\AA;\e) \; , \Eq(1.10) $$
and we look for a function $N_f(\AA;\e)$ such that the system
described by the hamiltonian \equ(1.10) is integrable, \ie
admits quasiperiodic motions. We will call {\sl counterterm}
the function $N_f(\AA;\e)$.
This can be regarded as a ``control theory theorem''
and it might to have some applications in
the costruction of stabilizing devices:
for instance one is interested to have
persistence of tori in stellarators used for plasma confinement
via the application of toroidal magnetic fields, (see, \eg,
[SHS], [HC]). Of course we expect that there are
perturbations in the motion of the particles which
should generate chaotic motions, but the following results shows
that by tuning in a suitable way proper
magnetic fields the motion could remain stable.

\*

The following result, conjectured in [G2] and proven in [E2],
[EV], will be shown to fit into our general scheme and proven again.

\*

\\{\bf 1.4.} {\cs Theorem.} {\it Given the hamiltonian \equ(1.1),
where $\HH_0(\AA)=\oo_0\cdot\AA$, with the vector
$\oo_0$ verifying the condition \equ(1.3), and $f(\aa,\AA;\e)$
is analytic in $\e$, with $|\e|\le\e_1$, for
some positive constant $\e_1$, and
$$ f(\aa,\AA;\e) = \sum_{\nn\in{\zzz}^\ell} e^{i\nn\cdot\aa}\,
f_\nn(\AA;\e) \; , \qquad \sup_{|\e|\le\e_1} \sup_{\AA\in D}
|f_\nn(\AA;\e)| < F\,e^{-\x|\nn|} \; , $$
then it is possible
to fix a function $N_f(\AA;\e)$ such that the hamiltonian
\equ(1.10) admits a family of motions starting at $\aa(0)=\V0$
and having the form
$$ \AA(t) = \AA_0 + \V H (\AA_0,\oo_0t;\e) \; , \qquad
\aa(t) = \oo_0t + \V h(\AA_0,\oo_0t;\e) \; , \Eq(1.11) $$
where:
\acapo
(1) $\V H(\AA,\pps;\e)$ and $\V h(\AA,\pps;\e)$ are analytic in $\pps$
with} $\hbox{Re}\pps\in{\TTT}^l$, {\it and} $|\hbox{Im}\pps|<\x$, {\it
and in $\AA\in W(\AA_0,\r_0)$ $\=$ $\{ \AA\in D : |\AA-\AA_0|<\r_0\}$;
\acapo
(2) $\V H(\AA_0,\oo_0t;\e)$ and $\V h(\AA_0,\oo_0t;\e)$
have vanishing average;
\acapo
(3) $\V H(\AA_0,\oo_0t;\e)$ and $\V h(\AA_0,\oo_0t;\e)$ are
analytic in $\e$, for $|\e|<\e_0$, with a suitable
$\e_0$ close to zero:
$$ \e_0 = \h_0\,\left[ C_0\,F\,\r^{-1}_0\,\e_1^{-1}\,\right]^{-1},
\Eq(1.12) $$
with $\h_0$ a dimensionless constant depending only on $\ell$,
$\x$ and $\t$;
\acapo
(4) $N_f(\AA;\e)$ is
analytic in $\AA\in W(\AA_0,\r_0)$ and in
$\e$, for $|\e|<\e_0$.}

\*

In general the analyticity
properties of $N_f(\AA;\e)$ are related to those of
$f(\aa,\AA;\e)$. For example, let us consider the
Birckhoff hamiltonian, for $\ell=2$,
$$ \HH(\aa,\AA) = \left[ \omega_0 A_1 + A_2 \right] +
\e \left[ A_2 + f_1(\aa_1)\,f_2(\aa_2) \right] \; , \Eq(1.13) $$
where the vector $\oo_0=(\omega,1)$ is a diophantine one. Then
the function $N_f(\AA;\e)$ is trivially defined as $N_f((A_1,A_2);\e)$
$=$ $\e A_2$, (and the resulting hamiltonian becomes integrable,
see [G7], [GM1], for a proof with the methods
used in the present paper),
so that it is an entire function of $\e$.
The original conjecture in [G2] was that, under the stronger
hypothesis that the perurbation was a polinomial in $\e$,
the function $N_f(\AA;\e)$ would be {\it entire} in $\e$.
In the case of \equ(1.13), it is not difficult
to verify that our methods, applied to that particular case,
give $N_f(\AA;\e)=\e A_2$; however, in general, we are not able to
prove or refute the conjecture in the stronger form,
since we have been successful only in proving an upper bound $C^k$,
for some positive constant $C$, to the $k$-th perturbative order.
However the validity of this stronger statement seems quite unlikely to us,
without extra assumptions.

\*

The paper is self-contained, and, although many ideas and
notations are inherited from [G7], [GM1] and [GM2], no one
of those works is required to be read in order to understand the
present one; on the contrary,
with respect to the previous papers, more
details are given, as, because of the greater
generality of the problem, the notations
often become more involved.
In \S 2, recursive formulae defining the coefficients of
the formal series expansion in powers of the perturbative parameter
of the solution of the equation of motions are given, and in
\S 3 a diagrammatic representation in terms
of {\sl tree diagrams} (or simply {\sl trees}) for such coefficients
is introduced. In \S 4 a multiscale decomposition of
the propagators is furnished, so that the construction
of the labeled trees is completed. In \S 5 the contributions
which are source of convergence troubles are identified,
while \S 6 and \S 7 are devoted
to the problem of dealing with such contributions; in
\S 6 and \S 7, it is shown
that the contributions arising from the resonances
can be split into two parts, the first of which vanishes when
the sum over all the trees is performed (see \S 6), while
the second one can be easily handled through dimensional
arguments (see \S 7): this completes the proof of
Theorem 1.2. In \S 8 the discussion is adapted in
order to prove Theorem 1.4.

In \S 9 a comparison with the existing literature on the matter
is proposed, and in particular we give a detailed translation of
Eliasson's original work in our formalism.

Some technical features of the proofs are discussed in Appendices
A1 and A2. In Appendix A3 we give a ``more natural'' multiscale
decomposition via a partition of unity with characteristic functions.
This is an alternative
approach to the one discussed in \S 2$\div$8, where the multiscale
decomposition is based on a partition of
unity built with smooth ($C^\io$) functions, with respect to which
it presents some technical intricacies, but it turns out to be more
suitable for the discussion of the breakdown of the
invariant tori in [GGM].

\vskip1.truecm

\centerline{\titolo 2. Recursive formulae}
\*\numsec=2\numfor=1

\\Let us consider first the notationally less involved case
in which the perturbation is linear in the perturbative parameter $\e$:
$$ f(\aa,\AA;\e)=\e\,f(\aa,\AA) \; , \qquad
F_0=\sup_{\AA\in W(\AA_0,\r_0)}\sup_{\aa\in{\ttt}^{\ell}}
f(\aa,\AA) \; , \qquad E=\max\{F_0,E_0\} \; . \Eq(2.1) $$
We shall see in Appendix A2 how
to extend the discussion to the more general case.

\*

Calling $\V H^{(k)}$ and $\V h^{(k)}$  the $k$-th order
coefficients of the Taylor expansion of $\V H$ and $\V h$ in powers of
$\e$, and writing the equations of motion as
$$ \eqalign{
{d\AA\over dt} & = -\e\dpr_\aa f(\aa,\AA) \; , \cr
{d\aa\over dt} & = \dpr_\AA \HH_0 + \e\dpr_{\AA} f(\aa,\AA) \; , \cr
} \Eq(2.2) $$
we get immediately
recursion relations for $\V H^{(k)},\V h^{(k)}$; for $k=1$:
$$ \eqalign{
& \oo_0\cdot\dpr_\pps\,\V H^{(1)} = - \dpr_\aa f \; , \cr
& \oo_0\cdot\dpr_\pps\,\V h^{(1)} = T(\AA_0) \V H^{(1)}
+ \dpr_\AA f \; , \cr} \Eq(2.3) $$
(being $T(\AA_0)$ defined before \equ(1.4)), and, for $k>1$:
$$ \eqalign{
\oo_0\cdot\dpr_\pps\,\V H^{(k)} = &
{\sum_{(k-1)}}^* (-\dpr_\aa) \sum_{p\ge 0}\sum_{q\ge 0}{ 1 \over p!\,q! }
\prod_{s=1}^p \left( \V h^{(k_s)} \cdot \dpr_\aa \right)
\prod_{r=1}^q \left( \V H^{(k_r')} \cdot \dpr_\AA\right)
f(\oo_0t,\AA_0) \; , \cr
\oo_0\cdot\dpr_\pps\,\V h^{(k)} = &
T(\AA_0) \V H^{(k)} +
{\sum_{(k)}}^* \dpr_\AA \sum_{q\ge 2} {1\over q!}
\prod_{r=1}^q \left( \V H^{(k_r')} \cdot \dpr_\AA \right)
\HH_0(\AA_0) \cr
+ &
{\sum_{(k-1)}}^* \dpr_\AA \sum_{p\ge 0}\sum_{q\ge 0}{ 1 \over p!\,q! }
\prod_{s=1}^p \left(\V h^{(k_s)} \cdot \dpr_\aa \right)
\prod_{r=1}^q \left( \V H^{(k_r')} \cdot  \dpr_\AA\right)
f(\oo_0t,\AA_0) \; , \cr} \Eq(2.4) $$
where the $\sum^*_{(K)}$, $K=k,k-1$, denotes summation over the integers
$k_{s}\ge1$, $k_{s}'\ge1$, with: $\sum_{s=1}^p k_{s}$
$+$ $\sum_{r=1}^{q} k_{r}'$ $=$ $K$, and the derivatives are
supposed to apply to the functions $f(\aa,\AA)$, and then
evaluated in $(\aa,\AA)=(\oo_0t,\AA_0)$. If $p=0$ or $q=0$,
(both cases simultaneously are not possible),
the corresponding product is meant as 1.

\*

\\{\bf 2.1.}
{\it Proof of the formal solubility of the recursive relations.}
We proceed inductively through the following steps.
\acapo
(1) From the equations of motion for the angular momenta, we obtain
immediately the first recursive relation in \equ(2.4). Then
suppose that $\V h^{(k)}$ has vanishing average in $\pps$,
for $1\le k< k_0$, and that $\V h^{(k)}$ and $\V H^{(k)}$
solve the equations of motion, (and therefore \equ(2.4)),
for $1\le k < k_0$. Note that for $k=1$ the statement
holds, (as can be trivially verified), provided
the $\V0$-th Fourier component of $\V H^{(1)}$ is suitably
chosen so that the right hand side of the second equation
in \equ(2.3) has vanishing $\V0$-th Fourier component.
\acapo
(2) Then the first equation in \equ(2.4)
can be solved for $k=k_0$, if the right hand side has vanishing
average.
This can be easily checked, as follows, (see [CG], App. A12,
for an analogous discussion; see also [CZ]).
Let be $Y(\pps)\=Y=(\V h(\AA_0,\pps;\e), \V H(\AA_0,\pps;\e))$,
where $\V h$ and $\V H$ are defined by the
formal series expansions, respectively, $h=\sum_k\V h^{(k)}\e^k$ and
$H=\sum_k\V H^{(k)}\e^k$, and let be $E$ the symplectic matrix.
Then, by assumption,
$$ {dY\over dt} \= (\oo_0\cdot\dpr_\pps)\,Y = (E\dpr\HH)(Y) \;
, \qquad \hbox{ up to order } k_0-1 \; , \Eq(2.5) $$
where $\dpr=(\dpr_\aa,\dpr_\AA)$, and $\HH(\aa,\AA)$ is the
hamiltonian \equ(1.1).
But, for any periodic function $Y(\pps)$, $\pps\in{\TTT}^\ell$,
(not necessarily the previously considered one),
one has
$$ \int_{{\ttt}^\ell} d\pps \, (\dpr_\pps Y )(\pps)\,\cdot
\left[E(\oo_0\cdot\dpr_\pps)Y\right](\pps) = \V0 \; , \qquad
\int_{{\ttt}^\ell} d\pps \, (\dpr_\pps Y )(\pps)\,\cdot
\left(\dpr\HH\right)(\pps) = \V0 \; , \Eq(2.6) $$
where the first identity can be easily obtained by integration
by parts and depends only on the periodicity of the
function $Y(\pps)$, while the second one is trivial,
the integrand being the gradient with respect to $\pps$
of the hamiltonian $\HH$.
Then, if $Y$ is the function $(\V h, \V H)$, the fact that
the equations of motion are satisfied up to order $k_0-1$,
(see \equ(2.5)), implies that the sum of the two identities
\equ(2.6), \ie
$$ \int_{{\ttt}^\ell} d\pps \, \left[ \dpr_\pps Y (\pps)\,\cdot
\left( E (\oo_0\cdot\dpr_\pps)\,Y + \dpr \HH(\pps) \right) \right]
= \V0 \; , \Eq(2.7) $$
to order $k_0$ gives simply
$$ \int_{{\ttt}^\ell} d\pps \, \left[ \dpr_\pps Y (\pps) \right]^{(0)}
\cdot \left[ \dpr \HH (\pps) \right]^{(k_0)} = \V0 \; , \Eq(2.8) $$
but the first term in \equ(2.8) is a constant,
since $Y^{(0)}(\pps)=(\pps,\AA_0)$, so that
we have that the average of the function $\dpr_\aa\HH$ is
vanishing to order $k_0$. Then the assertion follows immediately,
by noting that $[-\dpr_\aa\HH]^{(k_0)}$ is exactly the right
hand side of the first equation in \equ(2.4).
\acapo
(3) Then the first equation in \equ(2.4) yields a function
$\V H^{(k_0)}(\pps)$ which is defined up to the constant
$\V\m^{(k_0)}\=\V H^{(k_0)}_{\V0}$, (\ie the $\V0$-th Fourier
component),
which we call ``counterterm".
Such a constant, however, must be taken so that the equation for
the angle variables, \ie the second of \equ(2.4),
has zero average, in order to be soluble.
\acapo
(4) Hence the equation
for $\V h^{(k)}$ can be solved and its solution is
defined up to an arbitrary constant: such a constant can be chosen to be
vanishing, and the procedure can
be iterated. This shows that a formal solution of \equ(2.4)
can be obtained. \qed

\*

If we look at \equ(2.3) and \equ(2.4), we see that $\V H^{(k)}$
is given by a sole contribution which has always (at least) one
derivative with respect to $\aa$, whereas $\V h^{(k)}$ is the sum of two
or three contributions such that the first one
has again a derivative with respect to $\aa$ in front of all,
while the other ones have a derivative with respect to $\AA$.
Then we can introduce the following notation: $\V H^{(k)}$ is
given by a sum of terms which are of the form $H\ot h$, where $H$
denotes that they contribute to $\V H^{(k)}$ and
$h$ that the first derivative is with respect to the angle variables.
In the same way, $\V h^{(k)}$ is given by three sums (the
second one is absent if $k=1$) of terms which are,
in the first sum, of the form $h\ot h$, and, in the latter two,
of the form $h\ot H$. The terms of the
second and the third sums will be distinguished
by a label $\d$, which can be set $\d=0$ for the second one,
and $\d=1$ for the third one; for future convenience we assign
a label $\d=1$ also to the other terms $H\ot h$ and $h\ot h$.

\*

It can be convenient to write the above recursive formulae
in the Fourier space. Then, if we take into account also
the compatibility conditions required in order to make
the equations \equ(2.3), \equ(2.4) to be soluble
(see item (3) in \S 2.1), we easily find,
for $k=1$, from \equ(2.3),
$$ \eqalign{
\V h^{(1)}_{\nn} & = {  -i T(\AA_0)\, \nn \, f_\nn(\AA_0)
\over(i\oo_0\cdot\nn)^2 } +
{\dpr_\AA \, f_\nn(\AA_0) \over i\oo_0\cdot\nn } \; ,
\qquad \nn\neq\V0 \; , \cr
\V H^{(1)}_{\nn} & = { -i\nn \,  f_\nn(\AA_0) \over
i\oo_0\cdot\nn } \; , \qquad \nn\neq\V0 \; , \cr
\V H^{(k)}_{\V0} & \= \V\m^{(1)} =
- T^{-1}(\AA_0)\,\dpr_\AA \,f_{\V0}(\AA) \; , \cr}
\Eq(2.9)$$
and, for $k>1$ and $\nn\neq\V0$, from \equ(2.4),
$$ \eqalign{
(i\oo_0\cdot\nn)\,\V H^{(k)}_\nn = &
{\sum_{(k-1)}}^* (-i\nn_0) \sum_{p\ge 0} \sum_{q\ge 0} { 1 \over p!\,q! }
\prod_{s=1}^p \left( i\nn_0 \cdot \V h^{(k_s)}_{\nn_s} \right)
\prod_{r=1}^q \left( \V H^{(k_r')}_{\nn_r'} \cdot \dpr_\AA \right)
f_{\nn_0}(\AA_0) \; , \cr
(i\oo_0\cdot\nn)\,\V h^{(k)}_\nn = &
T(\AA_0) \V H^{(k)}_\nn +
{\sum_{(k)}}^* \dpr_\AA \sum_{q\ge 2} {1\over q!}
\prod_{r=1}^q \left( \V H^{(k_r')}_{\nn_r'} \cdot \dpr_\AA \right)
\HH_0(\AA_0) \cr
+ &
{\sum_{(k-1)}}^* \dpr_\AA \sum_{p\ge 0} \sum_{q\ge 0} { 1 \over p!\,q! }
\prod_{s=1}^p \left( i\nn_0 \cdot \V h^{(k_s)}_{\nn_s} \right)
\prod_{r=1}^q \left( \V H^{(k_r')}_{\nn_r'} \cdot \dpr_\AA \right)
f_{\nn_0}(\AA_0) \; , \cr} \Eq(2.10) $$
if, for $k>1$ and $\nn=\V0$, we set $\V H^{(k)}_{\V0} =\V\m^{(k)}$,
with
$$ \V\m^{(k)} =
{\sum_{(k-1)}}^* \left( - T^{-1}(\AA_0)\dpr_\AA \right)
\sum_{p\ge 0}\sum_{q\ge 0}{ 1 \over p!\,q! }
\prod_{s=1}^p \left( i\nn_0 \cdot \V h^{(k_s)}_{\nn_s} \right)
\prod_{r=1}^q \left( \V H^{(k_r')}_{\nn_r'} \cdot \dpr_\AA \right)
f_{\nn_0}(\AA_0) \; , \Eq(2.11) $$
where the $\sum^*_{(K)}$, $K=k,k-1$, denotes summation over the integers
$k_{s}\ge1$, $k_{r}'\ge1$, with: $\sum_{s=1}^p k_{s}$
$+$ $\sum_{r=1}^{q }k_{r}') =$ $K$, and over the integers $\nn_0$,
$\nn_s$, $\nn_r'$, with: $\nn_0+\sum_{s=1}^p\nn_s + \sum_{r=1}^q \nn_r'$
$=$ $\nn$, and in the second contribution of the second equation in
\equ(2.10) $\nn_0\=\V0$ identically.
The interpretation of the cases $q=0$ and $p=0$ is as in \equ(2.3).

Again, if we use the terminology introduced after \equ(2.3),
$\V H^{(k)}_\nn$ is given by a sum of terms of the form
$H\ot h$, and $\V h^{(k)}_\nn$ by three contributions which
are sums of terms
of the form $h\ot h$ and $h\ot H$, (and $\d=0,1$ in the second
case). The contribution to $\V\m^{(k)}$ can be interpreted
as a sum of terms of the form $\m\ot H$, and a label
$\d=1$ can be assigned to them.

\vskip1.truecm

\centerline{\titolo 3. Diagrammatic expansion}
\*\numsec=3\numfor=1

\\The equations \equ(2.8)$\div$\equ(2.10)
provide an algorithm to evaluate a formal power
series solution to our problem: in fact they allow us to
carry out a {\sl diagrammatic expansion} of $\V h^{(k)}_\nn$
and $\V H^{(k)}_\nn$: we simply ``i\-te\-ra\-te"
it until only $\V h^{(1)}_\nn$, $\V H^{(1)}_\nn$, $\nn\neq\V0$,
and $\V\m^{(1)}$ appear.

If we define the dimensionless quantity
$\V X^{(k)}_\nn(\z)$ as
$$ \V X^{(k)}_\nn(\z)=
\cases{ \V h^{(k)}_\nn \; , & if $\z=h$, $\nn\neq\V0$ , \cr
(C_0/J_m) \V H^{(k)}_\nn \; , & if $\z=H$, $\nn\neq\V0$ , \cr
(C_0/J_m) \V\m^{(k)} \; , & if $\z=\m$, $\nn=\V0$ , \cr} $$
then we shall show that it is
possible to write $\V X^{(k)}_\nn(\z)$ as sum of
contributions each of which can be graphically represented
as a tree diagram $\th$. In other words, we shall give some rules
in order to associate to a suitable diagram $\th$ a value
$\hbox{Val}(\th)$, and show that it turns out to be,
for each Fourier component,
$$ \V X^{(k)}_\nn(\z) = \sum_{\th\in\TT_k}
\hbox{Val}(\th) \; , \Eq(3.1) $$
where the sum is over all the trees contained in a certain
class $\TT_k$, (which will be defined later).

\vskip.7truecm

\\{\bf 3.1.} {\it Topological and semitopological trees.}
A tree diagram (or {\sl tree})
$\th$ will consist of a family of lines ({\sl branches} or
{\sl lines})
arranged to connect a partially ordered set of points ({\sl vertices}
or {\sl nodes}), with the higher vertices to the right. The branches are
naturally ordered as well; all of them have two vertices at their
extremes (possibly one of them is a top vertex), except the lowest
or {\sl first branch} which has only one vertex, the {\sl first vertex}
$v_0$ of the tree. The other extreme $r$ of the first branch will be called
the {\sl root} of the tree and will not be regarded as a vertex; we
shall call the first branch also {\sl root branch}. A possible
tree is represented in Fig.3.1.
%
%
\midinsert
\insertplot{240pt}{190pt}{
\ins{-35pt}{100pt}{$\hbox{root}$}
\ins{60pt}{95pt}{$v_0$}
\ins{152pt}{130pt}{$v_1$}
\ins{110pt}{60pt}{$v_2$}
\ins{190pt}{110pt}{$v_3$}
\ins{230pt}{170pt}{$v_5$}
\ins{230pt}{130pt}{$v_6$}
\ins{230pt}{95pt}{$v_7$}
\ins{230pt}{5pt}{$v_{11}$}
\ins{230pt}{30pt}{$v_{10}$}
\ins{200pt}{75pt}{$v_4$}
\ins{230pt}{75pt}{$v_8$}
\ins{230pt}{55pt}{$v_9$}
}{fig1}
\*
\didascalia{{\bf Fig.3.1.} A tree $\th$ with degree $d=11$.
Each line (branch) is supposed to carry an arrow
(which is not explicitly drawn) pointing to the root.
If we consider a vertex of the tree, \eg $v_3$, then we define
$\l_{v_3}$, or equivalently $v_1\ot v_3$, the line connecting
$v_3$ to $v_1$, and we write $v_1=v_3'$. The arrow, if drawn,
would point from $v_3$ to $v_1$, as one has to cross $v_1$
in order to reach the root from $v_3$.}
\endinsert
%

If $v_1$ and $v_2$ are two vertices of the tree we say that $v_1<v_2$ if
$v_2$ follows $v_1$ in the order established by the tree, \ie if
one has to pass $v_1$ before reaching $v_2$, while climbing the tree.
Since the tree is partially
ordered not every pair of vertices will be related by the order
relation, (which we are denoting $\le$): we say that
two vertices are comparable if they are related by the order relation.

Given a vertex $v$ we denote by $v'$ the vertex immediately
preceding $v$. We also imagine that each branch of
the tree carries an arrow pointing to the root (``gravity
direction'', opposite to the order): this means that if
a line connects two vertices, say $v'$ and $v$, with $v'<v$,
(so that $v$ {\sl follows} $v'$),
then the arrow points from $v$ to $v'$, and we say that the
line {\sl emerges from} $v$ and {\sl enters} $v'$.

Given a tree $\th$ with first vertex $v_0$, each vertex $v>v_0$ can
be considered the first vertex of the tree consisting of the vertices
following $v$: such a tree will be called a {\sl subtree} of $\th$.

Let us define the {\sl degree} of a tree as the number of vertices
of the tree. Obviously, the degree of a tree $\th$ counts also
the number of branches of $\th$: in fact there is a
correspondence 1-to-1 between vertices and branches,
if we associate to each vertex $v$ the branch $\l_v$ emerging from it.
We can also represent a branch $\l_v$ as $v'\ot v$,
(see Fig.3.1 for definiteness).

A group $\GG$ of transformations acts on the trees, generated by
the following operations: fix a node $v\in\th$ and permute the
subtrees emerging from it.
We call {\sl semitopological trees} the trees which are
superposable up to a continuous deformation of the branches
on the plane, and {\sl topological trees} when the same
happens modulo the action of the just defined group of
transformations. We shall denote by $\Theta^{(s)}$ the set
of semitopological trees, and by $\Theta^{(t)}$ the set
of the topological trees. For example, the trees
drawn in Fig.3.2
will be regarded as different as semitopological trees and
equivalent as topological trees.

%
%
\midinsert
\insertplot{220pt}{80pt}{
\ins{-10pt}{32pt}{$\hbox{root}$}
\ins{130pt}{32pt}{$\hbox{root}$}
\ins{35pt}{32pt}{$v_0$}
\ins{175pt}{32pt}{$v_0$}
\ins{50pt}{62pt}{$v_1$}
\ins{190pt}{20pt}{$v_1$}
\ins{75pt}{63pt}{$v_2$}
\ins{215pt}{63pt}{$v_2$}
\ins{75pt}{35pt}{$v_3$}
\ins{215pt}{33pt}{$v_3$}
\ins{75pt}{5pt}{$v_4$}
\ins{215pt}{5pt}{$v_4$}
}{fig2}
\*
\didascalia{{\bf Fig.3.2.} Two trees $\th_1$ and $\th_2$ of
degree $d=5$, which are different if regarded as semitopological
trees and identical if regarded as topological trees. In fact,
if we permute the subtrees emerging from the first vertex, we
obtain $\th_2$ from $\th_1$ and {\it viceversa}.}
\endinsert
%
%

Note that the number of topological trees and the number of
semitopological trees of degree $d$ can be bounded by
$2^{2d}$, (see, \eg, [HP]).

Given a vertex $v\in\th$ at which the tree $\th$ bifurcates
into $m_v$ subtrees among which there are only $\BB_v$
topologically different subtrees $\th_1$, $\ldots$, $\th_{\BB_v}$,
each of which is repeated $\NN_v(\th_i)$ times, and given a
function $F(\th)$ whose value depends only on the
topological tree, one has
$$ \sum_{\th\in\Theta^{(s)}} \prod_{v\in\th} {1\over m_v!} F(\th) =
\sum_{\th\in \Theta^{(t)}} \prod_{v\in\th}
\prod_{i=1}^{\BB_v}
{1\over \NN_v(\th_i)!} F(\th) \; , \Eq(3.2) $$
where we recall that $\Theta^{(s)}$ is the set of semitopological trees
and $\Theta^{(t)}$ is the set of topological trees, (see also [G8],
\S 5).

\*

\\{\bf 3.2.} {\it Numbered trees.}
It can be convenient to introduce also another kind of trees,
which we call {\sl numbered trees}, (following [G8]): they are
obtained by imagining to have a deposit of $d$ branches numbered
from 1 to $d$ and depositing them on the branches of a topological
tree with degree $d$. The numbered trees will be regarded as
identical if superposable by the action of a transformation
of the group $\GG$, in such a way
that all the numbers associated to the lines match.
One has, for a function
$F(\th)$ depending only on the topological trees,
$$ \sum_{\th\in\Theta^{(s)}} \prod_{v\in\th} {1\over m_v!} F(\th) =
{1\over d!}\sum_{\th\in \Theta^{(n)}} F(\th) \; , \Eq(3.3) $$
if $\Theta^{(n)}$ is the set of numbered trees.
The number of numbered trees of degree $d$ is bounded by $d!2^{2d}$.

To work with the numbered trees can be very convenient from a
combinatorial point of view: this will be particularly evident
in \S 6, where the cancellation mechanisms operating to
all perturbative orders will be investigated and will be shown
to be easily visualized in terms of numbered trees.
Nevertheless it is important to keep in mind that it is
absolutely equivalent to study the perturbative expansions in terms of
semitopological trees or in terms of numbered trees, and to
choose one of the two possibilities is only a matter of convenience.
So in \S 5, it will turn out to be easier to work with
semitopological trees.

\*

\\{\bf 3.3.} {\it Labeled trees.}
To each vertex we associate a finite set of labels,
defined as follows.
\acapo
(1) $d_v$ is the number of vertices $w$, such
that $w\ge v$, \ie the number of vertices of the subtree having
$v$ as first vertex, (and it is the degree of such a subtree);
\acapo
(2) $\d_v=0,1$;
\acapo
(3) $k_v$ is defined as $k_v=\sum_{w\ge v}\d_w$, and it is called
the {\sl order} of the subtree having $v$ as first vertex;
\acapo
(4) $\nn_v\in{\ZZZ}^\ell$ is called the {\sl mode label};
\acapo
(5) $\z^1_v$ and $\z^2_v$ can assume the symbolic values
$\z^1_v,\z_v^2=h,H,\m$;
\acapo
(6) $m_v$ is the number of branches entering the vertex $v$,
(if $v$ is a top vertex, then trivially $m_v=0$);
morover, let us define $p_v$ and $q_v$ the number of branches
entering $v$ and emerging from vertices $w$ (such
that $w'=v$) carrying, respectively,
a label $\z_w^1=h$ and a label $\z_w^1=H,\m$, so that
$m_v=p_v+q_v$.

If a branch $\l_v$ connects to a vertex $v'$ a vertex $v$, with
labels $\z^1_{v}$ and $\z^2_{v}$, we shall classify
the branch as a $\z_v^1\ot\z_v^2$ branch.
To each branch $\l_v$ we associate a {\sl momentum}
$\nn_{\l_v}\=\sum_{w\ge v}\nn_w$, and
a functional given by the product of an operator
$\OO_{\l_v}$ times a propagator $g(\oo\cdot\nn_{\l_v})$,
which are defined as follows:
$$ \matrix{
\hbox{operator} & \hbox{propagator} & \hbox{branch} \cr
& & \cr
C_0^2 \left[ i\nn_{v'}\cdot(-iT\,\nn_v) \right] &
[i\oo\cdot\nn_{\l_v}]^{-2} & h \ot h \cr
& & \cr
C_0 \left[ i\nn_{v'}\cdot(\dpr_{\AA_v})
\right] &
[i\oo\cdot\n_{\l_v}]^{-1} & h \ot H \cr
& & \cr
C_0 \left[ \dpr_{\AA_{v'}}\cdot(-i\nn_v) \right] &
[i\oo\cdot\nn_{\l_v}]^{-1} & H \ot h \cr
& & \cr
\dpr_{\AA_{v'}}\cdot (-T^{-1}\dpr_{\AA_v}) & 1 &
\m \ot H \cr} \Eq(3.4) $$
for all the branches distinct from the root branch, and
$$ \matrix{
\hbox{operator} & \hbox{propagator} & \hbox{branch} \cr
& & \cr
C_0^2 \left[ -iT\,\nn_v \right] &
[i\oo\cdot\nn_{\l_v}]^{-2} & h \ot h \cr
& & \cr
C_0 \left[ \dpr_{\AA_v} \right] &
[i\oo\cdot\n_{\l_v}]^{-1} & h \ot H \cr
& & \cr
C_0^2\,J_m^{-1} \left[ -i\nn_v \right] &
[i\oo\cdot\nn_{\l_v}]^{-1} & H \ot h \cr
& & \cr
C_0\,J_m^{-1} \left[ - T^{-1}\dpr_{\AA_v} \right]
& 1 & \m \ot H \cr} \Eq(3.5) $$
for the root branch; the first three terms
in \equ(3.4) and \equ(3.5) can occur only if $\nn_{\l_v}\neq\V0$,
while the fourth one only if $\nn_{\l_v}=\V0$. All the other
pairs $(\z_v^1,\z_v^2)$ are defined to give a vanishing contribution,
so that we can get rid of them.

\*

\\{\bf 3.4.} {\it Remark.}
Note that, in \equ(3.4),
the operators corresponding to the lines $h\ot H$ and $H\ot h$
are opposite to each other under the change of the vertices $v'$
and $v$.

\*

Then we multiply all the above operators $\OO_{\l_v}$ (we simply
regard as multiplication operators the factors in which
no derivative appears) to the function
$$ \prod_{v\in\th \atop \d_v=1} f_\nn(\AA_v)
\prod_{v\in\th \atop \d_v=0} \HH_0(\AA_v) \; , \Eq(3.6) $$
and evaluate the result at the points $\AA_v\=\AA_0$, $\forall v\in\th$.

We define $O_v$ the (tensor)
factor we can associate to each vertex, once the above operators
$\OO_{\l_w}$, $w\in\th$,
have been applied to the function \equ(3.6),
$$ \prod_{v\in\th} O_v = \prod_{v\in\th} \OO_{\l_v}
\prod_{v\in\th \atop \d_v=1} f_\nn(\AA_v)
\prod_{v\in\th \atop \d_v=0} \HH_0(\AA_v)
\Big|_{\AA_v=\AA_0} \; , \Eq(3.7) $$
where $O_v$ is given by
$$ \eqalign{
O_v = & \Big\{
\Big[ C_0^2\Big(-iT\,\nn_v\Big)\,\d_{\z_v^1,h}\,\d_{\z_v^2,h}
+ C_0\Big(\dpr_{\AA_v}\Big) \,\d_{\z_v^1,h}\,\d_{\z_v^2,H} +
C_0^2\,J_m^{-1}\Big(-i\nn_v\Big)\,\d_{\z_v^1,H}\,\d_{\z_v^2,h} \cr
& +
C_0\,J_m^{-1}\Big(-T^{-1}\,\dpr_{\AA_v}\Big)\,
\d_{\z_v^1,\m}\,\d_{\z_v^2,H} \Big] \cdot \cr
& \cdot \prod_{w \atop w'=v} \Big[ (i\nn_v)\,\d_{\z_w^1,h}
+ C_0^{-1}J_m\,(\dpr_{\AA_v})\Big( \d_{\z_w^1,H} + \d_{\z_w^1,\m} \Big)
\Big] \Big\} \; \cdot \cr
& \cdot \;
\Big[ f_{\nn_v}(\AA_v) \, \d_{\d_v,1} +
\HH_0(\AA_v) \, \d_{\d_v,0} \Big] \Big|_{\AA_v=\AA_0}
\; , \cr} \Eq(3.8) $$
being the product over all the vertices immediately
following $v$ (so that it is missing if $v$ is a top vertex).

\*

\\{\bf 3.5.} {\it Remark.}
Note that, unlike the operators, the factors $O_v$ do not
depend only on the line $\l_v$, but also on the lines
entering $v$, (which can carry some derivatives with respect
to the action variables acting on $f_{\nn_v}(\AA_v)$ or
$\HH_0(\AA_v)$). For this reason we prefer to associate the
factors to the vertices rather than to the lines, but, obviously,
this is somewhat arbitrary, as there is a correspondence 1-to-1
between vertices and branches, (see \S 3.1 and Fig.3.1).
 From \equ(3.8) we see that the derivatives with respect
to the action variable $\AA_v$,
$v\in\th$, collected together, give a tensor
$$ \Big[ (1- \d_{\z_v^2,H}) + \d_{\z_v^2,H} \,
\dpr_{\AA_v} \Big] \prod_{w \atop w'=v} \Big[
\dpr_{\AA_v} \Big( \d_{\z_w^1,H} + \d_{\z_w^1,\m} \Big)
\Big] \Big[ f_{\nn_v}(\AA_v) \, \d_{\d_v,1} +
\HH_0(\AA_v) \, \d_{\d_v,0} \Big] \Big|_{\AA_v=\AA_0} \; , $$
each of whose entries can be bounded by
$$ (q_v + \d_{\z_v^2,H})!\,E\,\r_0^{-(q_v+\d_{\z_v^2,H})} \; , $$
through the Schwarz's lemma, [Ti], \S 5.2, (which is in turn a trivial
application of the Cauchy formula).

\*

\\{\bf 3.6.} {\it Remark.} Note that two trees topologically or
semitopologically equivalent can become different as labeled
trees. They will be considered as identical only if,
when superposed (if topologically equivalent) or superposed
modulo a transformation of the group $\GG$ (if semitopologically
equivalent), all their labels match. If the trees are numbered,
they will be considered equivalent if also the numbers match.

\*

Now we have all the definitions and notations necessary
to represent $\V X(\z)$ as sum of values associated to
semitopological trees (in \S 6, we shall see how to
change the definition of tree value in order to express
$\V X(\z)$ in terms of numbered trees).

We can define the {\sl value of a labeled tree} $\th$ as
$$ \hbox{Val}(\th) = \prod_{v\in\th}
{O_{v}\over m_v!}\,g(\oo\cdot\nn_{\l_v}) \; , \Eq(3.9) $$
and the $\nn$-the Fourier component of the
function $\V X^{(k)}(\z)$ can be expressed as a sum
of the form \equ(3.1), where $\TT_k=\TT^{(s)}_k$,
if $\TT^{(s)}_k$ is the {\it collection of
all the possible not equivalent labeled semitopological
trees} with order $k_{v_0}=k$ and $\nn_{\-{v_0}}=\nn$,
if $v_0$ is the first vertex of the trees, (the notion
of equivalence being defined in Remark 3.6).

The proof of such an assumption can be easily obtained,
if we recall that any subdiagram emerging from a vertex $v\in\th$
is again a tree having $v$ as first vertex
and $v'$ as root, and look at the
recursive formulae \equ(2.8), \equ(2.9) and \equ(2.10).
Then the interpretation of all the labels listed in 3.3
becomes clear, as they can be related to the formulae
\equ(2.8)$\div$\equ(2.10) and to the notations introduced
at the end of \S 2 (for the labels $\d$ and the symbols
$\z_v^1\ot \z_v^2$). In particular the following result
can be easily proven to hold.

\*

\\{\bf 3.7.} {\cs Proposition.}
{\it The labels so defined have to satisfy the following
{\rm compatibilty condition}: if $\d_v=0$, then it is
$\z_v^1=h$, $\z_v^2=H$,
$p_v\=0$ and $m_v\=q_v\ge 2$.
Then it is easy to see that $d_v\le 2k_v-1$ for each $v$.
In particular $d\le 2k-1$.}

\*

\\{\bf 3.8.} {\it Proof of Proposition 3.7.}
We note from \equ(2.8) that $\d_v\=1$ if $v$ is a top vertex.
Then Proposition 3.7 follows immediately. \qed

\*

\*

Note that each non trivial (\ie not corresponding
to the case $\m\ot H$) propagator can be written as
$$ g(\oo\cdot\nn_{\l_v})={1\over[\oo\cdot\nn_{\l_v}]^{R_{\l_v}} }
\; , \Eq(3.10) $$
where $R_{\l_v}=1$ if $\l_v$ is a $H\ot h$ branch,
($\z_v^1=H$ and $\z_v^2=h$), or
a $h\ot H$ branch, ($\z_v^1=h$ and $\z_v^2=H$)),
$R_{\l_v}=2$ if $\l_v$ is a $h\ot h$ branch, ($\z_v^1=\z_v^2=h$).

\vskip1.truecm

\centerline{\titolo 4. Multiscale analysis of the tree values}
\*\numsec=4\numfor=1

\\We introduce a multiscale decomposition of the propagator.
Let $\chi(x)$ be a $C^\io$ not increasing
function such that $\chi(x)=0$, if $|x|\ge 2$ and $\chi(x)=1$ if $|x|\le 1$,
and let $\chi_n(x)=\chi(2^{-n}x)-\chi(2^{-(n-1)}x)$, $n\le 0$, and
$\chi_1(x)=1-\chi(x)$: such functions realize a $C^\io$ partition
of unity, for $|x|\in[0,\io)$, in the following way. Let us write
$$ 1=\chi_1(x)+\sum_{n=-\io}^0\chi_n(x)\=\sum_{n=-\io}^1\chi_n(x)
\; . \Eq(4.1) $$
Then we can decompose the propagator in the following way:
$$ g(\oo\cdot\nn_{\l_v})={1\over [i\oo\cdot\nn_{\l_v}]^{R_{\l_v}}}\=
\sum_{n=-\io}^1{\chi_n(\oo\cdot\nn_{\l_v})
\over [i\oo\cdot\nn_{\l_v}]^{R_{\l_v}}}\=
\sum_{n=-\io}^1 g^{(n)}(\oo\cdot\nn_{\l_v}) \; \Eq(4.2) $$
where $g^{(n)}(\oo\cdot\nn_{\l_v})$ is the ``propagator at scale $n$".
If $n<0$, $g^{(n)}(\oo\cdot\nn_{\l_v})$ is a $C^{\io}$
compact support function different from $0$ for
$2^{n-1}<|\oo\cdot\nn_{\l_v}|\le 2^{n+1}$,
while $g^{(1)}(\oo\cdot\nn_{\l_v})$ has
support for $1<|\oo\cdot\nn_{\l_v}|$. In the domain where
it is different from zero, the propagator verifies the bound
$$ \Big| {\partial^p\over\partial x^p} g^{(n)}(x)
\Big|_{x=\oo\cdot\nn_{\l_v}} \le a_{R_{\l_v}}(p)
\, 2^{-n(R_{\l_v}+p)} \; , \qquad p\in {\NNN} \; ,
\Eq(4.3) $$
where $a_{R_{\l_v}}(p)$ is a suitable constant,
such that $a_{R_{\l_v}}(0)=2^{R_{\l_v}}$,
which depends on the form of the function $\chi(x)$.
The constant $a_{R_{\l_v}}(p)$ has a bad dependence on $p$,
(since $g^{(n)}(x)$ is only a $C^{\io}$ function), but we shall
see that in our bounds $p$ does not increase ever beyond 2.

Proceeding as in quantum field theory, see [G5], given a tree $\th$
we can attach a {\sl scale label} $n_{\l_v}$ to each branch
$\l_v$ in $\th$, which is equal to the scale of the propagator
associated to the branch.

Looking at such labels we identify the connected
cluster $T$ of vertices which are
linked by a continuous path of branches with the same scale labels
$n_T$ or a higher one and which are maximal: we shall say that
{\sl the cluster $T$ has scale $n_T$}. Therefore an inclusion relation is
established between the clusters, in such a way that the innermost clusters
are the clusters with the highest scale, and so on.
Each cluster can have an arbitrary number of branches entering it,
({\sl incoming lines}), but only one branch exiting, ({\sl outgoing line});
we use the fact
that the branches carry an arrow pointing to the root: this gives a
meaning to the words ``incoming" and ``outgoing". We call
{\sl external lines} the lines which are either
outgoing or incoming.
A possible situation is described in Fig.4.1.

%
%
\midinsert
\vskip2.2truecm
\insertplot{320pt}{190pt}{
\ins{10pt}{83pt}{$\hbox{root}$}
\ins{90pt}{40pt}{$n_1$}
\ins{94pt}{129pt}{$n_2$}
\ins{150pt}{-20pt}{$n_0$}
\ins{205pt}{50pt}{$n_3$}
\ins{210pt}{130pt}{$n_4$}
\ins{232pt}{130pt}{$n_6$}
\ins{247pt}{60pt}{$n_5$}
\ins{70pt}{83pt}{$v_1$}
\ins{125pt}{50pt}{$v_2$}
}{fig3}
\vskip3.5truecm
\didascalia{{\bf Fig.4.1.} A tree with scale labels associated to
the lines. There are seven clusters $T_0$, $\ldots$, $T_6$
on scale, respectively, $n_0$, $\ldots$, $n_6$, which
satisfy the ordering relations: $n_0>n_1$, $n_0>n_2$, $n_0>n_3$,
$n_3>n_4$, $n_3>n_5$, and $n_4>n_6$.
If the external lines of the cluster $T_1$ carry
the same momentum (\ie $\nn_{v_1}+\nn_{v_2}=\V0$), then
$T_1$ is a resonance (see Definition 5.1), and we write
$T_1=V_1$. Note that there is always a maximal cluster
encircling all the tree, and that there is only
one outgoing line per cluster.}
\endinsert
%

The multiscale decomposition \equ(4.2) of the propagator
allows us to rewrite \equ(3.9)
$$ \hbox{Val}(\th)=\prod_{v\in\th} {O_{v}\over m_v!}\,
g^{(n_{\l_v})}(\oo\cdot\nn_{\l_v}) \; , \Eq(4.4) $$
and a formula like \equ(3.1) holds still, provided that
we count also the scale labels among the tree labels.

Obviously the choise of the partition of unity is not uniquely fixed:
a different possibility is envisaged in Appendix A3.
Of course the result we are looking for, \ie the proof of
Theorem 1.2, is independent on the particular partition,
but the technical features can be more or less suitable
for the discussion. For instance in [GGM] it turns out to
be more convenient to work with the partition which is
illustraded in Appendix A3.

\vskip1.truecm

\centerline{\titolo 5. Resonances and related problems}
\*\numsec=5\numfor=1

\\In this section, we confine ourselves to single out the
contributions wich can be source of problems and need a
more careful analysis. The discussion of such terms,
and the exhibition of the cancellation mechanisms which
have to be exploited in order to prove the convergence of
the perturbative series are differed to next sections.

\*

\\{\bf 5.1.} {\cs Definition (Resonance).} {\it Among the clusters we
consider the ones with the property that there is only one
incoming line, carrying the same momentum of the outgoing line,
and we define them {\rm resonances}. If $V$ is one such cluster
we denote by $\l_V$ the incoming line, and by $d(V)$ and $k(V)$,
respectively, the number of vertices contained in $V$
({\rm resonance degree}) and the quantity
$\sum_{w\in V}\d_w$ ({\rm resonance order}).
We call $n_{\l_V}$ the
{\rm resonance-scale}, and $\l_V$ a {\rm resonant line}:
if $n_V$ is the scale of the resonance as a cluster, \ie
the lowest scale of the line inside $V$, one has $n_V\ge n_{\l_V}+1$.}

\*

Given a tree $\th$, let us define $N_n(\th)$ the number of lines with
scale $n\le 0$, and $N^j_n(\th)$, $j=1,2$, the number of lines $\l$
with scale $n\le 0$ and $R_\l=j$.
Let us define also $\D(\th)$ the collection of vertices $v$'s in $\th$
(with $\d_v=1$) such that $\nn_v\neq\V0$, and $M(\th)$ the
quantity $M(\th)=\sum_{v\in\D(\th)}|\nn_v|$.

Then it is easy to check that the
scaling properties of the propagators and the definitions \equ(3.4),
\equ(3.5) and \equ(3.7) immediately imply
that the contribution to $\V X^{(k)}_\nn(\z)$
arising from a given tree $\th$ can be bounded as follows:
$$ \left| \hbox{Val}(\th) \right| \le
\CC^{k}\,e^{-\x \, M(\th)}
\prod_{n\le 0} 2^{-(2n N_n^2(\th)+n N_n^1(\th))}
\prod_{v\in\D(\th)}
{|\nn_v|^{p_v+1} \over p_v!} \; , \Eq(5.1) $$
({\sl dimensional bound}) for a suitable constant $\CC$, given by
$$ \CC=\left[ 2^3 \, J_M \, J_m^{-2} \, C_0^2 \, E \, \ell \,
\r^{-2} \right]^2 \; , \Eq(5.2) $$
where $C_0$ is the diophantine constant introduced in \equ(1.3),
$\r$ is introduced after \equ(1.7),
the eigenvalues $J_m$ and $J_M$ are defined in \equ(1.4) and
$E$ in \equ(1.7).

\*

\\{\bf 5.2.} {\it Proof of \equ(5.1) and \equ(5.2).}
In \equ(5.2), $2^3$ arises from the definition
of the compact support of the propagators, (which gives $a_{R_{\l_v}}
(0)\le a_2(0)=2^2$), and
from the fact that (1) to each vertex $v$
a factorial $(q_v+\d_{\z_v^2,H})!$ is associated
when the Cauchy formula is used in order to bound the
derivatives, (see Remark 3.5), (2) one has $p_v!\,q_v!\le m_v!$,
and (3) one can bound
$(q_v+\d_{\z_v^2,H})!/q_v!\le 2^{q_v}$, with $\sum_vq_v\le d_{v_0}$,
(which gives another $2$).

Moreover each line $\l_v$, $v_0<v\in\th$, has
associated a scalar product, which,
in according to the kind of branch $\l_v$ and
if we neglect the factorials, (which have been
already taken into account), can
be bounded, respectively, by
$$ \matrix{
C_0^2\,J_m^{-1}\,|\nn_{v'}|\,|\nn_v|\,E\,e^{-\x|\nn_v|}
& \quad h \ot h \cr
& \cr
C_0^2\,J_m^{-1}\,|\nn_{v'}|\,\ell\,\r^{-1}\,E\,e^{-\x|\nn_v|}
& \quad h \ot H \cr
& \cr
C_0^2\,J_m^{-1}\,\ell\,\r^{-1}\,|\nn_v|\,E\,e^{-\x|\nn_v|}
& \quad H \ot h \cr
& \cr
C_0^2\,J_M\,J_m^{-2}\,\ell\,\r^{-2}\,E\,e^{-\x|\nn_v|}
& \quad \m \ot H \cr} $$
and then each of the four above quantities can be bounded with
$$ J_M\,J_m^{-2}\,C_0^2\,\ell\,r^{-1}\,E\,e^{-\x|\nn_v|}\,
\max\{|\nn_{v'}|,1\}\,\max\{|\nn_{v}|,1\}\, e^{-\x|\nn_v|} \; , $$
where $r$ is defined after \equ(1.7),
$r^{-1}=\max\{1,\r^{-2}\}$;
an analogous bound holds for the root branch too, \ie
$$ J_M\,J_m^{-2}\,C_0^2\,r^{-1}\max\{|\nn_{v_0}|,1\}\,E\,
e^{-\x|\nn_{v_0}|}\; , $$
so that, if we recall Proposition 3.4, (which
allows us to bound $d\=d_{v_0}\le 2k-1$), then
\equ(5.2) immediately follows. \qed

\*

The concept of resonance is a very important one: in fact
it allows us to identify the terms which need an
improvement of the dimensional bound \equ(5.1) in order to prove
the convergence of the Lindstedt series. Indeed if there
were no resonances, the series would converge, as we can easily
show.

In order to obtain a bound on $\V X^{(k)}_\nn(\z)$, in general
we have to sum the values of all the labeled semitopological
trees. Such a sum can be arranged as
$$ \sum_{\th \in \TT^{(s)}_k} = \sum_{\th \in \Theta^{(s)} }
\sum_{\{\d_v\}_{v\in\th} \atop \sum_{v\in\th} \d_v = k}
\sum_{\{\z_v^1,\z_v^2\}_{v\in\th}}
\sum_{\{\nn_v\}_{v\in\th} } \sum_{\{n_v\}_{v\in\th} } \, $$
where the summations have to be performed by starting from
the rightmost one, and run over the sets defined in the
following way:
\acapo
(1) the first one is
over all the semitopological trees
with no labels;
\acapo
(2) the second one is over all the possible
assignments of the labels $\d_v$ to the vertices $v\in\th$
of a fixed unlabeled semitopological tree,
with the constraint $\sum_{v\in\th}\d_v=k$;
\acapo
(3) the third one is over all the possible
assignments of the labels $\z_v^1$, $\z_v^2$,
to the vertices $v\in\th$;
\acapo
(4) the fourth one is over the mode labels;
\acapo
(5) the last one is over the scale labels.

This exhausts all the possibilities, being all
the other labels uniquely determined by the just considered ones.

Then we perform the fifth sum, keeping the mode
labels (and all the other labels)
fixed, so obtaining $2^{2(2k-1)}$ terms.\footnote{${}^2$}{\nota
Because of the compact support of the propagators
(see comments between \equ(4.2) and \equ(4.3)), given a value
$\nn_\l$, there are only two consecutive scales
$n'=n,n+1$ such that $g^{(n')}(\oo\cdot\nn_{\l})$ is
different from zero, and there are at most $2k-1$ propagators
associated to the lines of the tree.
We note also that, given a tree, when all its labels are fixed
(in an arbitrary way),
most of the contributions we obtain are vanishing:
the mode labels fix uniquely the momenta running through the
lines of the tree, and, as a consequence of the previous paragraph,
only two scale labels associated to that line are possible. This simply
means that the sum over the trees in \equ(3.1) is restricted
over all the {\it compatible} trees. This property has
been taken into account in the labels counting in the text.}
Successively we write, as far the mode labels are concerned,
$$ \sum_{ \{\nn_v\}_{v\in\th} } = \sum_{\D(\th)}
\sum_{ M(\th) = |\D(\th)|}^{\io}
\sum_{ \{\nn_v\}_{v\in\D(\th)} \atop \sum_{v\in\th} |\nn_v| = M(\th) }
\Eq(5.3) $$
where $|\D(\th)|$ denotes the number of vertices
in $\D(\th)$.

We want to show that, if there are no resonances,
fixed $M(\th)$
in the rewriting \equ(5.3) of the sum in item (4),
then we obtain a quantity which depends in
a summable way on $M(\th)$, and the resulting expression can be
bounded by $C^k_2$ for some positive constant $C_2$.
In fact, if we are successful in proving such an estimate, then
the remaining sums, \ie the sums in items (1), (2) and (3),
are over $<2^{2(2k-1)}\cdot 2^{2k-1}\cdot 2^{2(2k-1)}$ terms,
(as there are four possible pairs $(\z_v^1,\z_v^2)$ and two
possible values $\d_v$, for each $v\in\th$,
and the number of unlabeled semitopological trees of
order $k$ is bounded by $2^{2(2k-1)}$), so that
an overall bound $G_0^k$ for some positive constant $G_0$
follows immediately, {\sl as far as the resonances are neglected},
(see (5.11) below).
The claimed estimate of the sum in item (4) is a consequence of the
following result.

\*

\\{\bf 5.3.} {\cs Lemma (Siegel-Bryuno's lemma).}
{\it The following bound holds for the number of lines $\l\in\th$
with scale $n_{\l}=n\le 0$ and $R_{\l}=j$, which
we denote by $N_n^j(\th)$:
$$ N_n^1(\th)+2N_n^2(\th) \le 8\,M(\th)2^{(n+2)/\t}
+ \sum_{T \atop n_T=n} \Big[ - 2
+ \sum_{j=1}^2 j\,m_T^j(\th) \Big] \; , \Eq(5.4) $$
where $m^j_T(\th)$ is the number of resonances $V$'s of $\th$
inside the cluster $T$,
having resonance-scale $n_{\l_V}=n_T$ and $R_{\l_V}=j$,
being $R_{\l_V}$ defined in \equ(3.10).}

\*

This is an adaptation of the Bryuno's proof, [B], of the
Siegel's lemma, [S1], as it is presented in [P\"o] and [G7]:
a proof is in Appendix A1.

\*

Therefore, fixed $M(\th)$, if we define $P(\th)=\sum_{v\in\th}p_v$,
we can bound
$$ \sum_{\{\nn_v\}_{v\in\D(\th)} \atop \sum_{v\in\th}|\nn_v|=M(\th)}
\Big[ \prod_{v\in\D(\th)} e^{-\x|\nn_v|} \,
{|\nn_v|^{p_v+1} \over p_v!} \Big] \le
e^{-\x M(\th)} \,
{ [M(\th)]^{P(\th)+(\ell+1)|\D(\th)|} \over [\ell\,|\D(\th)|]!\,
P(\th)!\,|\D(\th)|!}
\; , \Eq(5.5) $$
where $|\D(\th)|\le k$ and $P(\th)\le d < 2k$.

\*

\\{\bf 5.4.} {\it Proof of \equ(5.5).}
In the left hand side of \equ(5.5), if we set $D=|\D(\th)|$
and $P=P(\th)$, we can bound
$$ \eqalign{
\prod_{v\in\D(\th)}{|\nn_v|^{p_v+1}\over p_v!}
& \le
\prod_{v\in\D(\th)} |\nn_v|
\sum_{\sum_{v\in\D(\th)}p_v=P}
\prod_{v\in\D(\th)}{|\nn_v|^{p_v}\over p_v!}
\cr & \le
\left[ \sum_{\sum_{v\in\D(\th)} p_v=D}
\prod_{v\in\D(\th)} {|\nn_v|^{p_v} \over p_v!} \right]
\left[ \sum_{\sum_{v\in\D(\th)}p_v=P}
\prod_{v\in\D(\th)}{|\nn_v|^{p_v}\over p_v!} \right]
\cr & \le
{M^P\over P!}\,{M^D\over D!} \; , \cr} $$
where $M=M(\th)$ and in the sum in the first
brackets the $p_v$'s are only summation labels; moreover we have
$$ \sum_{\{\nn_v\}_{v\in\D(\th)} \atop \sum_{v\in\D(\th)}
|\nn_v|=M} 1 \le {M^{\ell\,D}\over [\ell\,D]!} \; , $$
so that \equ(5.5) follows. \qed

\*

Then we proceed in the following way in order to bound the small
divisors contribution. Let be $n_0$ a negative integer value to be
fixed later. We can write
$$ \prod_{n\le0}\prod_{j=1}^2 2^{-jnN_n^j(\th)}
\le 2^{-2n_0 k}
\prod_{n\le n_0} \prod_{j=1}^2 2^{-jnN_n^j(\th)} \; , \Eq(5.6) $$
where the scale label $n=1$ can be forgotten, since the
propagators on scale $n=1$ can be bounded by 1.
Then the product in the right hand side of \equ(5.6) can be bounded by
$$ \eqalign{
\prod_{n\le n_0} \prod_{j=1}^2 & 2^{-jnN_n(\th)} \cr \le
& \exp \Big[ 8\,\ln 2\,M(\th)\sum_{n=-\io}^{n_0} (-n2^{n/\t}) \Big]
\prod_{n\le0}\prod_{T \atop n_T=n} 2^{2n}
\prod_{j=1}^2 2^{-jnm_T^j(\th)}\; , \cr} \Eq(5.7) $$
where the product on $n$ in the right hand
side is extended also to the scale labels $n>n_0$:
obviously this could be avoided, but we keep it so
in order to not complicate the analysis of the cancellations,
since we are not looking for an optimal bound.
The sum in the
argument of the exponential, in the right hand side of \equ(5.7),
can be bounded by $c_{n_0}M(\th)\,2^{n_0/\t}$, for
some constant $c_{n_0}>0$ depending on $n_0$ and explicitly computable:
$$ c_{n_0}=8\,\ln 2 \sum_{p=0}^{\io} (p-n_0)\,2^{-p/\t} \; . $$
Therefore, fixed
the value $\x$, we can choose $n_0=n_0(\x)$, such that
$ c_{n_0(\x)}2^{n_0(\x)/\t} <\x/2$, so that we can bound
$$ \eqalign{
\prod_{v\in\D(\th)} & e^{-\x|\nn_v|} \,
{|\nn_v|^{p_v+1} \over p_v!}
\prod_{n\le0}\prod_{j=1}^2 2^{-jnN_n^j(\th)}
\cr & \le
e^{-\x M(\th)/2}\,
{ [M(\th)]^{P(\th)+(\ell+1)|\D(\th)|} \over
[\ell\,|\D(\th)|]!\,P(\th)!\,|\D(\th)|! }
\,2^{-2n_0(\x)k}
\prod_{n\le0}\prod_{T,n_T=n} 2^{2n} \prod_{j=1}^2
2^{-jnm_T^j(\th)} \; , \cr} \Eq(5.8) $$
(note that such a $n_0(\x)$ depends on $\t$ as $a_1\t\ln\t$, for
some constant $a_1<0$; if $\t>\ell-1$, then we
have $2^{-2n_0(\x)}\sim (\ell !)^{-a_2}$, for some positive
constant $a_2$). If we take into account that
$$ { [M(\th)]^{P(\th)+(\ell+1)|\D(\th)|} \over
[\ell\,|\D(\th)|]!\,P(\th)!\,|\D(\th)|! }
\le C_4^{-[P(\th)+(\ell+1)|\D(\th)|]}\,\exp[C_4\,M(\th)]
\le C_4^{-(\ell+3)k}\,\exp[C_4\,M(\th)] \; , \Eq(5.9) $$
we can choose $C_4$ small enough so that $C_4\le\x/4$,
and write, in \equ(5.3),
$$ \sum_{\D(\th)} = \sum_{D=0}^k \sum_{\D(\th) \atop
|\D(\th)|=D} = \sum_{D=0}^k \left( \matrix{ k \cr D \cr} \right)
= 2^k \; , \Eq(5.10) $$
so obtaining, {\sl if we
neglegt the resonances}, a well defined
expression which is summable
on $M(\th)$ and gives a bound
$G_0^k$ for some positive constant $G_0$,
whose (non optimal) value
can be deduced from the above discussion:
$$ G_0 = 2^{15}\,\CC\,(4/\x)^{\ell+3}\,2^{-2n_0(\x)} \; . $$
Then we have obtained a bound on $\V X_{\nn}^{(k)}(\z)$ of the form
$$ |\V X_{\nn}^{(k)}(\z)|
\le e^{-\x|\nn|/4}\,G_1\,G_0^k G_2^k\; . \Eq(5.11) $$
for some constant $G_1>0$, (the above discussion gives
$G_1=2^{-7}[1-\exp(-\x/4)]^{-1}$), and for $G_2=1$ if we neglect the
resonances.

However the presence of resonances has the effect that,
for each tree $\th$, we have to take into account also the factor
$$ \Big[
\prod_{n\le 0} \prod_{T \atop n_T=n} 2^{2n}\,
2^{-(2nm_T^2(\th)+nm_T^1(\th))} \Big] \; , \Eq(5.12) $$
arising from the resonant lines (see Definition 5.1).
It is possible to show that there are trees $\th$ such that
the dimensional bound \equ(5.1) gives a behaviour
$C^k(k!)^{\a}$, for some positive constants $C$ and $\a$,
(an explicit example can be found in [E1], \S II). By taking into
account the cancellations occurring between the various
tree values contributing to the same perturbative order,
it is possible to see that a bound \equ(5.11) is still possible,
for some constant $G_2>1$. To the proof of such an assertion next two
sections are devoted.

\vskip1.truecm

\centerline{\titolo 6. Approximate cancellations of the resonances}
\*\numsec=6\numfor=1

\\In this section, it will be more convenient to work with
the numbered trees, (see 3.2). In fact, if we recall
\equ(3.3) and we consider numbered trees, \equ(4.4) has
to be replaced with
$$ \hbox{Val}(\th)= {1\over d!}
\prod_{v\in\th} O_v\,g^{(n_{\l_v})}
(\oo\cdot\nn_{\l_v}) \; , \Eq(6.1) $$
and \equ(3.1) becomes
$$ \V X_\nn^{(k)}(\z)= \sum_{\th\in\TT^{(n)}_k}
\hbox{Val}(\th) \; , \Eq(6.2) $$
where the sum is over all the labeled numbered trees of order $k$.
Note that two trees have to be regarded as identical if they
are topologically equivalent, (\ie they are superposable
modulo a transformation of the group $\GG$ defined in 3.1),
and all their labels (included the numbers
associated to the branches) match, (see also Remark 3.5).
The advantage of dealing with the perturbative
expansion in terms of numbered trees is that in such a way each
tree is ``weighted'' with the same combinatorial factor.

\*

Let us introduce some notations to classify the resonances: this
will be useful in the following.

\*

\\{\bf 6.1.} {\cs Definition (Resonance factor).}
{\it Let be given a resonance $V$;
let $\l_V$ be the incoming line, as in Definition 5.1.
We denote by $w_0$ the vertex
from which the outgoing line of $V$ comes out,
and by $w_1$ the vertex from which the incoming line of $V$ comes
out, (then the vertex $w_1$ is outside the resonance,
but $w_0\in V$). Let us consider the labels
$\z_{w_0}^2$ and $\z_{w_1}^1$, which can assume only the values
$H$ and $h$, (by construction the value $\m$ is forbidden
in such cases).
We denote by $\PP(w_0,w_1')$ the (unique) path leading from
$w_0$ to $w_1'$, being $w_1'$ the vertex immediately preceding $w_1$,
(\ie $\l_V\=\l_{w_1}$ is the line $w_1'\ot w_1$).
Let us define the {\rm resonance factor}
$\VV^{n_{\l_V}}_{\z_{w_0}^2,\z_{w_1}^1}(\oo\cdot\nn_{\l_V})$
as the quantity
$$ \VV^{n_{\l_V}}_{\z_{w_0}^2,\z_{w_1}^1}(\oo\cdot\nn_{\l_V})=
\prod_{w\in V} O_{w} \prod_{\l\in V}
g^{(n_{\l})}(\oo\cdot\nn_\l) \; , \Eq(6.3) $$
where the first product is over all the $d(V)$ vertices inside
the resonance $V$ and the second one is over the $d(V)-1$ lines
internal to $V$. Let us define $V_0$ as
the collection of lines and vertices in $V$ external to
the maximal resonances contained inside $V$,
(we say that a line is in $V_0$ if both
its extremes are in $V$ and at least one of them
is in $V_0$).}

\*

We modify the rules how to
construct the trees by splitting each resonance factor
$\VV$ as $\VV=\LL\VV+(\openone - \LL)\VV$, where
$$ \eqalign{
\LL \VV^n_{h,h}(\oo\cdot\nn) & =
\VV^n_{h,h}(0) + [\oo\cdot\nn]\,\dot\VV^n_{h,h}(0) \; , \cr
\LL \VV^n_{H,h}(\oo\cdot\nn) & = \VV^n_{H,h}(0) \; , \cr
\LL \VV^n_{h,H}(\oo\cdot\nn) & = \VV^n_{h,H}(0) \; , \cr
\LL \VV^n_{H,H}(\oo\cdot\nn) & = 0 \; , \cr} \Eq(6.4) $$
where $\dot\VV^n_{\z_{w_0}^2,\z_{w_1}^1}(0)$ denotes
the first derivative of
$\VV^n_{\z_{w_0}^2,\z_{w_1}^1}$ with respect to $\oo\cdot\nn$, computed in
$\oo\cdot\nn=0$.
The operator $\LL$ will be called {\sl localization operator}.

\*

\\{\bf 6.2.} {\it Remark.}
Note that the resonance factors depend on
$\oo\cdot\nn$ only through the propagators.
Then, for each line $\l$ inside the resonance, the momentum
flowing in it is given by $\nn_\l\=\nn_\l^0+\e_\l\nn$,
where $\nn_\l^0$ is the sum of the mode labels corresponding to
the vertices following $\l$ but inside the resonance, and $\e_\l=1$
if $\l\in\PP(w_0,w_1')$, and $\e_\l=0$ otherwise.
Even if we set $\oo\cdot\nn=0$, (\ie $\oo\cdot\nn_\l=
\oo\cdot\nn_\l^0$ for each $\l$ inside the resonance), no too
small divisor appears because of the presence of the compact support
functions $\ch_{n_\l}(\oo\cdot\nn_\l)$, $n_\l>n$.

\*

Given a tree, on any cluster the $\LL$ or $\openone-\LL\=\RR$
operators apply. We want to show in this section
that the contributions arising from diagrams containing
resonances on which the $\LL$ operator applies add to zero,
so that we can rule out such contributions and consider only
trees with resonances on which the operator $\RR$ applies.
The latter kind of trees will be studied in \S 7, where the
convergence of the Lindstedt series will be eventually proven.

\*

\\{\bf 6.3.} {\cs Definition (Resonance family).}
{\it Given a tree $\th$ with some resonances, let us consider the
family of trees $\FF_V(\th)$ obtained from $\th$ in the following
way. Given a resonance $V$ in $\th$,
if $\z_{w_1}^1=h$, we add to $\th$ the trees we
obtain by detaching from the resonance the subtree with root
in $w_1'$, then reattaching it to all the remaining vertices $w\in V$
having $\d_w=1$ and external to the resonances internal to $V$;
if $\z_{w_0}^2=h$,
to the just considered trees we add all the trees we obtain
by detaching the outgoing line of the resonance from the vertex $w_0$,
then reattaching to all the remaining vertices $w\in V$
having $\d_w=1$ and external to the resonances internal to $V$. The
number of terms so obtained is $\tilde k(V)^2$,
(if $\z_{w_0}^2=\z_{w_1'}^1=h$), and $\tilde k(V)$, (if only
one of the two labels $\z_{w_0}^2$, $\z_{w_1'}^1$
assumes the value $h$), where
$\tilde k(V) \= k(V)-\sum_{V'\subset V}k(V')$, being the
sum extended over all the maximal resonances internal to $V$.
We call $\FF_V(\th)$ a {\rm resonance family} (associated to
the resonance $V$).}

\*

The simplest case of resonance family is drawn in Fig.6.1.

%
\midinsert
\*
\insertplot{400pt}{80pt}{
\ins{37pt}{80pt}{$\th_1$}
\ins{137pt}{80pt}{$\th_2$}
\ins{237pt}{80pt}{$\th_3$}
\ins{337pt}{80pt}{$\th_4$}
\ins{-10pt}{42pt}{$\hbox{root}$}
\ins{90pt}{42pt}{$\hbox{root}$}
\ins{190pt}{2pt}{$\hbox{root}$}
\ins{290pt}{2pt}{$\hbox{root}$}
\ins{28pt}{42pt}{$v_1$}
\ins{128pt}{42pt}{$v_1$}
\ins{228pt}{42pt}{$v_1$}
\ins{328pt}{42pt}{$v_1$}
\ins{45pt}{18pt}{$v_2$}
\ins{145pt}{18pt}{$v_2$}
\ins{245pt}{18pt}{$v_2$}
\ins{345pt}{18pt}{$v_2$}
\ins{83pt}{0pt}{$v_3$}
\ins{183pt}{40pt}{$v_3$}
\ins{283pt}{40pt}{$v_3$}
\ins{383pt}{0pt}{$v_3$}
}{fig4}
\vskip.6truecm
\didascalia{{\bf Fig.6.1.} The possible resonance families
$\FF_V(\th)$'s associated to resonance $V$'s with degree $d(V)=2$:
(1) if $\z_{w_0}^2=\z_{w_1}^1=h$, one has only one family
with four trees, $\FF_V(\th_1) = \{
\th_1$, $\ldots$, $\th_4 \}$,
(2) if $\z_{w_0}^2=H$ and $\z_{w_1}^1=h$,
one has two families with two trees, $\FF_{V_1}(\th_1) = \{ \th_1,$
$\th_2 \}$ and $\FF_{V_2}(\th_3) = \{ \th_3,$ $\th_4 \}$,
(3) if $\z_{w_0}^2=H$ and $\z_{w_1}^1=h$,
one has two families with two trees, $\FF_{V_1}(\th_1)
= \{ \th_1,$ $\th_4 \}$ and $\FF_{V_2}(\th_2) = \{ \th_2,$ $\th_3 \}$,
(4) if $\z_{w_0}^2=\z_{w_1}^1=H$, one has four one-tree families
$\FF_{V_i}(\th_i) = \{ \th_i \}$, $i=1,\ldots,4$.
Note that, unlike the labels $\z_{w_0}^2$ and $\z_{w_1}^1$,
the location of the vertices $w_0$ and $w_1'$ varies inside
the family $\FF_V(\th)$. For instance, in the case (1),
one has $w_0=v_1$ in $\th_1$ and $\th_2$, and $w_0=v_2$
in $\th_3$ and $\th_4$, and, analogously, $w_1'=v_1$
in $\th_2$ and $\th_3$, and $w_1'=v_2$ in $\th_1$ and $\th_4$.
Note also that, if $\th'\in\FF_V(\th)$, then $\FF_V(\th)=
\FF_V(\th')$: this simply means that a resonance family
can be defined with respect to any tree it contains.
For instance, in item (1), one can define the resonance
family as $\FF_V(\th_i)$, $\forall i=1,\ldots,4$.
The resonance $V$, containing the vertices
$v_1$ and $v_2$, has scale $n_V\ge n_{\l_V}+1$, if $n_{\l_V}$
is the scale of the resonant line $\l_V\=\l_{v_3}$ (which is
equal to the scale of the line entering the root): the line
connecting the vertices $v_1$ and $v_2$ is a line $v_1\ot v_2$
in $\th_1$ and $\th_2$, and is a line $v_2\ot v_1$ in $\th_3$
and $\th_4$.}
\endinsert
%

Note that each time we shift the outgoing line of a resonance,
we produce an {\sl apparently} deep change on the tree value. In fact
all the arrows superposed to the lines point
to the tree root, so that all the arrows of the lines
inside the resonance have to point to the vertex $w_0$ to
which the outgoing line is reattached,
(note that in general, given a tree $\th'\in\FF_V(\th)$,
we call $w_0$ the vertex from which the outgoing line of the
resonance emerges, and $w_1'$ the vertex which the incoming line
$w_1'\ot w_1$
enters: then the location of $w_0$ and $w_1$ depends
on the particular tree in $\FF_V(\th)$,
while the values $\z_{w_0}^2$ and $\z_{w_1'}$ are the
same for any tree contained in the resonance family, so that
their values do not depend on the vertices $w_0$ and $w_1'$;
see also Fig.6.1).
This means that some
arrows superposed to the branches inside the resonance change
their direction, and, correspondingly, the lines $h\ot H$
become lines $H\ot h$ and {\it viceversa}.
The reason why we say that the change is only apparently deep
will become clear in Lemmata 6.4 and 6.6 below.

In general, fixed
the value of the momentum $\nn_{\l}$ associated to the line $\l$,
only two scale labels $n_{\l}$ give a propagator $g^{(n_{\l})}
(\oo\cdot\nn_{\l})$ which is not vanishing, (see note 5).
It can happen that the shift of the incoming and outgoing lines
produces a change of the value of the momenta flowing through
the lines inside the resonance.
This means that, fixed the scale labels inside the resonance
$V$ of a tree $\th$,
some of the resonance factors associated to trees
in $\FF_V(\th)$ (hence obtained by shifting of external lines)
can vanish, as containing vanishing
propagators.

However we can proceed in a different (and more suitable way):
we ignore the fact that, fixed all the labels, there are
vanishing contributions, and we consider all the trees inside
the family $\FF_V(\th)$ as possible (\ie as they were
compatible trees). In fact we shall see below that
each (vanishing and not vanishing)
resonance factor $\VV$ can be written as sum of
two parts, $\VV=\LL\VV+\RR\VV$, such that (1) the first one,
$\LL\VV$, is exploited in order to
obtain a cancellation, while (2) the latter, $\RR\VV$, can be
easily bounded. The proof of assertion (1) is
given in the remaining part of this section and leads to
Corollary 6.8, and the to the proof of assertion (2)
next section is devoted. Ovbiously, when the resonance factor is
vanishing, this simply means that the two parts of the
decomposition are equal and opposite,\footnote{${}^3$}{\nota
The fact that the two parts into which a vanishing
resonance factor is decomposed are opposite, but
are bounded in a different way is not so surprising.
In fact we shall see in \S 7 that a gain $(2^{n_{\l_V}}/2^{n_V})^R$,
$R=1,2$, is obtained for the part $\RR\VV$ of the
resonance factor associated to a resonance $V$. If the
resonance factor $\VV$ is vanishing, then there has to be
some vanishing propagator $g^{(n_{\l})}(\oo\cdot\nn_{\l})$
corresponding to a line $\l$ inside $V$. If $\oo\cdot\nn_{\l}^0$
is far from the boundary of the support
of the $\ch_{n_{\l}}$ function, \ie centered near to $2^{n_{\l}}$,
(the momentum $\nn_{\l}^0$ is defined in Remark 6.2), then
$\oo\cdot\nn$ has to be large enough so that $\oo\cdot\nn_{\l}$
$=$ $\oo\cdot\nn_{\l}^0+\oo\cdot\nn$ is outside the interval
$[2^{n_{\l}-1},2^{n_{\l}+1}]$: in other words $\oo\cdot\nn$
has to be comparable with $\oo\cdot\nn_{\l}^0$, and so the
ratio $2^{n_{\l_V}}/2^{n_V}$ is bounded below by a constant and no true
gain is obtained with respect to the part $\LL\VV$ of the
decomposition. On the contrary, if $\oo\cdot\nn_{\l}^0$
is near to the boundary,
the considered ratio is small, but in such a case the function
$\chi_{n_{\l}}$ appearing in $\LL\VV$
is much lesser than 1, and the $\LL\VV$ is much
smaller than the dimensional bound can suggest.
In other words, in case (1) $\RR\VV$ is not truely
smaller than $\LL\VV$, in case (2) the $\LL\VV$ is
not truely larger than $\RR\VV$, and we can conclude
that, when the resonance factor is vanishing, the bounds
on the two parts into which it is decomposed are only
apparently different.}
so that the first one
is not really dangerous, and could be directly bounded without
using the cancellations. However one of the advantages of
the techniques we are using is that we can avoid
to distinguish between dangerous and not dangerous
contributions, and we can treat all resonances in the same way,
(see also the comments in \S 1).
Note that the number of extra trees we are counting by
considering also the vanishing contributions is bounded by
$C_3^k$, for some constant $C_3>0$.

\*

Then the following lemmata hold, which permit us to control
the differences between the values of the trees contained
in the same resonance family.

\*

\\{\bf 6.4.} {\cs Lemma.} {\it Consider a resonance
family $\FF_V(\th)$ obtained
from a tree $\th$ such that $V\in\th$ is a resonance, and
set $\nn=\V0$, if $\nn$ is the momentum flowing through
the branch entering the resonance ($\nn=\nn_{\l_V}$).
Then the product of the operator associated to any line $\l\in V$,
when applied to the function \equ(3.6),
times the corresponding propagator
assumes the same value for each tree $\th'\in\FF_V(\th)$.}

\*

\\{\bf 6.5.}
{\it Proof of Lemma 6.4.} That the product of the
operators times the propagators associated to
the lines in $\th$ do not change by setting $\nn=\V0$ and shifting the
incoming line $w_1'\ot w_1$, hence the point $w_1'$,
is trivial, and is due simply to the fact that (1) if
$\z_{w_1}^1=h$, the line $\l_{w_1}\=\l_V$ does not carry
any derivative acting on $f_{\nn_{w_1'}}(\AA_{w_1'})$, and
(2) if $\nn=\V0$,
the incoming line of the resonance does not contribute anymore
to the momenta flowing along the branches $\l\in V$, so that
no memory is left about the vertex which the line $\l_V$ enters,
(see Remark 6.2).

The independence of the product of the operators
times the propagators
on the location of the vertex $w_0$
from which the outgoing line of the resonance emerges
can be argued in the following way.
Note that, {\sl only} if $\z_{w_0}^2=h$,
the outgoing line $\l_{w_0}$
is shifted, and it is reattached {\sl only} to
vertices with $\d_w=1$, (by construction of the family $\FF_V(\th)$).

The arrows of the lines inside the resonance point all
to the tree vertex $w_0'$. If we detach the outgoing line from
$w_0\=w_2$, and reattach it to another vertex, say $w_3$, then
the direction of the arrows can change or not, in according to
the following rule: if the line either is contained simultaneously
in both paths $P(w_2,w_1')$ and $P(w_3,w_1')$, or is not
contained in none of the two paths, then the arrow
direction does not change, while, if the line is in only
one of the two paths, its direction changes.
Obviously, only the propagators and operators corresponding
to lines with reversed arrows can be different.

To treat such lines, we start by noting that each line
$v'\ot v$ inside the resonance $V$ realizes a partition
of the resonance into two subsets $W_v^1$ and $W_v^2$ such
that $W_v^1=\{w\in V: w \ge v \}$ and $W_v^2$ $=$ $V\setminus W_v^1$.
Since $\sum_{w\in V}\nn_w=\V0$, we have trivially
$\sum_{w\in W_v^1}\nn_w=-\sum_{w\in W_v^2}\nn_w$.
We define $\nn(v)$ $=$ $\sum_{w\in W_v^1}\nn_w$, and
we (arbitrarily) fix the set $W_v^1$ as the collection
of vertices $w\ge v$ in $V$ when the momentum flows from $v$
to $v'$, (the same convention will be followed henceforth).

Then if
the arrow direction of the line $\l_v$ changes, this means
that the line $\l_v$ should be denoted
by $\l_{v'}$, and the momentum flowing through it
(by setting $\oo\cdot\nn$ and recalling the definition of
$\nn_{\l}^0$ given in Remark 6.2)
is no more $\nn_{\l_v}^0=\nn(v)=\sum_{w\in W_v^1}\nn_w$,
but becomes $\nn_{\l_{v'}}^0$ $=$
$\sum_{w\in W_v^2}\nn_w\=-\nn(v)$,
so that it simply changes
its direction, \ie its sign.
Note that we are taking into account the fact
that, when the arrow changes its direction, then the momentum
flowing through the line $\l_v$ becomes the sum of the mode labels
associated to the vertices in $W_v^2$, which assumes now
the same r\^ole that $W_v^1$ had before the arrow reversal:
then if the arrow points from $v$ to $v'$, we have $\n_{\l_v}^0=\nn(v)$,
while if it points from $v'$ to $v$, we have $\nn_{\l_v}^0=-\nn(v)$.

Then, if we look at \equ(3.4), we see that, given a line $\l_v\in V$,
the case
$\z_{v}^1=\z_v^2=h$ can be dealt with by
considering the product of the operator times the propagator,
and by noting that
the matrix $T(\AA_0)$ is symmetric, so that the numerator
does not change by arrow reversal, and the same the denominator does,
depending quadratically on the momentum.

The other two cases $\z_{v}^1=h$, $\z_v^2=H$
and $\z_{v}^2=H$, $\z_v^1=h$ can be easily dealt with, by noting
that if we change the arrow direction, we change the sign
of the numerator, (see Remark 3.4),
and reverse the direction of the momentum,
so changing also the sign of the denominator: then the overall
sign remains the same.
\qed

\*

It can be useful to define the quantity
$\WW^n_{\z_{w_0}^2,\z_{w_1}^1}(\oo\cdot\nn)$ as
$$ \VV^n_{\z_{w_0}^2,\z_{w_1}^1}(\oo\cdot\nn)
= (-i\MM\,\nn_{w_0})\,\,(i\nn_{w_1'})\,
\WW^n_{\z_{w_0}^2,\z_{w_1}^1}(\oo\cdot\nn)
\; , \Eq(6.5) $$
where the matrix $\MM$ is defined to be $\MM=\openone$,
(if $\z_{w_0}^1=H$), or
$\MM=T(\AA_0)$, (if $\z_{w_0}^1=h$).

\*

\\{\bf 6.6.} {\cs Lemma.} {\it Consider a resonance
family $\FF_V(\th)$, and let $\VV^{n}_{\z_{w_0}^2,
\z_{w_1}^1}(\oo\cdot\nn)$ be the resonance
factor associated to the resonance $V$ in $\th$. Suppose that
$\z_{w_0}^2=\z_{w_1}^1=h$: this means that,
in according to which vertex $w_0$ the outgoing line
of the resonance comes out from and to which
vertex $w_1'$ the incoming line enters,
the resonance factor contains the two factors
$(-i\MM\nn_{w_0})$ and $i\nn_{w_1'}$. All the others
factors appearing in the $O_v$'s, $v\in V$,
are the same ones for all trees in $\FF_V(\th)$.
Then if we sum together all the contributions
{\rm to first order in $\oo\cdot\nn$}
arising from the family $\FF_V(\th)$, we
see that they differ only as far as the resonance factors
are concerned, so that we obtain a factorising term times
the following expression
$$ \eqalign{
\sum_{\th'\in\FF_V(\th)} & (\oo\cdot\nn)
\dot\VV^n_{\z_{w_0}^2,\z_{w_1}^1}(0) \cr
= &
\; (\oo\cdot\nn) \, \WW^{n}_{h,h}(0) \sum_{v\in V}
\Big[ -{ R_{\l_v} \over \oo\cdot\nn(v) }
+ { \dot \chi_{n_{\l_v}}(\oo\cdot\nn(v)) \over
\chi_{n_{\l_v}}(\oo\cdot\nn(v)) } \Big] \; \cdot \cr
& \cdot \; \Big[
\sum_{w_0 \in W_v^2 \cap V_0 \atop w_1' \in W_v^1 \cap V_0} -
\sum_{w_0 \in W_v^1 \cap V_0 \atop w_1' \in W_v^2 \cap V_0}
\Big]
(-i\MM\,\nn_{w_0}) (i\nn_{w_1'}) \; , \cr} \Eq(6.6) $$
where:
\acapo
(1) in the left hand side $\dot\VV_{\z_{w_0}^2,\z_{w_1}^1}(0)$
depends on the tree $\th'$, (though the dependence
is not explicitly shown);
\acapo
(2) the matrix $\MM$ is defined to be $\MM=\openone$ or $\MM=T(\AA_0)$;
\acapo
(3) $\dot
\chi_{n_{\l_v}}(\oo\cdot\nn(v))$ denotes the first derivative
of $\chi_{n_{\l_v}}(\oo\cdot\nn(v))$ with respect to its argument;
\acapo
(4) the quantity $\WW_{h,h}(0)$
is defined as in \equ(6.5), with $\z_{w_0}^2=\z_{w_1'}^1=h$; and
\acapo
(5) $\nn(v)\=\sum_{w\in W_v^1}\nn_w$; this means
that $\nn(v)=\sum_{v\le w\in V_0}\nn_w=\nn_{\l_v}^0$
only if $w_0\in W_v^2$, if $\nn_{\l_v}^0$ is defined as in Remark 6.2.}

\*

\\{\bf 6.7.}
{\it Proof of Lemma 6.6.} It is enough to note that only
the propagators of the lines inside $V$ can depend on $\oo\cdot\nn$,
so that, when we derive a term of the form
$\chi_{n_{\l}}(\oo\cdot\nn_{\l})\,
[i\oo\cdot\nn_{\l}]^{-R_{\l}}$
we obtain $-i\,R_{\l}\,\chi_{n_{\l}}(\oo\cdot\nn_{\l})\,
[i\oo\cdot\nn_{\l}]^{-R_{\l}-1}$
$+$ $\dot\chi_{n_{\l_v}}(\oo\cdot\nn(v))$
$[i\oo\cdot\nn_{\l}]^{-R_{\l}}$,
each time the line $\l$ is along the path $\PP(w_0,w_1')$
and zero otherwise. Then we sum together
all the contributions arising from the family $\FF_V(\th)$,
taking into account the fact that $\VV_{h,h}(0)$
assume the same value for all contribution,
(by Lemma 6.4), and rearrange the sum, by (1) fixing the line
and considering together all the trees in which the
path $\PP(w_0,w_1')$ contains that line, and
(2) exploiting the fact that the momentum $\nn_{\l_w}^0$
flowing through the line $\l_w$ is directed from $w_1$
toward $w_0$, so that it is $\nn(w)$ if $w_0\in W_v^2$
and $-\nn(w)$ otherwise. We are taking into account also the
fact that $\chi_{n_\l}(\oo\cdot\nn_{\l})$ is an even
function of its argument (see the definition given in \S 4),
so that the first derivative is odd.
The condition $w_0, w_1' \in V_0$
follows from the definition of the resonance family $\FF_V(\th)$.
\qed

\*

 From Lemma 6.4 and Lemma 6.6, we deduce immediately the
following corollary, which can be considered an extension
of the corresponding results of [G7], [CF1] and [GM2].

\*

\\{\bf 6.8.} {\cs Corollary.} {\it If we sum together all the values
corresponding to the trees $\th'$ contained in a
resonance family $\FF_V(\th)$,
then the contributions obtained by applying the $\LL$ operator
to the resonance factors identically vanish.}

\*

\\{\bf 6.9.}
{\it Proof of Corollary 6.8.} As we have said, the proof is an easy
consequence of Lemma 6.4 and Lemma 6.6.
If $\z_{w_2}^1=h$, we obtain a vanishing
contribution to first order, because we have cancellations
between the terms in $\FF_V(\th)$ we obtain by shifting the
incoming branch.
In fact by Lemma 6.4 the various contributions we obtain
differ only because of a factor $i\nn_{w_1'}$ associated
to the vertex $w_1'$:
when such an operation is performed we can choose as $w_1'$
vertices only the vertices $w\in V_0$ having $\d_w=1$, but
$$ \sum_{w\in V_0 \atop \d_w=1}\nn_w \= \sum_{w\in V}
\nn_w \=\V0\; , \Eq(6.7) $$
since $\nn_w\=\V0$ for $w$ with $\d_w=0$ and $\sum_{w\in V'}
\nn_w = \V0$ for any resonance $V'\subset V$,
so that $\sum_{w_1'\in V_0}\nn_{w_1'}=\V0$.

The same happens if $\z_{w_0}^2=h$, and the cancellations are
between the terms we obtain by shifting the outgoing branch.

If both $\z_{w_0}^2=h$ and $\z_{w_1}^1=h$,
we have a zero to second order, by Lemmata 6.4
and 6.6. In fact the zero to first order follows from
the above discussion, while the zero to second order
can be easily deduced from \equ(6.6): if we perform
explicitly the sums over the vertices $w_0$ and $w_1$,
we obtain, {\sl for any fixed $v$},
$[-i\MM\,(-\nn(v))]\,[i\nn(v)]$ $-$ $[-i\MM\,\nn(v)]\,
[i(-\nn(v))]$, which is trivially zero, (the matrix $\MM$
being defined after \equ(6.5)).

Then, if we recall the definition of the localization operator,
see \equ(6.4),
the statement of Corollary 6.8 follows immediately. \qed

\vskip1.truecm

\centerline{\titolo 7. Convergence of the Lindstedt series for KAM tori}
\*\numsec=7\numfor=1

\\In order to prove that $|\V X^{(k)}_\nn(\z)|\le
C^k$, for some constant $C$, we shall find convenient to modify
the definition of the functionals to associate to
the lines $h \ot H$.

Obviously the proof can be carried out
even if such a change is not introduced: however it will
turn out to simplify the analysis in a relevant way.
If we consider the graph rules given in \S 3 (see
in particular \equ(3.4)) and the definition of localization operator
in \equ(6.4), we can check that the are:
\acapo
(1) resonances such that $\z_{w_0}^2=\z_{w_1}^1=H$ (and with
external lines $\l$'s having by construction $R_{\l}=1)$,
for which no gain is obtained,
\acapo
(2) and resonances such that $\z_{w_0}^2=\z_{w_1}^1=h$ and
the incoming line is a $h\ot H$ line (\ie $\z_{w_1}^2=H$),
for which a larger than needed gain is obtained.

The reason why we say that the gain obtained for
resonances in item (2) is larger than how it is needed
can be understood by considering
how the cancellation mechanism described in \S 6 works.

Each time we have a resonance, there are two lines
$\l_{w_0}$ and $\l_{w_1}\=\l_V$ (the external lines of the resonance)
such that we can interpret the product of the corresponding
propagators as an ``effective propagator''
$[\oo\cdot\nn_{\l_V}]^{-R_{\l_{w_0}}-R_{\l_V}}$; if we have a chain
of resonances, \ie a sequence of resonances $V_1,\ldots,V_{\NN}$,
such that the incoming line of $V_i$, $i=1,\ldots,\NN-1$ is
the outgoing line of $V_{i+1}$, we can associate to it an
``effective propagator''
$$ [\oo\cdot\nn_{\l_{V_1}}]^{-R_{\l_{w_0}}}
\prod_{i=1}^{\NN} [\oo\cdot\nn_{\l_{V_1}}]^{-R_{\l_{V_i}}} \; , $$
where $\nn_{\l_{V_1}}=\ldots=\nn_{\l_{V_{\NN}}}$ and $\l_{w_0}$
is the outgoing line of the first resonance of the chain, \ie $V_1$.

Then the cancellation of the localized parts allows
us to extend the discussion in \S 5 in such a way to
cover also the resonances, if we can obtain
a quantity $O([\oo\cdot\nn_{\l_V}]^{R_{\l_V}})$,
up to factors which can be controlled,
from the bound on the resonance factor
associated to any resonance $V$,
once its localized part has been subtracted.

In fact problems arise when a chain of resonances appear
in a tree, so that there is a lot of ``repeated small
divisors'': but if a gain $O([\oo\cdot\nn_{\l_V}]^{R_{\l_V}})$
is obtained for each resonance $V$ in the chain, then
the corresponding effective propagator reduces itself to
$O([\oo\cdot\nn_{\l_{V_1}}]^{R_{\l_{w_0}}})$.

If we look at \equ(6.4), we see that the required
property is not fulfilled by the resonances in item (1) above.
On the other hand, the resonances
in item (2) have a gain which is
$O([\oo\cdot\nn_{\l_V}]^{R_{\l_V}+1})$.

However, if we consider a single
resonance of the kind in item (1), we can simply say that the effective
propagator is quadratic, and no problem arises {\sl unless
if a chain of such resonances on the same scale occur}.
But by construction this is possible only if resonances
as in item (2) are inserted, so that they provide
the extra required gain.

In order to simplify the analysis and take into account
directly the just described compensations along the chains,
we slightly change the definition of the functionals
associated to the lines $h\ot H$,
as anticipated in the beginning of this section.
Then the second row in \equ(3.2)
will be replaced with
$$ \matrix{
\hbox{operator} & \hbox{propagator} & \hbox{branch} \cr
& & \cr
C_0 \left[ i\nn_{v'}\cdot(\dpr_{\AA_v}) (i\oo\cdot\nn_{\l_v})
\right] & [i\oo\cdot\nn_{\l_v}]^{-2} & h\ot H \cr} \Eq(7.1) $$
(in such a way that the product of the operator times
the propagator remains the same, as it has to be).
Then the propagators will be always of the form \equ(3.8),
but $R_{\l_v}$ $=$ $1$ if $\z_v^1$ $=$ $H$ and $R_{\l_v}$ $=$ $2$ if
$\z_v^1$ $=$ $h$. Then \equ(5.12) has to be replaced by
$$ \Big[
\prod_{n\le 0} \prod_{T \atop n_T=n} 2^{2n}\,
2^{-(2nm_T^2(\th)+nm_T^1(\th))} \Big]
\prod_{v \in \th \atop \z_v^1=h, \z_v^2=H} 2^{n_{\l_v}} \; . \Eq(7.2) $$
while a bound $G_0^k$ still holds for the
resonanceless trees, with the same constant $G_0$,
because some lines which in \S 5
were counted among $m_T^1(\th)$ contribute now to $m_T^2(\th)$,
(they are exactly the resonant lines $h\ot H$).

The definition \equ(7.1) allows us to
change the definition of the $\LL$ operator and
substitute the equations \equ(6.4) with the following ones:
$$ \eqalign{
\LL \VV^n_{h,h}(\oo\cdot\nn) & =
\VV^n_{h,h}(0) + [\oo\cdot\nn]\,\dot\VV^n_{h,h}(0) \; , \cr
\LL \VV^n_{H,h}(\oo\cdot\nn) & =
\VV^n_{H,h}(0) + [\oo\cdot\nn]\,\dot\VV^n_{h,h}(0) \; , \cr
\LL \VV^n_{h,H}(\oo\cdot\nn) & = \VV^n_{h,H}(0) \; , \cr
\LL \VV^n_{H,H}(\oo\cdot\nn) & = \VV^n_{H,H}(0)
\; , \cr} \Eq(7.3) $$
such that a result analogous to Corollary 6.8 applies to them.
More precisely one can easily prove the following result.

\*

\\{\bf 7.1.} {\cs Corollary.}
{\it If the operators and the propagators associated to the
lines $h\ot H$ are defined as in \equ(7.1),  while the
other ones as in \equ(3.4), and if the action of the
$\LL$ operator on the resonance factors is given by \equ(7.3),
then, when we sum together all the values corresponding to
trees $\th'$ contained in a resonance family $\FF_V(\th)$,
the contributions we obtain by applying the $\LL$
operator to the resonance factors identically vanish.}

\*

\\{\bf 7.2.} {\it Proof of Corollary 7.1.}
With respect to the previous situation, only the
cases with $\z_{w_0}^2=H$ behave in a different way, because
of the factor $[i\oo\cdot\nn_{\l_{w_0}}]$ appearing in
the operator associated to the line $\l_{w_0}$, (which
has to be a line $h \ot H$): then we have a zero to one
order higher for $\LL\VV^n_{H,h}(\oo\cdot\nn)$ and
$\LL\VV^n_{H,H}(\oo\cdot\nn)$, $\nn=\nn_{\l_V}$. \qed

\*

Therefore we can rule out again all the contributions
in which the $\LL$ operator applies to any resonance,
and we are left with resonances on which only the $\RR$
operator can act. Obviously, in \equ(7.2) the last
product has to be replaced with
$$ \prod_{v \notin W_0(\th) \atop \z_v^1=h, \z_v^2=H} 2^{n_{\l_v}}
\; , \Eq(7.4) $$
if $W_0(\th)$ is defined as the set of the vertices in $\th$ from which
a resonance outgoing line comes out: in fact in such cases,
(\ie $v\in W_0$),
the factor $[i\oo\cdot\nn_{\l_v}]$ is used in order to
implement the cancellation of the resonance factors.

\*

It is convenient to write the effect of $\RR$ on a resonance $V$ as
$$ \eqalign{
\RR\VV^n_{\z,\z'}(\oo\cdot\nn) & = (\oo\cdot\nn) \ig_0^1 dt\;
\dot\VV^n_{\z,\z'}(t\oo\cdot\nn) \qquad
\qquad (\hbox{first order zero}) \; , \cr
\RR\VV ^n_{\z,\z'}(\oo\cdot\nn) & = (\oo\cdot\nn)^2 \ig_0^1 dt\;t\;
\ddot\VV^n_{\z,\z'}(t\oo\cdot\nn)
\qquad (\hbox{second order zero}) \; , \cr}
\Eq(7.5) $$
where $\ddot\VV^n_{\z,\z'}$ denotes the second derivative with respect to
the variable $\oo\cdot\nn$.

As there are resonances enclosed in other resonances the above formula
can suggest that there are propagators derived up to $\approx k$
times, if $k$ is the order of the graph. This would be of course
a source of problems, as $a_{R_{\l_V}}(p) > p!$, where $a_{R_{\l_V}}(p)$
is defined in \equ(4.3). However it is not so: in fact the propagators
are derived at most two times. This can be seen as follows.
Let $n$ be the resonance-scale of the
maximal resonance $V$, and recall that $V_0$
is the collection of lines and vertices in $V$ not contained
in any resonance internal to $V$,
(see Definition 6.1).
Then we can write $\RR\VV(\oo\cdot\nn_{\l_V})$, (we do not
write explicitly the labels of the resonance factor), as
$$ \RR \Big( \prod_{\l\in V_0}
g^{(n_{\l})}(\oo\cdot\nn_\l)
\prod_{V'\subset V}
[\RR\VV (\oo\cdot\nn_{\l_{V'}})] \prod_{v\in V_0} O_{v} \Big)
\; , \Eq(7.6) $$
being the second product over the resonances $V' \subset V$ which
are maximal; in \equ(7.6), for any resonance $V' \subseteq V$,
$\RR\VV(\oo\cdot\nn_{\l_{V'}})$ can be written either
as in \equ(7.5) or as a difference
$\RR\VV(\oo\cdot\nn_{\l_{V'}})$ $=$
$\VV(\oo\cdot\nn_{\l_{V'}})-\LL\VV(\oo\cdot\nn_{\l_{V'}})$,
in according to which expression turns out to be more convenient to
deal with.

Then the first step is to write the action of $\RR$ on the maximal
cluster as in \equ(7.5), leaving the other terms
$\RR\VV (\oo\cdot\nn_{\l_{V'}})$ written
as differences: so \equ(7.6) can be written by the Leibniz's rule
as a sum of terms, and the derivatives of $\RR$ apply either on
some propagator $g^{(n_\l)}$ or on some
$\RR\VV (\oo\cdot\nn_{\l_{V'}})$. In the end there
can be either no derivative, or one derivative, or two derivatives
applied on each $\RR\VV (\oo\cdot\nn_{\l_{V'}})$.
If only one derivative acts on
$\RR \VV(\oo\cdot\nn)$, $\nn=\nn_{\l_{V'}}$, and the zero is
of the second order,
then we write, when such a term is not vanishing,
$$ \dpr \RR \VV(\oo\cdot\nn)=\dpr \VV(\oo\cdot\nn)-
\dot\VV(0)=(\oo\cdot\nn)\ig_0^1 dt\, \ddot \VV(t\oo\cdot\nn) \; , $$
because the derivative with respect to
$\oo\cdot\nn$ is equal to the derivative with respect to
$\oo\cdot\nn_{\l_{V}}$, while if two derivatives act on
$\RR\VV (\oo\cdot\nn_{\l_{V'}})$, then we write
$$ \dpr^2 \RR \VV(\oo\cdot\nn)= \ddot \VV(\oo\cdot\nn) \; . $$
The case of a first order zero is easier,
and can be discussed in the same way.
Then no more than two derivatives can act on each resonance $V'$
in any case, and the procedure can be iterated, since the resonances
$V'$ can be dealt with as the resonance $V$.

The effect of the $\RR$ operator is to obtain a gain factor either
$2^{2+n-n'}$ or $2^{2+n-n'}2^{2+n'-n{'}{'}}$,
where $n'$ and $n{'}{'}$ are
the scales of two lines $\l'$ and $\l{'}{'}$ contained in some
clusters $T'$ and $T{'}{'}$ inside $V$,
$n$ is the resonance-scale, and the
factor $2$ is due to the support properties
of the propagators; the line $\l{'}{'}$ can
coincide with $\l'$, or also be absent, if there is
a first order zero. So we
can rewrite, \eg, the first factor as $2^{2+n-n'}=2^22^{n-n_1}\ldots
2^{n_q-n'}$, where $n_i$ is the scale of the cluster $T_i\supset T_{i+1}$,
with $T_0=V$ and $T_{q+1}=T'$. Analogous considerations hold for $n{'}{'}$,
so that we can conclude that:
\acapo
(1) no more than two derivatives can
ever act on any propagators;
\acapo
(2) a gain
$2^{R_{\l_{V'}}(2+n_{\l_{V'}}-n_{V'})}$ is obtained for any resonance
$V'\subseteq V$;
\acapo
(3) the
total number of terms generated by the derivation operations is
bounded by $k(V)^2$, if $k(V)$ is the order of the resonance $V$,
(see Definition 5.1), as $\sum_{V'\in\th}\tilde k(V') = k(V)$.

Therefore, for the value of the diagram formed by the resonance
plus its incoming line, we find the bound
$$ \eqalign{
2^{-R_{\l_V}n_V} \Big[ 2^{4k}\,\tilde\CC^k
\prod_{v \notin W_0 \atop \z_v^1=h, \z_v^2=H} & 2^{n_{\l_v}}
\prod_{n\ge n_V}2^{-(2nN_{n}^2+nN_{n}^1)}\Big] \cdot \; \cr &
\cdot \; \Big[ \prod_{n\ge n_V} \prod_{T \atop n_T=n}\prod_{j=1}^2
\prod_{i=1}^{m_T^j(\th)}\,2^{R_{\l_{V_i}}(n-n_{V_i}) }
\Big] \; , \cr} \Eq(7.7) $$
where $n_V$ is the scale of the resonance, $R_{\l_V}=1,2$
and the second square bracket is the part coming from the resummations,
and follows from the above discussion about the gain factors. The constant
$\tilde\CC$ differs from $\CC$ in \equ(5.2) as it takes into account
the bound on the derivatives of the propagators: we can set
$\tilde\CC=\CC\,e^2\,[a_2(2)2^{-2}]^2$,
as the sum over all the outer resonances $V$'s
of the factors $[2k(V)]^2$ can be bounded by $e^{2k}$, and
$a_R(p)\le a_2(2)$, for any $R=1,2$, and $0\le p \le 2$,
(and the factor $2^{-2}$ simplifies $a_2(0)$ in $\CC$).

\*

Then if we recall \equ(5.11), we see that the $j m_T^j(\th)$
is taken away by the first factor in $\,2^{R_{\l_{V_i}}n}$
$2^{-{R_{\l_{V_i}}}n_{V_i}}$, being $n=n_{\l_{V_i}}$, while the
remaining $\,2^{-R_{\l_{V_i}}n_{V_i}}$ are compensated by factors
furnished by the clusters counting. In particular
we can get rid of the factor $2^{-R_{\l_V}n_V}$ in \equ(7.7).
This completes the discussion of the resonances: then Theorem 1.2
is proven, in the case in which the perturbation
is of the form \equ(2.1), and the estimate \equ(1.9)
follows, with $E$ defined in \equ(2.1).
The extension of the proof to the general
case is given in Appendix A2, and yields \equ(1.9)
with $E$ defined in \equ(1.7).

\vskip1.truecm

\centerline{\titolo 8. Renormalizability of the isochronous hamiltonians}
\*\numsec=2\numfor=1

\\As done for Theorem 1.2,
let us consider first the notationally less involved
case \equ(2.1). The general hamiltonian appearing
in the statement of Theorem 1.4 will be studied
in Appendix A2.

Let us write the solution of the equations
of motions, (if there are any), as in \equ(1.11), where
$\AA_0$ is a constant vector in $D$ and the set $W(\AA_0,\r_0)$,
defined in item (1) of Theorem 1.4,
is contained in $D$;
then let us call $\V H^{(k)}$ and $\V h^{(k)}$  the $k$-th order
coefficients of the (formal)
Taylor expansion of $\V H$ and $\V h$ in powers of
$\e$.

\*

We look for a solution of the equation of motion corresponding
to the hamiltonian \equ(1.10)
of the form \equ(1.11), with $\V h(\oo_0t,\AA_0;\e)$
and $\V H(\oo_0t,\AA_0;\e)$ being quasiperiodic
functions with vanishing average.

The equations of motion are
$$ \eqalign{
{d\AA\over dt} & = -\e\dpr_\aa f(\aa,\AA) \; , \cr
{d\aa\over dt} & = \oo_0 + \e\dpr_{\AA} f(\aa,\AA)
- \dpr_\AA N_f (\AA;\e) \; . \cr
} \Eq(8.1) $$
Then we get immediately
recursion relations for $\V H^{(k)},\V h^{(k)}$:
$$ \eqalign{
\oo_0\cdot\dpr_\pps\,\V H^{(k)} = &
{\sum_{(k-1)}}^* (-\dpr_\aa)
\Big[ \sum_{p\ge0}\sum_{q\ge0}{ 1 \over p!\,q! }
\prod_{s=1}^p \left( \V h^{(k_s)} \cdot \dpr_\aa \right)
\prod_{r=1}^q \left( \V H^{(k_r')} \cdot \dpr_\AA\right)
\Big] f(\oo_0t,\AA_0) \; , \cr
\oo_0\cdot\dpr_\pps\,\V h^{(k)} = &
{\sum_{(k-1)}}^* \dpr_\AA \Big[ \sum_{p\ge0}\sum_{q\ge0}
{ 1 \over p!\,q! }
\prod_{s=1}^p \left(\V h^{(k_s)} \cdot \dpr_\aa \right)
\prod_{r=1}^q \left( \V H^{(k_r')} \cdot  \dpr_\AA\right)
\Big] f(\oo_0t,\AA_0) \cr
- & {\sum_{(k)}}^* \dpr_\AA \Big[ \sum_{q\ge0} {1\over q!}
\prod_{r=1}^q \left( \V H^{(k_r')} \cdot \dpr_\AA \right)
\Big] N_f^{(k_0)}(\AA_0)  \; , \cr} \Eq(8.2) $$
where the $\sum_{(k-1)}^*$ denotes summation over the integers
$k_{s}\ge1$, $k_{s}'\ge1$, with: $\sum_{s=1}^p k_{s}$
$+$ $\sum_{r=1}^{q} k_{r}'$ $=$ $k-1$,
the $\sum_{(k)}^*$ denotes summation over the integers
$k_{s}\ge1$, $k_{s}'\ge1$, $k_0\ge1$, with: $k_0$ $+$
$\sum_{s=1}^p k_{s}$
$+$ $\sum_{r=1}^{q} k_{r}'$ $=$ $k$,
and the derivatives are
supposed to apply to the functions $f(\aa,\AA)$, $N_f^{(k_0)}(\AA)$
and then
evaluated in $(\aa,\AA)=(\oo_0t,\AA_0)$.
The terms in brackets corresponding to the values $p, q=0$ have
to be interpreted as 1. In the first two brackets $p$ and $q$
can not be simultaneously vanishing for $k>1$; in the third
line of \equ(8.2) $q=0$ yields $k_0=k$.

\*

\\{\bf 8.1.}
{\it Proof of the formal solubility of the recursive relations.}
We proceed inductively as in \S 2.1, the only difference being that,
when we impose that the right hand side of the second equation
in \equ(8.2) has vanishing average, then we obtain some constraint
on the $N^{(k)}_f(\AA_0)$ coefficients.

In fact, from the equations of motion \equ(8.1) one obtains
immediately the equations in \equ(8.2). Then assume that,
for a suitable choise of the coefficients
$N^{(k)}_f(\AA_0)$, $k<k_0$, the functions $\V H$ and $\V h$,
up to order $k_0-1$,
(1) have vanishing average on $t$ (\ie in $\pps$), and
(2) solve the equations of motion.
We show that, under
such an asumption, it is possible to prove that it is
possible to choose $N_f^{(k_0)}(\AA_0)$ in such a way that
$\V H^{(k_0)}$ and $\V h^{(k_0)}$ solve (formally) the equations
of motions \equ(8.2) for $k=k_0$, and both functions have
vanishing average, \ie their $\V0$-th Fourier components
are identically vanishing:
$\V H^{(k_0)}_{\V0}=\V h^{(k_0)}_{\V0}=\V0$.

The proof of such an assertion can be carried out as in \S 2.1.
Therefore we do not give explicitly the details: we confine
ourselves to note that the right hand side of the first
equation in \equ(8.2) is automatically with vanishing
average, as a consequence of the assuntions on the functions
$\V H^{(k)}$ and $\V h^{(k)}$ for $k<k_0$. The
$\V0$-th Fourier component of the right
hand side of the second equation can be set
equal to zero, by suitably fixing the value of $N_f^{(k_0)}(\AA_0)$
in such a way that it exactly cancels the $\V0$-th Fourier
component of the remaining expression. We stress here that,
at this level, we obtain simply a formal identity between
two expressions: this yields that a formal perturbative
expansion is possible, but no further information is
provided, and the convergence of the perturbative series
has to be proven yet.

Once $N_f^{(k_0)}(\AA_0)$ is fixed, the equations \equ(8.2)
can be solved, and $\V H^{(k_0)}$ and $\V h^{(k_0)}$
are obtained up to a constant, which, in both cases, can be
chosen to be zero, so that the procedure can be iterated.
\qed

\*

Then it is easy to check that, in the Fourier space, we obtain,
for $\nn\neq\V0$,
$$ \eqalign{
(i\oo_0\cdot\nn)\,\V H^{(k)}_\nn = &
{\sum_{(k-1)}}^* (-i\nn_0) \Big[ \sum_{p\ge0} \sum_{q\ge0} {1\over p!\,q!}
\prod_{s=1}^p \left( i\nn_0 \cdot \V h^{(k_s)}_{\nn_s} \right)
\prod_{r=1}^q \left( \V H^{(k_r')}_{\nn_r'} \cdot \dpr_\AA \right)
\Big] f_{\nn_0}(\AA_0) \; , \cr
(i\oo_0\cdot\nn)\,\V h^{(k)}_\nn = &
{\sum_{(k-1)}}^* \dpr_\AA \Big[ \sum_{p\ge0}\sum_{q\ge0}{1\over p!\,q!}
\prod_{s=1}^p \left( i\nn_0 \cdot \V h^{(k_s)}_{\nn_s} \right)
\prod_{r=1}^q \left( \V H^{(k_r')}_{\nn_r'} \cdot \dpr_\AA \right)
\Big] f_{\nn_0}(\AA_0) \cr
- & {\sum_{(k)}}^{*} \dpr_\AA \Big[ \sum_{q\ge0} {1\over q!}
\prod_{r=1}^q \left( \V H^{(k_r')} \cdot \dpr_\AA \right)
\Big] N_f^{(k_0)}(\AA_0)  \; , \cr} \Eq(8.3) $$
and, for $\nn=\V0$,
$$ \eqalign{
N_f^{(k)}(\AA_0) = &
{\sum_{(k-1)}}^* \Big[
\sum_{p\ge0}\sum_{q\ge0}{ 1 \over p!\,q! }
\prod_{s=1}^p \left( i\nn_0 \cdot \V h^{(k_s)}_{\nn_s} \right)
\prod_{r=1}^q \left( \V H^{(k_r')}_{\nn_r'} \cdot \dpr_\AA \right)
\Big] f_{\nn_0}(\AA_0) \cr
- & {\sum_{(k)}}^{*} \Big[ \sum_{q>0} {1\over q!}
\prod_{r=1}^q \left( \V H^{(k_r')} \cdot \dpr_\AA \right)
\Big] N_f^{(k_0)}(\AA_0)  \; , \cr} \Eq(8.4) $$
where the $\sum^*_{(k-1)}$ denotes summation over the integers
$k_{s}\ge1$, $k_{r}'\ge1$, with: $\sum_{s=1}^p k_{s}$
$+$ $\sum_{r=1}^{q }k_{r}') =$ $k$, and over the integers $\nn_0$,
$\nn_s$, $\nn_r'$, with: $\nn_0+\sum_{s=1}^p\nn_s + \sum_{r=1}^q \nn_r'$
$=$ $\nn$, while $\sum_{(k)}^{*}$
denotes summation over the integers
$k_{s}\ge1$, $k_{r}'\ge1$, $k>k_0\ge 1$ with: $k_0$ $+$
$\sum_{s=1}^p k_{s}$
$+$ $\sum_{r=1}^{q }k_{r}'$ $=$ $k$, and over the integers $\nn_0$,
$\nn_s$, $\nn_r'$, with: $\nn_0+\sum_{s=1}^p\nn_s + \sum_{r=1}^q \nn_r'$
$=$ $\nn$, and the constraint $q>0$
implies that the case $k_0=k$ has to be excluded.
As in \equ(8.2) the case $p=q=0$ in the first two lines
of \equ(8.3) and in the first line of \equ(8.4) is
possible only if $k=1$.

\*

\\{\bf 8.2.} {\it Remark.} Obviously the equation for $N^{(k)}_f$
involves its derivative with respect to the action variable, \ie
$\dpr_{\AA}N^{(k)}_f(\AA_0)$, but we can trasfer the equation
directly on the function $N^{(k)}_f$, because we can define it
up to a constant (in $\aa$ and $\AA$), without loss of
generality.

\*

For example, if $k=1$, we have from \equ(8.3) and \equ(8.4)
$$ \eqalign{
\V H^{(1)}_\nn & = {-i\nn_v \, f_\nn(\AA_0) \over i\oo_0\cdot\nn}
\; , \qquad \nn\neq\V0 \cr
\V h^{(1)}_\nn & = {\dpr_\AA f_\nn(\AA_0) \over i\oo_0\cdot\nn}
\; , \qquad \nn\neq\V0 \cr
N^{(1)}_f & = f_{\V0} (\AA_0) \; , \cr} \Eq(8.5) $$
where the last equation fixes the value of the first
coefficient of the series expansion of $N_f(\AA_0;\e)$.

\*

If we compare the equations \equ(8.4) and \equ(8.5) with
the equations \equ(2.9)$\div$\equ(2.11) in \S 2, we see
that it is still possible to carry out a
diagrammatic expansion of the coefficients
$\V h^{(k)}_\nn$, $\V H^{(k)}_\nn$ and $N_f^{(k)}$.

We give first the rules how to construct the diagrams in general.
First of all we introduce the labels to associate to the tree vertices.
Obviously we have to take into account that some vertices $v\in\th$
can have associated a factor $N^{(k_v)}_f$,
for some $k_v\ge 1$, instead of
a factor $f_{\nn_v}$: the meaning of the label $k_v$ is now
very different from that of the label introduced in item (3)
of \S 3.3 (there it denoted the order of the subtree
having $v$ as first vertex, here it is independent on such a subtree).
Let us introduce a label $\d_v$, such that
$\d_v=1$ if a factor $f_{\nn_v}$
is associated to $v$, and $\d_v=0$ otherwise: in the latter case,
also a label $k_v\ge1$ is assigned to $v$.
The terminology
is somewhat reminiscent of that used in \S 3, where $\d_v=0$
denoted a factor $\HH_0$, which, like $N_f^{(k_v)}$, depends
only on the action variables, but note that, whereas in \S 3.3
$\d_v=0$ implied that no $\e$ was associated to that vertex,
now $\d_v=0$ implies that a factor $\e^{k_v}$ is associated
to $v$.

Then we have the following collection of labels.
\acapo
(1) $\d_v=0,1$;
\acapo
(2) $k_v\=1$ if $\d_v=1$, and $k_v\ge1$ if $\d_v=0$;
\acapo
(3) $\nn_v\in{\ZZZ}^{\ell}$ is the mode label;
\acapo
(4) $\z_v^1$ and $\z_v^2$ can assume the symbolic values
$\z_v^1,\z_v^2=h,H,N$;
\acapo
(5) $m_v=p_v+q_v$ is defined as in item (6) in \S 3.3.

With respect to (3.4) and (3.5) of \S 3, now we have
the following functionals associated to the tree branches.
$$ \matrix{
\hbox{operator} & \hbox{propagator} & \hbox{branch} \cr
& & \cr
0 & 0 & h \ot h \cr
& & \cr
C_0 \left[ i\nn_{v'}\cdot(\dpr_{\AA_v})
\right] &
[i\oo\cdot\n_{\l_v}]^{-1} & h \ot H \cr
& & \cr
C_0 \left[ \dpr_{\AA_{v'}}\cdot(-i\nn_v) \right] &
[i\oo\cdot\nn_{\l_v}]^{-1} & H \ot h \cr
& & \cr
0 & 0 & N \ot H \cr} \Eq(8.6) $$
for all the branches distinct from the root branch, and
$$ \matrix{
\hbox{operator} & \hbox{propagator} & \hbox{branch} \cr
& & \cr
0 & 0 & h \ot h \cr
& & \cr
C_0 \left[ \dpr_{\AA_v} \right] &
[i\oo\cdot\n_{\l_v}]^{-1} & h \ot H \cr
& & \cr
C_0^2\,\left[ -i\nn_v \right] &
[i\oo\cdot\nn_{\l_v}]^{-1} & H \ot h \cr
& & \cr
1 & 1 & N \ot H \cr} \Eq(8.7) $$
for the root branch.

Then all the above operators are applied to the function
$$ \prod_{v\in\th\atop\d_v=1} f_{\nn_v}(\AA_v)
\prod_{v\in\th\atop\d_v=0} [-N_f^{(k_v)}(\AA_v)] \; , \Eq(8.8) $$
which replaces \equ(3.6), so obtaining a factor $O_v$ for
each vertex $v$, analogously to what happened in \S 3.
The explicit expression of $O_v$ can be obtained from
the above discussion, by reasoning as in \S 3, (see
in particular \equ(3.8)).

The following result (analogous to Proposition 3.7 in \S 3)
holds.

\*

\\{\bf 8.3.} {\cs Proposition.} {\it Let us consider a tree of order $k$;
then
$$ \sum_{v\in\th \atop \d_v=1} 1 + \sum_{v\in\th\atop\d_v=0}
k_v = k \; , \Eq(8.9) $$
and, if $\d_v=0$, $\z_v^1=h$ and $\z_v^2=H$, and
$\z_w^1=H$ and $\z_w^2=h$ for any $w\in\th$ such that $v=w'$.}

\*

\\{\bf 8.4.} {\it Proof of proposition 8.3.}
The properties stated in Proposition 8.3 are immediate
consequences of the definitions and of \equ(8.3)$\div$\equ(8.5). \qed

\*

We can not introduce the dimensionless variable $\V X_\nn^{(k)}$
of \S 3, because now it would be $J_m=0$. This does not means
that it is impossible to introduce dimensionless variables
(of course), but simply the definition given in \S 3 was
suitable only in that case (system verifying the anisochrony
condition).

\*

\\{\bf 8.5.} {\it Remark.}
Note that, if we define $T(\AA_0)$ as the matrix of the
second derivatives of the free hamiltonian, then it is
trivially $T(\AA_0)\=0$, (see also the previous remark).
Therefore, if we compare \equ(8.6) and \equ(8.7) above
with the corresponding ones in \S 3, we see that no
matrix $T(\AA_0)$, $T(\AA_0)^{-1}$ appears now, as it
has to be if we want that the formulae we write are
meaningful.

\*

We note that, as it is possible to check, some
contributions corresponding to trees
carrying labels $\d_v=0$ on some vertices
can cancel exactly, when
summed together to some other tree values.

To see this, let us consider a tree
$\th_1$ of order $k$, with a vertex $v$ carrying a label $\d_v=0$
and having associated a factor $N^{(k_v)}_f$. Now let us consider
also the tree $\th_2$ which
differ from $\th_1$ because:
(1) the vertex $v$ carries a label $\d_v=1$, and
(2) other subtrees $\th_1'$, $\ldots$, $\th_s'$
(which were absent in $\th_1$)
emerge from $v$, such that the sum of their orders add to $k_v$;
all the other labels are the same in $\th_1$ and $\th_2
\setminus \cup_{j=1}^s \th_j'$.
If we expand $N^{(k_v)}_f$ in $\th_1$ according to the graph rules, then,
among the several possible trees,
{\it we can obtain from $\th_1$ also a tree which has exactly
the same shape and the same labels as $\th_2$, hence
the same value, but with opposite sign},\footnote{${}^4$}{\nota
Because of the sign minus before $N_f^{(k_v)}$ in
\equ(8.8). Note also that the cancellation we are
considering occurs only if the combinatorial weights
of the two trees are equal: this happens only if no subtree
among $\th_1'$, $\ldots$, $\th_s'$ is equivalent to some other
subtree emerging from $v$ and apearing both in $\th_1$ and $\th_2$.}
so that the two values cancel each other.

In this way one can cancel also the
values of some trees with resonances,
if (1) the line $\l_v$ carries the same momentum of one of
the lines entering $v$, (2) all the lines of the other subtrees
emerging from $v$ have a higher scale label, and (3)
the tree representation of $N_f^{(k_v)}$ does not destroy
the inclusion relations between the clusters.
But it is important the fact
that the corresponding resonance factors are of the
form $\VV_{H,H}^n$, and we recall from \S 6 (see in particular
the fourth term in \equ(6.4)) that {\it no cancellation was required}
in such a case.

Therefore the cancellations occurring in according to the
just described mechanism are not quite dangerous, and
can be easily forgotten: in other words
there is no cancellations overlapping and we can reason as
if no cancellation involving the factors $N^{(k_v)}_f$
can happen, and only the cancellations described in
\S 6 are taken into account.

\*

We can now proceed as in the proof of Theorem 1.2,
with the propagators
and operators defined as in \equ(8.6) and \equ(8.7) above.

Then the results of the previous sections apply.
Note that it is not really
necessary to repeat the discussion (if we are
not looking for optimal bounds): it is enough to
realize that now the diagrammatic rules are simplified
with respect to the case studied before. In fact we
can simply (1) write down the perturbative expansion by
using the rules of \S 3 and \S 4; (2) get rid of the $h \ot h$ lines
by imposing that each time such a line appears we obtain a
vanishing contribution; (3) analogously get rid of
the lines $N\ot H$ which are not root branches; (4)
replace the operator in the fourth row
in (3.5) with the identity $\openone$;
(5) the factors $N^{(k_v)}_f$ associated to the
vertices $v\in\th$ with $\d_v=0$ can be expanded
and graphically represented as trees in according to \equ(8.4),
so giving a factorizing term which can be treated
in the same way. In particular
it is easy to check that a bound like \equ(1.12) follows,
if we set $F\e_1^{-1}=F_0$.

This concludes the discussion in the case in which the
interaction is as in \equ(2.1). The general case will
be discussed in Appendix A2.

\vskip1.truecm

\centerline{\titolo 9. A comparison with other
proofs of the convergence }

\centerline{\titolo of the Lindstedt series}
\*\numsec=9\numfor=1

\\After Eliasson works, [E1], [E2] and [E3], somewhat hard
to read (the fundamental paper [E1] has not been yet published),
simplified and clearer proofs of the convergence of
the Lindstedt series have been proposed in several papers,
all quoted in the introduction. This section is devoted to a
comparison between such proofs and the one discussed in
the present paper.

\*

\\{\bf 9.1.} The first proof is the Eliasson's one, [E1].
In [E1], \S II, Eliasson writes a tree expansion for the invariant tori,
(although the trees are called {\sl simple index sets}
which is called {\sl tree structure} in [E2]), and
realizes that some terms to order $k$ in this expansion
are of order $O( C^k (k!)^\a )$, for some positive constants
$\a$ and $C$ depending on the particular contribution studied:
then he concludes that the Lindstedt series
would converge to an analytic function only if there are
``very sharp compensations of signs between different terms of these
series expansion", [E1], page 3. However Eliasson {\it does not look
for an identification of terms between which there are compensations}:
on the contrary he replaces the Lindstedt series with a
different series whose convergence follows {\it without the necessity
of exploiting any cancellation}. Since the sum of the
series does not verify the equations of motions,
he introduces in it a free parameter to be fixed so that
the series verifies such equations: if this can be accomplished,
then the resulting series has to
coincide with the Lindstedt series.

We first propose here a translation of Eliasson's work in our
formalism, up to a minor technical modification which will
be explained below.
We consider, following Eliasson's spirit, the quantities
$\V h^*(\pps)=\sum_{\nn\in {\zzz}^l}e^{i\nn\cdot\pps} \V h_\nn^*$
and $\V H^*(\pps)=\sum_{\nn\in {\zzz}^l} e^{i\nn\cdot\pps}\V H_{\nn}^*$,
defined as
$$ \V X_\nn^* (\z) =\sum_k\sum_{\theta\in\TT^*_k} \hbox{Val }(\theta)
\; , \qquad \z \in \{h,H,\m\} \; , \Eq(9.1) $$
(compare with \equ(3.1)), where
$\hbox{Val}(\theta)$ is defined as in (4.4) and $\TT^*_k$ is the set of
the labeled semitopological trees such that {\sl on the resonances
only the $\RR$ operators apply}, \ie tree values
containing factors $\LL\VV^n_{\z_{w_0}^2,\z_{w_1}^1}(\oo\cdot\nn)$
are discarded.

We know from the analysis in \S 7 that the convergence
of the above series follows simply from the Siegel-Bryuno's bound,
while, in order to prove the convergence of the original series
one has to exploit the cancellation mechanisms discussed in \S 6.

The series $\V h^*(\vec\psi),\V H^*(\vec\psi)$ are not the original
series for the tori, hence they will not verify the equations of motion
\equ(2.2); however they obey to
a very simple modification of them. If
$\AA^*$ and $\aa^*$ are given by \equ(1.11) with $\V h$ and
$\V H$ replaced with $\V h^*$ and $\V H^*$, one has
$$ \eqalign{
{d\V H ^*\over dt} =  & [-\e\dpr_{\aa^*} f(\aa^*,\AA^*)]
- \DD_{h,h}\,\V h^* - \DD_{h,H}\,\V H^* \; , \cr
{d\V h^*\over dt} = & [\dpr_{\AA^*} \HH_0 + \e\dpr_{\AA^*} f(\aa^*,\AA^*)]
- \DD_{H,h} \, \V h^* \; , \cr} \Eq(9.2) $$
where, using our notations, one can easily check that the
$\ell\times\ell$ matrices $\DD_{i,j}$ are given by the
sums is over all the possible localized resonance factors
$\LL\VV_{i,j}(-i\oo\cdot\dpr_{\pps})$,
running the sum on the perturbative order and over trees.

The proof of the above equation is quite simple. Let
us express $\V h^*$ and $\V H^*$ as sums over trees, and let
$v_0$ be the the first vertex, as in Fig.3.1.
A tree in which $v_0$ does not belong to
a resonant cluster contributes to the series expansion of the
square brackets in the rigth hand side of \equ(9.2); if $v_0$ is in
a resonant cluster, if we recall that $\RR=\openone - \LL$,
we can split the value of the corresponding tree into
two parts: the term in which $-\LL$ is applied on the
resonance factor contribute to the terms outside the square brakets,
while the term in which $\openone$ is applied on the resonance factor
is included in the square brackets contribution.

It is a remarkable fact that $\AA^*,\aa^*$ obey to equations
``so similar'' to the original equations of motion \equ(2.2).
With the notations of \S 2.1, we can rewrite \equ(9.2) as
$$ {dY^* \over dt} = (E\dpr\HH)(Y^*)
+ E D(Y^* - Y^{(0)}) \; , \Eq(9.3) $$
where $Y^*=(\V h^*, \V H^*)$, $Y^{(0)}=(\pps,\AA_0)$
and $D$ is the $2\ell\times2\ell$ matrix with entries
$D_{ij}=(-1)^{\d_{i,H}}$ $\DD_{i,j}$,
for $(i,j)\neq(H,H)$, and $D_{H,H}(0)=0$.

The series $\V h^*,\V H^*$ are not, in principle,
the solutions of the equations of motion $dY/dt$ $=$
$(E\dpr\HH)(Y)$, unless one proves
that the sum of the localized resonance factors vanishes
(\ie $D\=0$ identically).
This is exactly what was done in \S 6;
on the other hand we could proceed in a different more abstract way,
much closer to Eliasson's approach.

We can suppose that up to order $k_0-1$, $D\=0$ and
$Y^*=Y$, if $Y$ is the solution of the equation
$dY/dt=(E\dpr\HH)(Y)$. Then \equ(9.3) is satisfied also to
order $k_0$, so that $Y^*=Y$ up to order $k_0$. From
the fact that \equ(2.6) holds for any periodic function
and from the inductive assumption,
\equ(2.8) still holds and, as in \S 2.1, implies that the
average of the function $[\dpr_{\aa}\HH]^{(k_0+1)}$ vanishes.
Then $D_{11}^{(k_0)}\V h^{(1)}_{\V0} +
D_{12}^{(k_0)}\V H^{(1)}_{\V0}=\V0$,
and  the arbitrarity of $\V h^{(1)}_{\V0}$,
(see \S 2.1, item (4)), yields
(1) $D_{11}^{(k_0)}=0$, and (2) $D_{12}^{(k_0)}\V H^{(1)}_{\V0}=\V0$.
It turns out that the latter identity yields $D_{12}^{(k_0)}=0$.
Accepting the last statement we have also $D_{21}^{(k_0)}=0$,
as a consequence of Lemma 6.4. Since $D_{22}\=0$, see \equ(6.4),
this proves that $D\=0$.

It remains to check that $D_{12}^{(k_0)}\V H^{(1)}_{\V0}=\V0$
implies $D_{12}^{(k_0)}=0$. To see this, let us write the
perturbation $f\=f(\aa,\AA)$ as $f = f_0+\tilde f$, where
$$ f_0 = f_{\V0}(\AA) \; , \quad \tilde f =
\sum_{\nn\neq\V0} e^{i\nn\cdot\aa}\,f_{\nn}(\AA) \; , $$
and let us consider $f_0$ and $\tilde f$ as
two independent functions. Then, if we study the dependence on $f$,
we have, essentially by definition,
$D^{(k_0)}_{12}(f)$ $\=$ $D^{(k_0)}_{12}(f_0,\tilde f) $ $=$
$A(f_0,\tilde f) + B(\tilde f)$, where $A(f_0,\tilde f)$ depends
explicitly on $f_0$ and vanishes for $f_0=0$, while
$\V H^{(1)}_{\V0}(f_0)\=\V\m^{(1)}(f_0)$ depends only on $f_0$,
(see the third equation in \equ(1.9)), and
$\V\m^{(1)}(f_0)\neq\V0$ for $f_0\neq 0$, $\V\m^{(1)}(0)=\V0$.
Then $D_{12}^{(k_0)}\V H^{(1)}_{\V0}=\V0$ can be written
$\left[ A(f_0,\tilde f) + B(\tilde f) \right]
\V\m^{(1)}(f_0) = 0$, which holds
for any $f_0$. If $f_0\neq 0$, then $A(f_0,\tilde f)+B(\tilde f)$
$=$ $0$, and since $B(\tilde f)$ does not depend on $f_0$, one
has $A(f_0,\tilde f)=B(\tilde f)=0$ identically in $\tilde f$, \ie
$B(\tilde f)\=0$. This means that $D_{12}^{(k_0)}(f)=0$ for any $f$.

\*

Eliasson's discussion is very similar to the just described one,
with the following differences (which seem to us to be unessential).
First he performs a change of coordinates (simply a rescaling)
on the unperturbed hamiltonian, so that the graph rules are
slightly different from the ones described in \S 3; furthermore
he does not introduce any multiscale decompositions,
and this makes the discussion of the convergence of the series
more involved: he defines ``resonance"
a generic pair of vertices such that the lines emerging from
them carry the same momentum, ([E1], \S IV), and then
he has to solve the problem of the ``overlapping divergences",
so distinguishing among several types of resonances,
such that only the ``critical ones'' cannot be controlled
through the Siegel-Bryuno's lemma.
The series he studies are {\it not} exactly the above series
$\V h^{*}$ and $\V H^{*}$, rather they are series
for two functions $\V h^{**}$ and $\V H^{**}$, defined as in \equ(9.1),
but with the difference that the localization operator is defined
by \equ(6.4) as far as the first three lines are concerned,
whereas the last line is replaced with
$\LL \VV^n_{H,H}(\oo\cdot\nn)=\VV^n_{H,H}(0)$.\footnote{${}^5$}{\nota
It can also be worth to remark that, in the definition of
the series $\V h^*(\pps)$ and $\V H^*(\pps)$, (see \equ(9.1)),
the quantities which are subtracted from the resonance factors
(and which are precisely the localized resonance factors)
cannot at all be identified with the resonance factors of
some other trees, (as sometimes Eliasson's argument has been
erroneously interpreted): simply the contributions
which are responsible for the apparent divergence of
the perturbative series are deleted in order to obtain a series
whose convergence can be proved without exploiting cancellations.}

Therefore, instead of the series which solve the equations \equ(9.2),
Eliasson obtains somewhat different quantities,
$\V h^{**}$ and $\V H^{**}$, which can be thought
as obtained by the above $\V h^*$ and $\V H^*$, by replacing
$\partial_{\AA}\partial_{\AA}f_0$ with
$\partial_{\AA}\partial_{\AA}f_0+M$, being
$M$ a $\ell\times\ell$ matrix whose expression is
left free and has to be determined. It is then easy to check that
the following equation holds for $\AA^{**},\aa^{**}$:
$$ \eqalign{
{d\V H ^{**}\over dt} & = [-\e\dpr_{\aa^{**}} f(\aa^{**},\AA^{**})]
- \DD_{h,h} \, \V h^{**} - \DD_{h,H} \, \V H^{**} \; , \cr
{d\V h^{**}\over dt} & = [\dpr_{\AA^{**}} \HH_0 +
\e\dpr_{\AA^{**}} f(\aa^{**},\AA^{**})]
- \DD_{H,h} \, \V h^{**} -
\left( \DD_{H,H} + M \right) \V H^{**} \; . \cr} \Eq(9.4) $$
Then, by using an argument based on the symplectic
structure of the problem, similar
to the one used by Poincar\'e to prove the formal
existence of the Lindstedt series, Eliasson proves that it is possible
to choose $M$ as an analytic function of $\e$ so that all terms after
the first square brakets in \equ(9.4) are vanishing, and,
because of the uniqueness of the solution,
$\V h^{**}=\V h, \V H^{**}=\V H$, ([E1], \S VI): then
the convergence of the Lindstedt series is proven.
The choice of $M$ has the effect that $\VV_{H,H}$ is
not really renormalized, (\ie we could define
$\LL \VV_{H,H}(i\oo\cdot\nn)=0$ as in \equ(6.4)).
In our formalism, this corresponds to the fact that,
if Eliasson's localization is used, all
the sums of localized resonance factors
in \equ(9.4) give a vanishing contribution, except
the sum corresponding to the resonances with
$\z_{w_0}^2=\z_{w_1}^1=H$: this means that {\it he subtracts
also a contribution, which in fact is not vanishing, but which
is known, from the analysis of \S 7, not to give problems}.
Then he has to introduce a parameter (the matrix $M$), in order
to recover such a contribution (which, of course, from
our point of view, should {\it not}
have been subtracted to begin with): all the other localized resonance
factors are automatically vanishing.
The symplectic argument is not just the same we reproduced
above: Eliasson's original argument is very quickly sketched in [E1]
and it  is similar the Poincar\'e's argument for proving the formal
existence of the Lindstedt series, [P], Vol. II, \S 126,
while in the above discussion we followed an argument similar
to the proof of the formal solubilty in \S 2.1, based on [CZ], [CG].

In [E2], Eliasson applies the same methods in order to prove
the conjecture by Gallavotti stated in the introduction
of the present paper, Theorem 1.4.

\*

\\{\bf 9.2.} We stress that the possibility of fixing the
the initial data, in order to obatin
a formal power expansion by solving the equations of motion,
is standard, and was well known already to Lindstedt, [L],
Newcomb, [Ne],  and Poincar\'e, [P], Vol. II.
The first work in which Eliasson ideas have been resumed is
a paper by Feldman and Trubowitz, [FT], where
the graph representation of the solution of the equations of motion
in terms of trees is illustrated (following [E1] and [E3]), and
the analogy with quantum field theory
is pointed out as to the introduction of the counterterms
in order to obtain the formal solubility of the equations of motion,
(this corresponds to \S 2.1 in the present paper).
The problem of the convergence of the perturbative
series is not touched, (on this point the authors refer to
the classical proofs and to Eliasson's work).
The possibility of interpreting the Lindstedt series
as formal perturbative series of a euclidean field theory
on the torus $\TTT^{\ell}$ is pointed out in [G9], where an action
giving rise to the Lindstedt series is proposed (together with a
convergence proof). The idea is further developed in [GGM].

\*

\\{\bf 9.3.} In an effort to understand Eliasson's work his
ideas have been applied to simplified models in which
the free hamiltonian is quadratic in the action variables,
and the interaction potential is taken to be action-independent
and either (1) a trigonometric polynomial
in the angles, or (2) or an analytic function in the actions.

The analysis of the model (1) is carried out
in [G7], with the further simplification that the potential
is an {\it even} trigonometric polynomial and the rotation vectors
satisfy a property stronger than the usual diophantine one.
The second hypothesis is completely relaxed in [GG].
A discussion of the model (2) can be found in [CF1].

\*

The cancellation mechanism
between the resonances described in the present paper
is exactly the same as in [G7] and [CF1],
as far as the first order zero is concerned;
the second order zero is ensured in [G7] from
the parity properties of the interaction potential
(a situation very often realized in physics), while in [CF1]
a more subtle anlysis of the cancellations is required
in order to discuss the not even case.
In this regard, we mention, as a relevant fact,
that we have taken from [CF1] the idea to shift also
the exiting line of the resonances in order to see the
cancellations in the not even case.
The main difference between [G7] and [CF1] is in the
definition of resonance (and therefore in the
way the related problems are solved).
However both definitions in the end turn out
to be somewhat equivalent, because the ultimate aim
is to assure that some quantities (essentially
the quantities which correspond to our resonance factors),
considered as functions of the
scalar product between $\oo$ and
the momentum flowing through the incoming line of
the resonance, say $\oo\cdot\nn_{\l_V}$,
do not go out from their analyticity domain,
when shifting the external lines of the resonances.
This is implemented in [CF1] by
requiring that the quantity $|\oo\cdot\nn_{\l_V}|$
corresponding to the momentum $\nn_{\l_V}$ flowing through
the resonant line cannot be
smaller than a prefixed fraction of the quantity $|\oo\cdot\nn_{\l}|$
corresponding to
the momentum of any line $\l$ inside a resonance, so that
the shift of the external lines of the resonances
``does not modify too much'' the small divisors. In [G7]
the same is accomplished by requiring that
the number of lines inside the resonance
are bounded by a suitable constant. In [G7] use is made also
of the {\sl strong diophantine property}, \ie $\oo$ is such that
$\min_{0\ge p\ge n}|C_0|\oo\cdot\nn|-2^{-p}|>
2^{-(n+1)}$ for $0<|\nn|\leq (2^{n+3})^{-\t^{-1}}$:
in this way one can forbid that two
quantities $|\oo\cdot\nn_{\l}|$ and $|\oo\cdot\nn_{\l'}|$
with different scales are too much near to each other;
the last condition was eliminated in [GG] through a suitable
choice of the unity partition used in the multiscale
decomposition of the propagator.
Both constructions reflect the fact that only resonances $V$
such that the scales $n_{\l_V}$ and $n_V$ are very different
can give problems (because it is only when this happens
that dangerously small divisors can occur) and cannot be treated by
the Siegel-Bryuno's lemma, or any variant of it.
We can note that in [G7] and [GG] the analysis is somewhat simpler
with respect to [CF1], as far as what concerns the problem of
singling out the contributions needing a more
careful discussion (in order to show the cancellations).
For instance, as far as the non overlapping of resonances is concerned,
this is implied by the definition of {\sl critical resonance} in [CF1]
(see Proposition 5.2 in [CF1]),
while it is automatically satisfied by the construction
of clusters in [G7] and [GG], (and, obviously, in the successive related
papers, including the present one).

\*

With respect to the just described works, the technique
used in the present paper (which is taken from [GM1] and
[GM2], where simplified models are studied)
is closer to quantum field methods,
and in fact reduces the proof of the KAM theorem to the study
of a renormalizable field theory, thus allowing us to use all
the powerful ideas which have been developed so far
in order to treat such kind of problems.
We think that the reason why the proof can be given in a form
as simple as in [G7], [GM1] (for
semplified models) is due precisely to the use of such ideas.

Furthermore some simplifications can be obtained
in the proof: first the
resonances are defined {\it tout court}
as the clusters such that
there is only one entering line, and it carries the
same momentum of the exiting one (there is no need to
distinguish between resonances, $\l$-resonances and
critical resonances, as in [CF1]: obviously this
corresponds to the fact that a gain is truly necessary
only for the last ones)
and the problem of exiting from certain analyticity domains
never arise. This property is easily obtained through the
introduction of the compact support functions
realizing the partition of the unity (see \S 4 and App. A3), and it can
be understood in the following way.

If the notations in \S 6 are adopted, for each resonance $V$,
one has that the sum over all the trees contained in the resonance family
$\FF_V(\th)$ produces a quantity which, considered as a function
of $\oo\cdot\nn_{\l_V}$, vanishes to first and second order;
then, if such a function is analytic in $\oo\cdot\nn_{\l_V}$
in a ball centered on the origin,
the Schwarz's lemma for analytic functions can be
used in order to obtain bounds on the small divisors.
This is what was done in [G7], [GG] and [CF1].
{}From a technical point of view, this yields that the
quantities $\oo\cdot\nn_{\l_V}$ appearing in the small
divisors have to be dealt with as parameters which can
assume values larger than their true values, and, if
we have several resonances contained in each other, a very
careful analysis of the holomorphy domains is needed.
For details we refer to the quoted papers and to [GG]
in particular (where the problem is discussed to mucxh extent):
the main point here is that the technical intricacies are
a consequence of the analyticity request.
In this paper we do not require analyticity, but we want
only a bound on the second derivatives of the small divisors
(see \S 7), once we have proven that no higher order derivative
appears ({\it ibidem}). This means that we do not need a bound
on the values of the small divisors in a neighbourhood of the origin,
but {\it only} on the their first and second derivatives
appearing in the interpolation formulae \equ(7.5): in other words
we only require that the resonance factors are twice differentiable.
See also the comments between Definition 6.3 and Lemma 6.4.
The deep reason and the drawback
of this semplification is that it follows from an overcompensation,
\ie we collect together terms producing more cancellations
than it would be necessary in order to make the series convergent;
the overcompensation is paid by worse final estimates
for the convergence radius.

\*


\vskip1.truecm

\\{\bf Acknowledgements.}
This work springs from the purpose
of extending to the most general case the results found in simplified
models in previous papers, and can be considered as a development
of the original ideas contained in the paper [G7] of Giovanni
Gallavotti, which we thank for having introduced us to a quantum
field theory approach to KAM theorem and for many
enlightening discussions.
We thank Antonio Giorgilli for a useful discussion about the
work [GL], which we do not discuss here, as it is a new proof
of KAM theorem in Kolmogorov's spirit, so no directly related
to the works we have analyzed in \S 9.
One of us (G.G.) thanks IHES for partial support and hospitality,
while part of this work was done. This work is part of the research
program of the European Network on ``Stability and Universality
in Classical Mechanics'', \# ERBCHRXCT940460.


\vskip1.truecm

\centerline{\titolo Appendix A1.
Resonant Siegel-Bryuno's bound.}
\*

\\Given a tree $\th$, if we are interested only in
the momentum and mode labels, the case
in which either there is a line entering a vertex $v$ and
carrying a zero momentum or there is no line at all
behave exactly in the same way, as far the momenta of
the vertices $w<v$ are concerned.
This means that \equ(5.6)
can be written as a product of $1+\NN_{\V0}(\th)$ factorising
terms, if $\NN_{\V0}(\th)$ is the number of lines $\l$
in $\th$ (other than the branch root) such that $\nn_{\l}=\V0$.

\*

Therefore we can
confine ourselves to the case in which there is no line
carrying a vanishing momentum.

\*

We prove by induction on the tree order that,
if $N_n^*(\th)$ is defined as the number
of non resonant lines in $\th$ carring a scale label $\le n$,
then $N_n^*(\th)\le 2M(\th)2^{(n+2)/\t}-1$, if $N_n(\th)\neq0$.

Let $\th$ be a tree of order $k$.
If $\th$ has the root line with scale $>n$ then calling
$\th_1,\th_2,\ldots,\th_m$ the subtrees of $\th$ emerging from the
first vertex of $\th$ and with $M(\th_j)>2^{-(n+2)/\t}$ lines, it is
$N^*_n(\th)=N^*_n(\th_1)+\ldots+N^*_n(\th_m)$ and the statement
is inductively implied from its validity for $k'<k$ provided it is true that
$N^*_n(\th)=0$ if $M(\th)<2^{-(n+2)/\t}$, which is is certainly the case.

In the other case, (\ie if the root branch has scale label $\le n$),
it is $N^*_n(\th)\le 1+\sum_{i=1}^mN^*_n(\th_i)$, and if
$m=0$ the statement is trivial, or if $m\ge2$ the statement is again
inductively implied by its validity for $k'<k$.

If $m=1$ we once more have a trivial case unless it is
$M(\th_1)>M(\th)-2^{-1}2^{-(n+2)/\t}$. But in the latter case,
it turns out that the root line
of $\th_1$ is either a resonant line or it has scale $>n$.

Accepting the last statement we have: $N^*_n(\th)=1+N^*_n(\th_1)=
1+N^*_n(\th'_1)+\ldots+N^*_n(\th'_{m'})$, with $\th'_j$'s being the $m'$
subtrees emerging from the first vertex of $\th'_1$ with
$M(\th_j')>2^{-(n+2)/\t}$: this is so because the root line of $\th_1$ will
not contribute its unit to $N^*_n(\th_1)$. Going once more through the
analysis the only non trivial case is if $m'=1$ and in that case
$N^*_n(\th'_1)=N^*_n(\th{'}{'}_1)+\ldots+N^*_n(\th{'}{'}_{m{'}{'}})$,
\etc., until we reach either a trivial case or a tree $\tilde\th$ such that
$M(\tilde\th)<M(\th)-2^{-1}2^{-n/\t}$.

It remains to check that, if $M(\th_1)>M(\th)-2^{-1}2^{-(n+2)/\t}$, then the
root line of $\th_1$ has scale $>n$, unless it is entering a resonance.

Suppose that the root line of $\th_1$ has scale $\le n$ and is not
entering a resonance. Note that $|\oo\cdot\nn_{\l_{v_0}}|\le\,2^{n+1}$,
$|\oo\cdot\nn_{\l_{v_1}}|\le\,2^{n+1}$,
if $v_0$ and $v_1$ are the first vertices of $\th$ and $\th_1$
respectively.  Hence $\d\=|(\oo\cdot(\nn_{\l_{v_0}}-
\nn_{\l_{v_1}})|\le2\,2^{n+1}$ and
the diophantine assumption implies that $|\nn_{\l_{v_0}}-
\nn_{\l_{v_1}}|>
(2^{n+2})^{-1/\t}$, or $\nn_{\l_{v_0}}=\nn_{\l_{v_1}}$.
The latter case being
discarded as we are not considering the
resonances,\footnote{${}^6$}{\nota
Note that $M(\th)-M(\th_1) < 2{-1}2^{-(n+2)/\t}$ implies
that $|\oo\cdot\l|\ge 2^{1+(n+2)/\t}$ for all the lines
$\l$ preceding the root line and not contained in $\th_1$, so that
the set composed by such lines, if
$\nn_{\l_{v_0}}=\nn_{\l_{v_1}}$, is a resonance
on scale $n'>n$.}
it follows
that $M(\th)-M(\th_1)<2^{-1}\,2^{-(n+2)/\t}$ is inconsistent:
it would in fact imply that $\nn_{\l_{v_0}}-\nn_{\l_{v_1}}$
is a sum of $k-k_1$ vertex modes such that
$|\nn_{\l_{v_0}}-\nn_{\l_{v_1}}|< 2^{-1}2^{-(n+2)/\t}$,
hence $\d>2^{n+3}$ which is contradictory with the above opposite
inequality.

\*

Analogously, we can prove that, if $N_n(\th)>0$,
then the number $p_n(\th)$ of clusters of scale $n$ verifies the bound
$p_n(\th) \le 2 M(\th)2^{(n+2)/\t})-1$. In fact this is true
for a tree $\th$ such that $M(\th)\le 2^{(n+2)/\t}$. Otherwise, if the
first tree vertex $v_0$ is not in a cluster of scale $n$, it is
$p_n(\th)=p(\th_1)+\ldots+p_n(\th_m)$, with the above notation,
and the statement follows by induction. If $v_0$ is in a cluster
on scale $n$ we call $\tilde\th_1, \ldots, \tilde\th_m$ the
subdiagrams emerging from the cluster containing $v_0$ and
such that $M(\th_j)>2^{-(n+2)/\t}$, $j=1,\ldots,m$: it will be
$p_n(\th)=1+p(\tilde\th_1)+\ldots+p_n(\tilde\th_m)$. Again
we can assume $m=1$, the other cases being trivial. But in
such a case there will be only one branch entering the cluster $T$
on scale $n$ containing $v_0$ and it will have a momentum of
scale $n'\le n-1$. Therefore the cluster $T$ must contain vertices
such that at least $\sum_{v\in T}|\nn_v|>
2^{-(n+2)/\t}$ vertices, (otherwise, if $\l$ is a line on scale $n$
contained in $T$, and $\nn_\l^0$ is the sum of the mode labels
corresponding to the vertices following $v_0$ but inside $T$, we would have
$|\oo\cdot\nn_\l|\le 2^{n+1}$ and, simultaneously,
$|\oo\cdot\nn_\l|\ge 2^{n+3}-2^{n-1}>2^{n+2}$, which would lead to a
contradiction). This means that $M(\th_1)\le M(\th) -2^{-(n+2)/\t}$.

\*

 From the above proven results, \equ(5.4) follows, if we note
that $\sum_{T,n_T=n}1=p_n(\th)$.


\vskip1.truecm

\centerline{\titolo Appendix A2. Relaxing of the hypothesis \equ(2.1)}
\*\numsec=2\numfor=1

\\If $f(\aa;\AA;\e)$ is analytic in $\e$, then we can write
$$ f(\aa,\AA;\e) = \sum_{k=1}^{\io} f^{(k)}(\aa,\AA)\,\e^k
\; , \Eqa(A2.1) $$
where $|f^{(k)}|\le F\,\e_1^{-k}$ in a domain $|\e|\le\e_1$.

Let us consider first the hamiltonian in Theorem 1.2, and trees such
that $\d_v=1$ $\forall v\in\th$. Then
a diagrammatic expansion is still possible, and the only difference is
that now to order $k$ we have to consider all the possible graphs
with $p$ vertices, $p=1,\ldots,k$, such that (1) to each vertex $v$
a factor $f^{(k_v)}_{\nn_v}$ is associated, and (2) the $k_v$'s
labels have to satisfy the constraint $\sum_{v\in\th}k_v=k$.

\*

For each tree
a bound $C_1^p$ can be obtained: this can be easily
argued from the discussion for the
interaction \equ(2.1), and the constant $C_1$ is the
same one (up to the factor $F$, which now is
missing as the factors $f_{\nn_v}$ are replaced by the new ones
$f_{\nn_v}^{(k_v)}$), times the product $\prod_{v\in\th}F\,\e_1^{-k_v}$.
Then we have to consider all the
possible ways to assign the factors $f^{(k_v)}_{\nn_v}$, \ie
the $k_v$ labels, to the vertices of the tree, which gives a sum
$$ \sum_{p=1}^k \sum_{k_1+\ldots+k_p=k \atop \{k_i \ge 1 \}_{i=1}^p }
C_1^p\,[F\,\e_1^{-1}]^k \le [F\,\e_1^{-1}]^k
\sum_{p=1}^k { C_1^p \, k^p \over p!} \le
[e\,C_1\,F\,\e_1^{-1}]^k \; , \Eqa(A2.2) $$
so that, like in the case discussed previously, (interaction
of the form \equ(2.1)), we find again a bound $C_2^k$, where
now $C_2=C_0\,e\,\e_1^{-1}$, if $C_0$ was the value
previously obtained. If there are also
vertices $v\in\th$ with $\d_v=0$, then the previous discussion
has to be restricted to the vertices having $\d_v=1$,
and the same result holds.
This concludes the proof of Theorem 1.2.

\*

In the case of Thorem 1.4, we can repeat the same analysis,
by restricting it again to the $k$ vertices $v$'s having $\d_v=1$,
(as the other ones correspond to factors $\HH_0$), and the same
result is obtained.


\vskip1.truecm

\centerline{\titolo Appendix A3. A partition of unity via characteristic
functions}
\*\numsec=3\numfor=1

\\Besides the partition of unity described in \S 4, there are
other possibilities that could be envisaged. A very natural one is
discussed in this Appendix.

Let us define
$$ \c_n(x) = \theta (|x|-2^{n-1}) - \theta(|x|-2^n) \; ,
\quad n \le 0 \; , \qquad \ch_1(x) = \theta (|x|-1) \; , \Eqa(A3.1) $$
where $\theta(x)$ is the Heaviside function, defined as
$$ \theta(x) = \cases{ 1 \; , & if $x > 0 \; , $ \cr
1/2 \; , & if $x=0 \; , $ \cr
0 \; , & if $x<0 \; . $\cr} \Eqa(A3.2) $$
Then, for any $x\in{\RRR}$, we have
$$ \sum_{n=-\io}^1 \ch_n(x) = 1 \; , \Eqa(A3.3) $$
so that we can define the ``propagator at scale $n$'' as
$$ g^{(n)}(\oo\cdot\nn_{\l}) = { \c_n(\oo\cdot\nn_{\l}) \over
(\oo\cdot\nn_{\l})^{R_{\l}}} \; , \Eqa(A3.4) $$
which replaces \equ(4.2).

\*

Then the discussion of \S 5 remains unchanged, the only
(irrelevant) difference being that the sum in item (5) before
\equ(5.3) is over $\le 2^{2k-1}$ terms. In fact,
if the momentum $\nn_{\l}$ in \equ(A3.2) is fixed,
there is only one scale
$n$ such that the propagator is not vanishing, (see note 5
for analogous considerations for the partition \equ(4.1)),
except the case in which $\oo\cdot\nn_{\l}$ is a diadic
point $2^n$, so that two successive scales
are possible, but then \equ(A3.1) and \equ(A3.2)
give a factor $1/2$ for each scale.

\*

The discussion in \S 6 about the approximate cancellations
can be easily adapted. Looking at \equ(6.4)
and \equ(6.6), one could think that problems arise from
derivatives of the functions \equ(A3.1), since delta
functions appear:
$$ \dot\c_n(x) = \d(|x|-2^{n-1})-\d(|x|-2^n) \; . \Eqa(A3.5) $$
In fact, we can rule out all contributions containing any
derivatives $\dot \c_{n_{\l_v}}(\oo\cdot\nn_{\l_v})$,
because the corresponding localized resonance factor
either is vanishing or gives a vanishing contribution when the sum over
the scales is performed.

This is a property which follows from the fact that,
from definition \equ(A3.5),
the derivatives of the characteristic functions in \equ(A3.1)
can be different from zero {\sl only} if
some $\oo\cdot\nn_\l$ falls on the boundary of some
diadic interval $[2^{n_\l-1},2^{n_\l}]$,
say or $2^{n}$, if $n=n_{\l}$ or $n=n_{\l}-1$.

But in such a case, by starting from the outermost (\ie maximal)
resonances for which this happens,
we consider together the values of the two trees in which
the scale label of that line is $n$ and $n+1$. Then, by denoting
as usual $n_V$ the scale of the resonance $V$ as a cluster
and $n_{\l_V}$ the resonance-scale, if one has not
$\oo\cdot\nn_{\l}=2^{n_\l-1}=2^{n_V-1}=2^{n_{\l_V}}$,
both cases $n_{\l}=n,n+1$ are compatible
with the resonance structure, and we see that the
two values we obtain by (1) considering the derivative
of the $\ch_{n_{\l}}$ and (2) collecting together the $\LL$ and $\RR$
parts of the resonance factors for all resonances inside
$V$, differ only because the first one contains a delta
$-\d(|\oo\cdot\nn(v)| -2^n)$, whereas the latter
contains the same delta, {\sl but with opposite sign},
\ie $\d(|\oo\cdot\nn(v)|-2^n)$, (see \equ(A3.5)), being
all the other factors equal. Therefore the two
tree values are opposite, and, when summed together, cancel
exactly.

On the contrary, in the case
$\oo\cdot\nn_{\l}=2^{n_\l-1}=2^{n_V-1}=2^{n_{\l_V}}$,
if we assign the scale label $n+1$ to the line $\l$, then
we destroy the cluster structure, and we have no more a
resonance. But of course we can again define the localized part
of the quantity obtained from the resonance factor by
shifting by 1 the scale of $\l$: then such a part cancels with
the localized part of the resonance factor corresponding to $V$,
(exactly as before), while the remaining part can be easily
handled as the $\RR$ part of the resonance factor.

Once the maximal clusters are treated,
we pass to the next-to-maximal resonances, \ie to the maximal
resonances contained inside the maximal ones, and we
study in the same way the localized parts.
And so on until the innermost resonances are dealt with.

\*

A similar analysis is required in order to
adapt the discussion in \S 7, where some
$\dot \c_{n_{\l}}(\oo\cdot\nn_{\l}^0 + t\oo\cdot\nn)$ and
$\ddot \c_{n_{\l}}(\oo\cdot\nn_{\l}^0 + t\oo\cdot\nn)$ appear
($\nn_{\l}^0$ is defined in Remark 6.2, and $t$
is the interpolation parameter introduced in \equ(7.5)).

When $t$ varies in $[0,1]$,
$\oo\cdot\nn_{\l}^0 + t\oo\cdot\nn$ varies in
$[\oo\cdot\nn_{\l}^0, \oo\cdot\nn_{\l}^0 + \oo\cdot\nn]$.
Then we can proceed in the following way, by starting
from the maximal resonances as before and
by considering all the possible labels assignments
inside the resonance. In particular in this way, for each line $\l\in
V$, we sum all the scales $n_{\l}>n_{\l_V}$,
so that the functions \equ(A3.1) give a function
$\th(|\oo\cdot\nn_{\l}|-2^{n_{\l_V}})$.
Then the derivative of such a function can give a
delta, only if $\oo\cdot\nn_{\l}^0 + t\oo\cdot\nn$
falls on $2^{n_{\l_V}}$ for some $t\in[0,1]$.
But if this occur, then we can perform the integration
on $t$: no gain is obtained in such an operation, but no gain
is really needed in such a case (which would correspond
to have a scale label $n_{\l_V}+1$ on $\l$).
Then we pass to the next-to-maximal resonances, we apply
the $\LL$ and $\RR$ operators to the
corresponding resonance factors, and we proceed in the same way,
in order to study the $\RR$ part of the resonance factor.
And so on, until all the resonances are studied.

At this point, we can repeat the discussion in \S 7, and the
same results can be obtained.

\vfill
\eject

\pagina


\centerline{\titolone References}
\vskip1.truecm

\halign{\hbox to 1.2truecm {[#]\hss} &
        \vtop{\advance\hsize by -1.25 truecm \\#}\cr

A1& {V.I. Arnol$'$d:
Small divisor problems in classical and celestial mechanics,
{\it Uspekhi Mat. Nauk} {\bf 18}, No. 6, 91-192 (1963);
english translation in {\it Russ. Math. Surv.} {\bf 18},
No. 6, 85--191 (1963). }\cr
A2& {V.I. Arnol$'$d:
Small divisors II. Proof of a A.N. Kolmogorov's theorem on
conservation of conditionally periodic motions under small perturbations
of the hamiltonian,
{\it Uspekhi Mat. Nauk} {\bf 18}, No. 5, 13--40 (1963);
english translation in {\it Russ. Math. Surv.} {\bf 18},
No. 5, 9--36 (1963). }\cr
Ba& {F. Bacon: {\sl Essays}, LVIII; see also J.L. Borges:
{\sl El inmortal}. }\cr
B& {A.D. Bryuno:
The analytic form of differential equations I,
{\it Trans. Moscow Math. Soc.} {\bf 25},
131--288 (1971); II,
{\it Trans. Moscow Math. Soc.} {\bf 26}, 199--239 (1972). }\cr
CF1& {L. Chierchia, C. Falcolini:
A direct proof of a theorem by Kolmogorov in hamiltonian systems,
{\it Ann. Scuola Norm. Super. Pisa Cl. Sci. Ser. IV} {\bf 21},
No. 4, 541--593 (1995). }\cr
CF2& {L. Chierchia, C. Falcolini:
Compensations in small divisors problems,
preprint, Roma (1994),
archived in {$\tt mp\_arc@math.utexas.edu$}, \#94-270,}\cr
CG& {L. Chierchia, G. Gallavotti:
Drift and diffusion in phase space,
{\it Ann. Inst. H. Poincar\'e Phys. Th\'eor.}
{\bf 60}, No. 1, 1--144 (1994). }\cr
CZ& {L. Chierchia, E. Zehnder:
Asymptotic expansions of quasi--periodic motions,
{\it Ann. Scuola Norm. Super. Pisa Cl. Sci. Ser. IV} {\bf 16},
-- (1989). }\cr
DS& {E.I. Dinaburg, Ya.G. Sinai:
The one dimensional Schr\"odinger equation with a quasi periodic
potentional,
{\it Func. Anal. and its Appl.} {\bf 9}, 279--289 (1975). }\cr
E1& {L.H. Eliasson:
Absolutely convergent series expansions
for quasi-periodic motions, Report 2--88,
Department of Mathematics, University of Stockholm (1988). }\cr
E2& {L.H. Eliasson:
Hamiltonian systems with linear normal
form near an invariant torus, in {\sl Nonlinear Dynamics},
Ed. G. Turchetti, Bologna Conference, 30/5 to 3/6 1988,
World Scientific, Singapore (1989). }\cr
E3& {L.H. Eliasson:
Generalization of an estimate of small divisors by Siegel, in
{\sl Analysis, et cetera},
Ed. E. Zehnder \& P. Rabinowitz,
book in honor of J. Moser, Academic press (1990). }\cr
EV& {J. Ecalle, B. Vallet:
The trimmed and corrected forms of resonant vector fields or
diffeomorphisms, preprint, Paris (1994). }\cr
FT& {J. Feldman, E. Trubowitz:
Renormalization on Classical Mechanics and Many Body Quantum Field Theory,
{\it J. Anal. Math.} {\bf 58}, 213--247 (1992). }\cr
G1& {G. Gallavotti:
{\sl Elements of Mechanics}, Springer, New York (1983),
translation of {\sl Meccanica Elementare},
Boringhieri, Torino (1980). }\cr
G2& {G. Gallavotti:
A criterion of
integrability for perturbed harmonic
oscillators. ``Wick Or\-der\-ing'' of the
perturbations in classical mechanics and
invariance of the frequancy spectrum,
{\it Comm. Math. Phys.} {\bf 87}, 365--383 (1982). }\cr
G3& {G. Gallavotti:
Perturbation theory for classical hamiltonian systems,
in {\sl Scaling and Self-Similarity in Physics},
Ed. J. Fr\"olich, Birkh\"auser, Boston (1983). }\cr
G4& {G. Gallavotti:
Quasi--integrable mechanics sysyems, in
{\sl Critical Phenomena, Random Systems, Gauge Theories},
Les Houches, Session XLIII
(1984), Vol. II, 539--624, Ed. K. Osterwalder \& R. Stora,
North Holland, Amsterdam (1986). }\cr
G5& {G. Gallavotti:
Renormalization theory and ultraviolet stability
for scalar fields via renormalization group methods,
{\it Rev. Modern Phys.} {\bf 57}, 471--572 (1985). }\cr
G6& {G. Gallavotti:
Classical Mechanics and Renormalization Group,
Lectures notes at the 1983 Erice summer school,
Ed. A. Wightman \& G. Velo, Plenum, New York (1985). }\cr
G7& {G. Gallavotti:
Twistless KAM tori,
{\it Comm. Math. Phys.} {\bf 164}, 145--156 (1994). }\cr
G8& {G. Gallavotti:
Twistless KAM tori, quasi flat
homoclinic intersections, and other cancellations in the perturbation
series of certain completely integrable hamiltonian systems. A review,
{\it Rev. Math. Phys.} {\bf 6}, No. 3, 343--411 (1994). }\cr
G9& {G. Gallavotti:
Perturbation Theory,
in {\sl Mathematical Physics towards the XXI Century}, 275--294,
Ed. R. Sen \& A. Gersten, Ben Gurion University Press,
Ber Sheva (1994). }\cr
GG& {G. Gallavotti, G. Gentile:
Majorant series convergence for twistless KAM tori,
{\it Ergodic Theory Dynam. Systems}, to appear (1995). }\cr
GGM& {G. Gallavotti, G. Gentile, V. Mastropietro:
KAM theorem and quantum field theory, preprint, Bures sur Yvette, (1995),
archived in {$\tt mp\_arc@math.utexas.edu$}, \#95-151. }\cr
Ge1& {G. Gentile:
A proof of existence of whiskered tori
with quasi flat homoclinic intersections
in a class of almost integrable hamiltonian systems,
{\it Forum Math.}, to appear (1995). }\cr
Ge2& {G. Gentile:
Whiskered tori with prefixed frequencies and Lyapunov spectrum,
{\it Dynam. Stability of Systems}, to appear (1995). }\cr
GM1& {G. Gentile, V. Mastropietro:
KAM theorem revisited,
preprint, Roma (1994),
ar\-chiv\-ed in {$\tt mp\_arc@$} {$\tt math.$}
{$\tt utexas.$}{$\tt edu$}, \#94-403. }\cr
GM2& {G. Gentile, V. Mastropietro:
Tree expansion and multiscale analysis for KAM tori,
pre\-print, Roma (1994), archived in {$\tt mp\_arc@math.$}
{$\tt utexas.$} {$\tt edu$}, \#94-405. }\cr
GL& {A. Giorgilli, U. Locatelli:
Kolmogorov theorem and classical perturbation theory,
preprint, Milano (1995). }\cr
HC& {J.D. Hanson, J.R. Cary:
Elimination of stochasticity in stellarators,
{\it Phys. Fluids} {\bf 27}, 767--769 (1984). }\cr
HP& {F. Harary, E. Palmer:
{\sl Graphical Enumerations},
Academic Press, New York (1973). }\cr
K&  {A.N. Kolmogorov:
On the preservation of conditionally periodic motions,
{\it Dokl. Akad. Nauk} {\bf 96}, 527--530 (1954);
english translation in G. Casati, J. Ford:
{\it Stochastic behavior in classical and quantum hamiltonian},
Lectures Notes in Physics {\bf 93}, Springer, Berlin (1979). }\cr
L& {A. Lindstedt:
{\it M\'emoires de l'Acad\'emique de Saint-P\'etersbourg}
(1882), quoted in H. Poin\-ca\-r\'e:
{\sl Les m\'ethodes nouvelles de la m\'ecanique c\'eleste},
Vol.II, Gauthier-Villars, Paris (1893), {\it partim}. }\cr
M1& {J. Moser:
On invariant curves of an area preserving mapping of the annulus,
{\it Nachr. Akad. Wiss. G\"ottingen Math.--Phys.
Kl. II}, 1--20 (1962). }\cr
M2& {J. Moser:
A rapidly convergent iteration method and nonlinear
differential equations II,
{\it Ann. Scuola Norm. Super. Pisa Cl. Sci. Ser. III},
{\bf 20}, 499--535 (1966). }\cr
M3& {J. Moser:
Convergent series expansions for quasi--periodic motions,
{\it Math. Ann.} {\bf 169}, 136--176 (1967). }\cr
M4& {J. Moser:
On the construction of almost periodic solutions for ordinary
differential equations, {\it Proc. Int. Conf. of functional
analysis and related topics}, Tokyo, 60--67 (1969). }\cr
M5& {J. Moser:
{\sl Stable and Random Motions in Dynamical Systems,
with special emphasys on Celestial Mechanics},
Annals of Mathematical Studies 77, Princeton University Press,
Princeton (1973). }\cr
MS& {J. Moser, C.L. Siegel:
{\sl Lectures on Celestial Mechanics},
Springer, Berlin (1971). }\cr
N& {J. Nash:
The imbedding of Riemannian manifolds,
{\it Ann. of Math.} {\bf 63}, 20--63 (1956). }\cr
Ne& {S. Newcomb:
{\sl Smithsonian contributions to knowledge} (1874),
quoted in H. Poin\-ca\-r\'e:
{\sl Les m\'ethodes nouvelles de la m\'ecanique c\'eleste},
Vol.II, Gauthier-Villars, Paris (1893), {\it partim}. }\cr
P& {H. Poincar\'e:
{\sl Les m\'ethodes nouvelles de la m\'ecanique c\'eleste},
Gauthier-Villars, Paris, Vol. I (1892), Vol. II (1893),
Vol. III (1899). }\cr
P\"o& {J. P\"oschel:
Invariant manifolds of complex analytic mappings,
in {\sl Critical Phenomena, Random Systems, Gauge Theories},
Les Houches, Session XLIII (1984), Vol. II, 949--964, Ed.
K. Osterwalder \& R. Stora, North Holland, Amsterdam (1986). }\cr
%
%
R1& {H. R\"ussmann:
\"Uber der Normalform analytischer Hamiltonscher
Differentialgleichungen in der N\"ahe einer Gleichgewichtl\"osung,
{\it Math. Ann.} {\bf 169}, 55--72 (1967). }\cr
R2& {H. R\"ussmann:
\"Uber invariante Kurven differenzierbaren Abbildungen eines
Kreisringes,
{\it Nachr. Akad. Wiss. G\"ottingen Math.--Phys. Kl. II},
67--105 (1970). }\cr
R3& {H. R\"ussmann:
One dimensional Schr\"odinger equation,
{\it Ann. New York Acad. Sci.} {\bf 357}, 90--107 (1980). }\cr
S1& {C.L. Siegel:
Iterations of analytic functions,
{\it Ann. of Math.} {\bf 43}, No.4, 607--612 (1943). }\cr
S2& {C.L. Siegel:
\"Uber die Existenz einer Normalform analytischer Hamiltonscher
Differentialgleichungen in der N\"ahe einer Gleichgewichtl\"osung,
{\it Math. Ann.} {\bf 128}, 144--170 (1954). }\cr
SHS& {R.M. Sinclair, J.C. Hosea, G.V. Sheffield:
A method for mapping a toroidal magnetic field by storage
of phase stabilized electrons,
{\it Rev. Sci. Instrum.} {\bf 41}, 1552--1559 (1970). }\cr
T& {W. Thirring:
{\sl Course in Mathematical Physics},
Springer, New York, Vol. 1 (1978), Vol. 2 (1979),
Vol. 3 (1981), Vol. 4 (1983),
translation of {\sl Lehrbuch der Mathematischer
Physik},
Springer, Wien, Vol. 1 (1977), Vol. 2 (1978), Vol. 3 (1979),
Vol. 4 (1980). }\cr
Ti& {E.C. Titchmarsh: {\sl The theory of functions},
Oxford University Press, London (1932). }\cr
W& {K.T.W. Weierstrass: Letters published in
{\it Act. Math.} {\bf 35}, 29--65 (1911), quoted in
J. Moser:
{\sl Stable and Random Motions in Dynamical Systems,
with special emphasys on Celestial Mechanics},
Annals of Mathematical studies 77, Princeton University Press,
Princeton (1973), Ch. I, \S 2. }\cr }
%

\vfill
\eject

%
%
\font\titolone=cmbx12 scaled\magstep 2
\nopagenumbers
\null
\vskip2.truecm
\centerline{\titolone Methods for the analysis}
\vskip.2truecm
\centerline{\titolone of the Lindstedt series for KAM tori}
\vskip.2truecm
\centerline{\titolone and renormalizability
in classical mechanics}
\vskip.4truecm
\centerline{\titolo A review with some applications}
\vskip2.truecm
\centerline{{\titolo G. Gentile}$^{\dagger}$,
{\titolo V. Mastropietro}$^{\dagger\dagger}$}
\*

\0$\dagger$ IHES, 35 Route de Chartres, 91440 Bures sur Yvette, France.

\0$\dagger\dagger$
Dipartimento di Matematica, Universit\`a di Roma II
``Tor Vergata", 00133 Roma, Italia.
\vskip3.truecm
\centerline{\cs Abstract}
\*
\\This paper consists in a unified exposition of
methods and techniques of the renormalization group approach
to quantum field theory applied to classical mechanics,
and in a review of results:
(1) a proof of the KAM theorem,
by studing the perturbative expansion (Lindstedt series) for the
formal solution of the equations of motion;
(2) a proof of a conjecture by Gallavotti about the
renormalizability of isochronous hamiltonians,
\ie the possibility to add a term depending only on the actions
in a hamiltonian function not verifying the anisochrony condition
so that the resulting hamiltonian is integrable.
Such results were obtained first by Eliasson;
however the difficulties arising in the study of the perturbative
series are very similar to the problems which one has to deal with
in quantum field theory, so that the use the methods which have been
envisaged and developed in the last twenty years exactly in order to
solve them allows us to obtain unified proofs, both conceptually
and technically.
In the final part of the review,
the original work of Eliasson is analyzed and exposed in detail; its
connection with other proofs of the KAM theorem
based on his method is elucidated.
%
%

\vfill
\eject
\pagina

%
%
\nopagenumbers
\null
\vskip1.truecm
\centerline{\titolone Figures}
\vskip2.truecm
%
%
\midinsert
\*
\hrule{}
\*
\insertplot{240pt}{190pt}{
\ins{-35pt}{100pt}{$\hbox{root}$}
\ins{60pt}{95pt}{$v_0$}
\ins{152pt}{130pt}{$v_1$}
\ins{110pt}{60pt}{$v_2$}
\ins{190pt}{110pt}{$v_3$}
\ins{230pt}{170pt}{$v_5$}
\ins{230pt}{130pt}{$v_6$}
\ins{230pt}{95pt}{$v_7$}
\ins{230pt}{5pt}{$v_{11}$}
\ins{230pt}{30pt}{$v_{10}$}
\ins{200pt}{75pt}{$v_4$}
\ins{230pt}{75pt}{$v_8$}
\ins{230pt}{55pt}{$v_9$}
}{fig1}
\*
\didascalia{{\bf Fig.3.1.} A tree $\th$ with degree $d=11$.
Each line (branch) is supposed to carry an arrow
(which is not explicitly drawn) pointing to the root.
If we consider a vertex of the tree, \eg $v_3$, then we define
$\l_{v_3}$, or equivalently $v_1\ot v_3$, the line connecting
$v_3$ to $v_1$, and we write $v_1=v_3'$. The arrow, if drawn,
would point from $v_3$ to $v_1$, as one has to cross $v_1$
in order to reach the root from $v_3$.}
\*
\hrule{}
\*
\endinsert

%
\midinsert
\*
\hrule{}
\*
\insertplot{220pt}{80pt}{
\ins{-10pt}{32pt}{$\hbox{root}$}
\ins{130pt}{32pt}{$\hbox{root}$}
\ins{35pt}{32pt}{$v_0$}
\ins{175pt}{32pt}{$v_0$}
\ins{50pt}{62pt}{$v_1$}
\ins{190pt}{20pt}{$v_1$}
\ins{75pt}{63pt}{$v_2$}
\ins{215pt}{63pt}{$v_2$}
\ins{75pt}{35pt}{$v_3$}
\ins{215pt}{33pt}{$v_3$}
\ins{75pt}{5pt}{$v_4$}
\ins{215pt}{5pt}{$v_4$}
}{fig2}
\*
\didascalia{{\bf Fig.3.2.} Two trees $\th_1$ and $\th_2$ of
degree $d=5$, which are different if regarded as semitopological
trees and identical if regarded as topological trees. In fact,
if we permute the subtrees emerging from the first vertex, we
obtain $\th_2$ from $\th_1$ and {\it viceversa}.}
\*
\hrule{}
\*
\endinsert

%
\midinsert
\*
\hrule{}
\*
\vskip3.2truecm
\insertplot{320pt}{190pt}{
\ins{10pt}{83pt}{$\hbox{root}$}
\ins{90pt}{40pt}{$n_1$}
\ins{94pt}{129pt}{$n_2$}
\ins{150pt}{-20pt}{$n_0$}
\ins{205pt}{50pt}{$n_3$}
\ins{210pt}{130pt}{$n_4$}
\ins{232pt}{130pt}{$n_6$}
\ins{247pt}{60pt}{$n_5$}
\ins{70pt}{83pt}{$v_1$}
\ins{125pt}{50pt}{$v_2$}
}{fig3}
\vskip3.5truecm
\didascalia{{\bf Fig.4.1.} A tree with scale labels associated to
the lines. There are seven clusters $T_0$, $\ldots$, $T_6$
on scale, respectively, $n_0$, $\ldots$, $n_6$, which
satisfy the ordering relations: $n_0>n_1$, $n_0>n_2$, $n_0>n_3$,
$n_3>n_4$, $n_3>n_5$, and $n_4>n_6$.
If the external lines of the cluster $T_1$ carry
the same momentum (\ie $\nn_{v_1}+\nn_{v_2}=\V0$), then
$T_1$ is a resonance (see Definition 5.1), and we write
$T_1=V_1$. Note that there is always a maximal cluster
encircling all the tree, and that there is only
one outgoing line per cluster.}
\*
\hrule
\*
\endinsert
%

%
\midinsert
\*
\hrule{}
\*
\insertplot{400pt}{80pt}{
\ins{37pt}{80pt}{$\th_1$}
\ins{137pt}{80pt}{$\th_2$}
\ins{237pt}{80pt}{$\th_3$}
\ins{337pt}{80pt}{$\th_4$}
\ins{-10pt}{42pt}{$\hbox{root}$}
\ins{90pt}{42pt}{$\hbox{root}$}
\ins{190pt}{2pt}{$\hbox{root}$}
\ins{290pt}{2pt}{$\hbox{root}$}
\ins{28pt}{42pt}{$v_1$}
\ins{128pt}{42pt}{$v_1$}
\ins{228pt}{42pt}{$v_1$}
\ins{328pt}{42pt}{$v_1$}
\ins{45pt}{18pt}{$v_2$}
\ins{145pt}{18pt}{$v_2$}
\ins{245pt}{18pt}{$v_2$}
\ins{345pt}{18pt}{$v_2$}
\ins{83pt}{0pt}{$v_3$}
\ins{183pt}{40pt}{$v_3$}
\ins{283pt}{40pt}{$v_3$}
\ins{383pt}{0pt}{$v_3$}
}{fig4}
\vskip.6truecm
\didascalia{{\bf Fig.6.1.} The possible resonance families
$\FF_V(\th)$'s associated to resonance $V$'s with degree $d(V)=2$:
(1) if $\z_{w_0}^2=\z_{w_1}^1=h$, one has only one family
with four trees, $\FF_V(\th_1) = \{
\th_1$, $\ldots$, $\th_4 \}$,
(2) if $\z_{w_0}^2=H$ and $\z_{w_1}^1=h$,
one has two families with two trees, $\FF_{V_1}(\th_1) = \{ \th_1,$
$\th_2 \}$ and $\FF_{V_2}(\th_3) = \{ \th_3,$ $\th_4 \}$,
(3) if $\z_{w_0}^2=H$ and $\z_{w_1}^1=h$,
one has two families with two trees, $\FF_{V_1}(\th_1)
= \{ \th_1,$ $\th_4 \}$ and $\FF_{V_2}(\th_2) = \{ \th_2,$ $\th_3 \}$,
(4) if $\z_{w_0}^2=\z_{w_1}^1=H$, one has four one-tree families
$\FF_{V_i}(\th_i) = \{ \th_i \}$, $i=1,\ldots,4$.
Note that, unlike the labels $\z_{w_0}^2$ and $\z_{w_1}^1$,
the location of the vertices $w_0$ and $w_1'$ varies inside
the family $\FF_V(\th)$. For instance, in the case (1),
one has $w_0=v_1$ in $\th_1$ and $\th_2$, and $w_0=v_2$
in $\th_3$ and $\th_4$, and, analogously, $w_1'=v_1$
in $\th_2$ and $\th_3$, and $w_1'=v_2$ in $\th_1$ and $\th_4$.
Note also that, if $\th'\in\FF_V(\th)$, then $\FF_V(\th)=
\FF_V(\th')$: this simply means that a resonance family
can be defined with respect to any tree it contains.
For instance, in item (1), one can define the resonance
family as $\FF_V(\th_i)$, $\forall i=1,\ldots,4$.
The resonance $V$, containing the vertices
$v_1$ and $v_2$, has scale $n_V\ge n_{\l_V}+1$, if $n_{\l_V}$
is the scale of the resonant line $\l_V\=\l_{v_3}$ (which is
equal to the scale of the line entering the root): the line
connecting the vertices $v_1$ and $v_2$ is a line $v_1\ot v_2$
in $\th_1$ and $\th_2$, and is a line $v_2\ot v_1$ in $\th_3$
and $\th_4$.}
\*
\hrule
\*
\endinsert
%

\ciao